\documentclass[a4paper,11pt]{article}
\usepackage{jheppub} 
\usepackage{lineno}

\usepackage{amssymb}
\usepackage{amsmath}
\usepackage{graphicx}
\usepackage{mathtools}
\usepackage{epsfig}
\usepackage{hyperref}
\usepackage{breakurl}
\usepackage{bm}
\usepackage[usenames,dvipsnames]{color}
\usepackage{tikz}
\usepackage{todonotes}
\usepackage{cancel}
\usepackage[labelformat=simple]{subcaption}
\renewcommand\thesubfigure{(\arabic{subfigure})}
\usepackage[utf8]{inputenc}
\usepackage[section]{placeins}
\usetikzlibrary{snakes}
\usetikzlibrary{decorations}
\usetikzlibrary{trees}
\usetikzlibrary{decorations.pathmorphing}
\usetikzlibrary{decorations.markings}
\usetikzlibrary{external}
\usetikzlibrary{intersections}
\usetikzlibrary{shapes,arrows}
\usetikzlibrary{arrows.meta}
\usetikzlibrary{calc}
\usetikzlibrary{shapes.misc}
\usetikzlibrary{decorations.text}
\usetikzlibrary{backgrounds}
\usetikzlibrary{fadings}
\usetikzlibrary{tikzmark,calc,arrows,shapes,decorations.pathreplacing}
\usetikzlibrary{patterns.meta}
\makeatletter

\def\simgt{\mathrel{\lower2.5pt\vbox{\lineskip=0pt\baselineskip=0pt
           \hbox{$>$}\hbox{$\sim$}}}}
\def\simlt{\mathrel{\lower2.5pt\vbox{\lineskip=0pt\baselineskip=0pt
           \hbox{$<$}\hbox{$\sim$}}}}

\def\tree{{\rm tree}}

\makeatother

\newcommand{\be}{\begin{equation}}
\newcommand{\ee}{\end{equation}}

\newcommand{\Fig}[1]{Fig.~\ref{#1}}
\newcommand{\Eq}[1]{Eq.~\eqref{#1}}
\newcommand{\Eqs}[2]{Eqs.~\eqref{#1} and~\eqref{#2}}
\newcommand{\Sec}[1]{Sec.~\ref{#1}}

\newcommand{\App}[1]{App.~\ref{#1}}

\newcommand{\bra}[1]{\langle #1 |}
\newcommand{\ket}[1]{| #1 \rangle}

\renewcommand{\imath}{\mathrm{i}}

\newcommand{\nhatbf}{\widehat{\mathbf{n}}}
\newcommand{\khat}{\widehat{\mathbf{k}}}

\def\topbotatom#1{\hbox{\hbox to 0pt{$#1\bot$\hss}$#1\top$}} 

\DeclareMathOperator*{\sumint}{ 
\int\kern-.55cm\sum}

\usepackage{dsdshorthand}

\newcommand\wL{\mathbf{L}}
\newcommand\wS{\mathbf{S}}

\newcommand\scri{\mathscr{I}}

\newcommand\MK[1]{{\color{violet}{\bf MK: #1}}}
\newcommand\EH[1]{{\color{orange}{\bf EH: #1}}}

\usepackage[abs]{overpic}

\usetikzlibrary{positioning,decorations.pathreplacing}
\tikzset{snake it/.style={decorate, decoration=snake}}

\definecolor{energycolor}{RGB}{230,50,10}

\tikzset{
  energy/.style={->,
  energycolor,
  decoration={
      snake,
      amplitude=1pt,
      segment length=6pt,
      post length=1pt
    },
  decorate
  }
}

\input{tikz_figures.tex}

\newif\ifincludefigures
\includefigurestrue      

\begin{document}


\title{Energy Correlators in Perturbative Quantum Gravity}

\author[1]{Enrico Herrmann,}
\affiliation[1]{
	Mani L. Bhaumik Institute for Theoretical Physics,
	University of California at Los Angeles,\\
	Los Angeles, CA 90095, USA}
\emailAdd{eh10@g.ucla.edu} 

\author[2]{Murat Kologlu,}
\affiliation[2]{Department of Physics, 
    Yale University, 
    New Haven, CT 06511}
\emailAdd{murat.kologlu@yale.edu}

\author[2]{Ian Moult}
\emailAdd{ian.moult@yale.edu}

\abstract{Despite tremendous progress in our understanding of scattering amplitudes in perturbative (super-) gravity, much less is known about other asymptotic observables, such as correlation functions of detector operators. 
In this paper, we initiate the study of detector operators and their correlation functions in perturbative quantum gravity. 
Inspired by recent progress in field theory, we introduce a broad class of new asymptotic observables in gravity. 
We outline how correlation functions of detector operators can be efficiently computed from squared, state-summed amplitudes, allowing us to harness the wealth of perturbative scattering amplitude data to explore these observables.
We then compute the two-point correlator of energy detectors in the annihilation of two scalars into gravitons, in Einstein gravity minimally coupled to a massive scalar field. 
We study the kinematic limits of this correlator, finding that it is finite in the collinear limit, and exhibits a soft divergence in the back-to-back limit, as expected from the understanding of the factorization of gravitational amplitudes in the soft and collinear limits.
Our results offer a first exploration into the structure of detector operators and their correlators in perturbative quantum gravity, and we outline numerous directions for future study.
}

\maketitle

%
\section{Introduction}
\label{sec:intro}
%

Gravity famously admits only asymptotic observables. The most basic, and extensively studied, is the S-matrix. The past decades have seen tremendous advances in our ability to compute perturbative scattering amplitudes in (super-) gravity, see e.g.~\cite{Bern:1998xc,Abreu:2020lyk,Bern:2018jmv}, driven primarily by generalized unitarity \cite{Bern:1996fj,Bern:1994cg,Bern:1994zx}, and the color-kinematics duality \cite{Bern:2008qj, Bern:2010yg}. These new techniques have enabled a variety of previously impossible calculations with applications, ranging from explorations of the UV structure of gravitational theories \cite{Bern:1998ug,Bern:2015xsa,Bern:2017puu,Bern:2018jmv,Bern:2023zkg}, to precision calculations for gravitational wave experiments e.g.~\cite{Bern:2019crd,Kalin:2020fhe,Driesse:2024feo,Bjerrum-Bohr:2022blt, Kosower:2022yvp}. For a summary of further advances in the field of scattering amplitudes, see e.g.~the reviews \cite{Travaglini:2022uwo,Bern:2019prr,Elvang:2013cua}.

Due to the asymptotic nature of observables in gravity, the infrared structure of the theory plays a central role. Despite the long history of studies of the infrared structure of gravity, including the soft \cite{Weinberg:1965nx,Gross:1968in}, collinear \cite{Akhoury:2013yua,Akhoury:2011kq}, and Regge \cite{tHooft:1987vrq,Amati:1987wq,Giddings:2011xs,Giddings:2010pp} limits, there has been tremendous recent progress in the understanding of the infrared structure of gravity, including uncovering interesting relations to asymptotic symmetries and memory effects \cite{Cachazo:2014fwa,Strominger:2014pwa,Pasterski:2015tva,Strominger:2017zoo}. Infrared limits of the theory are interesting at both the classical and quantum level. At the classical level, they play an important role in observables relevant for gravitational wave detectors \cite{Gonzo:2020xza,Veneziano:2022zwh,Bini:2024rsy,Elkhidir:2024izo}. At the quantum level, they provide examples of observables that can be computed independently of any UV completion, treating gravity as an effective theory \cite{Donoghue:1995cz,Donoghue:1994dn,Donoghue:2017pgk}.

In addition to scattering amplitudes, another class of asymptotic observables in quantum field theory (QFT), are correlation functions of detector operators. These observables have a long history in collider physics, where they were initially used to characterize asymptotic energy flux in QCD \cite{Basham:1978zq,Basham:1979gh,Basham:1977iq,Basham:1978bw}. In a modern context, they were repopularized by Hofman and Maldacena \cite{Hofman:2008ar}, who highlighted them as an interesting observable in generic QFTs. The most famous example are correlation functions $\langle \Psi | \mathcal{E}(\nhatbf_1) \mathcal{E}(\nhatbf_1) \cdots \mathcal{E}(\nhatbf_k)   |\Psi \rangle$ of the average null energy (ANE) operator
\begin{align}
\label{eq:energy_detector_t_r_coords}
\mathcal{E}(\nhatbf) = \int^\infty_{0} dt\, \lim_{r\to \infty} r^{d-2} \,    \nhatbf^i T_{0i}(t,r\, \nhatbf)\,,
\end{align}
which are referred to as \emph{energy correlators}. Detector operators, and their correlators, have recently experienced a surge of interest driven primarily by two directions. On the one hand, detector operators (also known as light-ray operators) have come to play a central role in our understanding of conformal field theories (CFTs) where they provide the analytic continuation in spin of local operators \cite{Kravchuk:2018htv}, organizing them into Regge trajectories. On the other hand, recent advances have enabled calculations of higher-loop \cite{Belitsky:2013ofa, Dixon:2018qgp, Luo:2019nig,Henn:2019gkr}, and higher-point detector correlators in perturbative gauge theories \cite{Chen:2019bpb, Chicherin:2024ifn, Yan:2022cye,Yang:2022tgm} opening up numerous new applications at collider experiments \cite{Chen:2020vvp, Komiske:2022enw, Lee:2022ige, Andres:2022ovj,Yang:2023dwc,Holguin:2023bjf,Chen:2023zlx,CMS:2024mlf,CMS-PAS-HIN-23-004,ALICE:2024dfl,Tamis:2023guc}.  Despite this recent surge of activity in the study of detector correlators in QFT, they remain almost completely unexplored in gravity. There is the interesting but \emph{distinct} case of relating AdS gravitational physics to energy correlators in the holographically dual boundary CFT, starting from constraints on CFT from gravity in~\cite{Hofman:2008ar}, which paved the way for constraints on gravity from CFT~\cite{Kologlu:2019bco,Belin:2019mnx,Caron-Huot:2020adz}. The leading quantum-gravity corrections in AdS to energy correlators in the dual CFT were recently computed in \cite{Chen:2024iuv}. However, direct explorations of detector correlators \emph{in a gravitational theory itself} are rare. There are some notable exceptions in the context of classical gravity. Based on the framework developed by Kosower, Maybee, and O'Connell \cite{Kosower:2018adc} which expresses classical scattering observables in terms of weighted cross sections, or event shapes, Refs.~\cite{Herrmann:2021lqe,Herrmann:2021tct} computed the integrated energy loss and the deflection of Schwarzschild black hole scattering which has subsequently been applied to the gravitational waveform \cite{Cristofoli:2021vyo, Herderschee:2023fxh}. The pioneering work of Refs.~\cite{Gonzo:2020xza, Gonzo:2023cnv} explored detectors associated with Lorentz charges in gravity, their relation to light-ray operators, and computed their correlation function in the classical limit. In the classical limit, higher-point correlation functions of observables, such as energy detectors, factorize $\langle \cE \cE \rangle_{{\rm cl.}} = \langle \cE \rangle_{{\rm cl.}} \langle\cE\rangle_{{\rm cl.}}$ which leads to highly nontrivial relations between scattering amplitudes \cite{Cristofoli:2021jas}. In this paper, we go beyond the classical limit and explore correlation functions of gravitational detectors taking into account quantum-mechanical correlations in flat space perturbative quantum gravity.

Correlation functions of detector operators exhibit a number of properties which make them particularly appealing to explore in perturbative quantum gravity. First, they are well defined in theories that admit neither an S-matrix nor local operators, for example CFTs coupled to gravity. While in a CFT, the use of detector operators as an observable to constrain the theory \cite{Hofman:2008ar, Maldacena:2011jn, Zhiboedov:2013opa, Cordova:2017zej} is \emph{natural}, in a CFT coupled to gravity, it is \emph{essential}. Second, correlation functions of detector operators can be computed using perturbative scattering amplitudes, allowing us to re-purpose the wealth of known scattering amplitudes in (super-) gravity to generate concrete theoretical data to explore the structure of these observables. Third, their operator definition allows them to be studied using numerous techniques that are familiar for local operators, including operator product expansions \cite{Hofman:2008ar,Kologlu:2019mfz}, renormalization \cite{Caron-Huot:2022eqs}, and symmetry algebras \cite{Cordova:2018ygx, Belin:2020lsr, Korchemsky:2021htm}. This feature puts asymptotic measurements on a rigorous operator footing, which has proven extremely useful in the field theory context. 

%
\begin{figure}[ht]
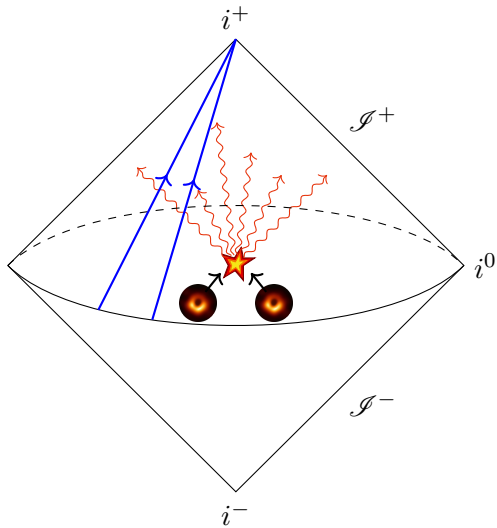

    \centering
    \penroseDiagBlackHoleAnnihilation
\caption{\label{fig:intro} 
The collision of two `black holes' producing a final state of gravitons. Correlation functions of detector operators at $\mathscr{I}^+$ (shown in blue) characterize correlations in the radiation, and provide an interesting observable in perturbative quantum gravity.}
\end{figure}
%

Driven by the intriguing properties of detector operators and their correlators in QFT, this paper explores their counterparts in perturbative quantum gravity. Since this topic holds relevance both to members of the CFT community who study detector operators, as well as to members of the amplitudes community who may investigate the perturbative structure of detector correlators, we aim to provide a concise review of what has been learned about detector operators in QFT, and why they provide a natural language in which to formulate questions about asymptotic observables in perturbative quantum gravity. As we will illustrate, many of the relevant techniques developed in perturbative QFT can be directly extended to perturbative quantum gravity. 

After reviewing salient features of light-ray and detector operators in QFT, we begin the exploration of correlation functions of detector operators in perturbation theory by generalizing a class of detector operators, which measure the ``energy-to-some-power", to perturbative quantum gravity. This broad class of detectors has played an important role in understanding detector operators and their renormalization in field theory, since they do not correspond to conserved fluxes. We anticipate that they will play an equally important role in perturbative quantum gravity. 

We present a detailed description of how correlation functions of detector operators can be computed from squared amplitudes, and review aspects of the construction of amplitudes in perturbative gravity from the double copy in Sec.~\ref{sec:amplitude_review}. In this paper we consider an explicit example of a massive scalar $\Phi$ minimally coupled Einstein-Hilbert gravity with the action 
\be
\label{eq:grav_action}
S_{{\rm EH} + \Phi}= \int d^dx \sqrt{-\mathfrak{g}} \left( 
      \frac{1}{16 \pi G_N} R 
    + \frac{1}{2} \Phi\left( \Box - m^2 \right)  \Phi
\right)\,,
\ee 
where $\mathfrak{g}_{\mu \nu}$ is the metric expanded in perturbations around flat space  $\mathfrak{g}_{\mu\nu} = \eta_{\mu\nu} + \kappa\,  h_{\mu\nu}$ ($\kappa^2= 32 \pi G_N$), $\mathfrak{g}$ is its determinant and $R$ is the corresponding Ricci-scalar. In the classical limit, this action serves as starting point for the description of binary Schwarzschild black hole encounters \cite{Cheung:2018wkq} and is relevant for QFT based approaches to precision predictions for phenomenologically relevant gravitational wave signals. These computations have attracted significant interest due to their promise of meeting the precision needs for upcoming gravitational wave detectors, see e.g.~\cite{Buonanno:2022pgc} and references therein. Technically, extracting classical physics from QFT based computations is greatly simplified by the use of powerful effective field theory techniques, see e.g.~\cite{Beneke:1997zp, Goldberger:2004jt, Bern:2019crd}.  

Here we focus on quantum gravitational aspects of detectors. For technical simplicity, we compute the two-point energy correlator in a state of gravitons produced through the \emph{annihilation} of two massive scalars. This setup is shown schematically in \Fig{fig:intro}. To our knowledge this is the first correlator of detector operators computed in quantum gravity. We obtain a simple analytic result, allowing us to study its kinematic limits, and compare them with those in perturbative gauge theory. We also compute the two-point correlator of detector operators with more general energy weights in the collinear limit. We believe that the perturbative data generated in our calculation is a concrete first step in the exploration of light-ray operators in perturbative quantum gravity.

The remainder of this paper is structured as follows: In \Sec{sec:gr_detectors}  we review basic properties of light-ray and detector operators in QFT, generalize some of them to gravity, and highlight some of the open questions we are interested in addressing in perturbative quantum gravity.  In \Sec{sec:eec_def}  we review how to compute correlation functions of detector operators from perturbative scattering amplitudes. In \Sec{sec:amplitude_review} we discuss the calculation of the relevant squared, state-summed amplitudes that will be required for our detector correlator calculation. In \Sec{sec:eec_5pt} we present the explicit calculation of the energy-energy correlator in Einstein gravity, and analyze its kinematic limits. We conclude and discuss a number of directions for future study in \Sec{sec:concl}.

\paragraph{Guide to the reader}~\\
We aimed to make the presentation of our work self contained, and accessible to experts in both CFTs and perturbative scattering amplitudes. As such, we have included reviews of both the light-ray operator perspective, as well as the construction of the relevant perturbative scattering amplitudes using the color-kinematic duality. We hope that these will be useful in bridging these communities. Readers familiar with these topics, or who are only interested in the final results, can skip the review sections without disrupting the flow of the paper. The final results are presented in a self-contained manner in \Sec{sec:eec_5pt}.

%
\section{Detector operators in QFT and gravity}
\label{sec:gr_detectors}
%

As highlighted in the introduction, perturbative calculations of correlators of detector operators have a long and illustrious history originating from the study of QCD at particle colliders. A recent advancement that has attracted greater interest from a formal audience is the formalization of general asymptotic detector \emph{operators}, within the framework of light-ray operators \cite{Kravchuk:2018htv, Caron-Huot:2022eqs}. These light-ray operators provide an organizing structure for families of operators in CFT, thus finding applications well beyond the collider physics context.

Although in this initial paper we will focus on concrete perturbative calculations of energy correlators in gravity, our ultimate goal is to develop an equivalent operator level understand of detector operators, their algebras, and their OPEs, in perturbative (and ultimately non-perturbative) quantum gravity. We therefore review in some detail this modern perspective on detector operators.  Our hope is that this section may also be useful for bridging the gap between the CFT community that studies light-ray operators, and experts in perturbative scattering amplitudes.

\subsection{Review: detector operators in QFT}
\label{sec:QFT_detectors}

We start by reviewing detector operators in QFT from the modern perspective of light-ray operators, without turning on gravitational interactions. This will serve as a guide in the pursuit of interesting questions in gravitational theories. This section is written with a CFT in mind for simplicity, but most results hold or easily generalize in a weakly-coupled QFT. For examples of studies of light-ray operators with non-vanishing $\beta$-functions, see \cite{Dixon:2019uzg, Chen:2023zzh}. The generalization to more general nonperturbative QFT is very interesting, albeit not as well understood, and we leave it outside the scope of our discussion here.

\subsubsection{Energy detector operators}

Let us begin by reviewing one of the most well-studied example of a detector operator in a QFT: an energy detector.\footnote{The very same operator is often called the ``ANE'' or ``ANEC'' operator as it measures the averaged null energy in a particular direction, and plays a central role in its namesake averaged null energy condition.} Consider a calorimeter cell in an idealized collider experiment. The calorimeter cell sits asymptotically far away at a particular angular position $\nhatbf$ on the celestial sphere $S^{d-2}$ surrounding the experiment. In this idealized context, we imagine it is placed there for all time, collecting radiation escaping to asymptotic infinity along its angular direction, and recording its energy. One way to define it is as the following limit~\cite{Korchemsky:1999kt, Hofman:2008ar, Belitsky:2013ofa, Belitsky:2013bja, Belitsky:2013xxa}
\begin{align}
\label{eq:Energy_detector_definition_in_terms_of_stress_tensor}
\cE(\nhatbf) \propto \int^{\infty}_0 dt \lim_{r\to \infty} r^{d-2}\, \nhatbf^i \, T_{0i} (t, r\, \nhatbf),
\end{align}
where the energy-momentum tensor $T_{\mu\nu}$ is understood to be normal-ordered and $t$ represents the working time of the detector. Note that the energy detector annihilates the vacuum $\cE|\Omega\> =0$. In this form, the action of the energy flow operator on an external on-shell state of $m$ massless particles, represented by $\ket{X} = \ket{k_1,\ldots,k_m}$ is given by
\begin{align}
\label{eq:energy_detector_on_asymptotic_states}
    \cE(\nhatbf) \ket{X} = \sum^m_{i=1} E_i\, \delta^{d-2}(\Omega_{\khat_i} - \Omega_{\nhatbf}) \ket{X}\,.
\end{align}
The four-momenta $k^\mu_i = (E_i,\textbf{k}_i)$ are related to the unit directions $\khat_i \equiv \frac{\textbf{k}_i}{|\textbf{k}_i|}= \frac{\textbf{k}_i}{E_i}$ of the momentum.  The unit vector $\nhatbf$ ($\nhatbf^2=1$) represents the location of the detector on the celestial sphere $S^{d-2}$. See Appendix \ref{app:coordinate_systems} for details on our notation, conventions, and some coordinate choices. The action of the energy-flow operator on on-shell states will become relevant in our discussion of the relationship between correlation functions of detector operators and weighted cross sections involving on-shell scattering amplitudes. 

For an illustrative example, one can compute the energy detector in free massless scalar theory (see e.g.~\cite{Bauer:2008dt,Caron-Huot:2022eqs}) in terms of the mode expansion of the scalar field $\f$,
\be
\label{eq:scalar_mode_expansion}
	\f(x)=\int \hat{d}^dp \, \hat{\delta}^+(p^2)\p{a^\dagger(p) e^{-ip\cdot x}+a(p)e^{ip\cdot x}}\, .
\ee 
Following~\cite{Caron-Huot:2022eqs}, we use the relativistic normalization where $a(p)$ is a Lorentz scalar, with commutator $[a(p),a^\dagger(p')]= 2p^0  \hat{\delta}^{d-1}(\bp-\bp')$. We find it convenient to define the shorthand notation for the measure and various delta functions
\begin{align}
\hat{d}^dx  & \equiv \frac{d^d x}{(2\pi)^d}\,,
\qquad 
\hat{\delta}^{d}(x) \equiv (2\pi)^d \delta^{d}(x)\,,
\qquad 
\hat{\delta}^+(p^2) = \theta(p^0) \hat{\delta}(p^2)\, ,
\end{align} 
where the Heavyside-theta function $\theta(p^0)$ ensures the propagation of positive energy quanta. One can insert this expansion into the stress-energy tensor,\footnote{See Ref.~\cite{Gonzo:2020xza} for a generalization to polynomial non-derivative interactions between the scalars.}
\begin{align}
T_{\mu \nu} = (\partial_\mu \f)(\partial_\nu \f) - \frac{1}{2} \eta_{\mu \nu} (\partial_{\alpha} \f)(\partial^\alpha \f)\,,
\end{align}
and evaluate~\eqref{eq:Energy_detector_definition_in_terms_of_stress_tensor} via a stationary-phase method (see e.g.~\cite{Bauer:2008dt,Gonzo:2020xza}). One finds\footnote{
We use the shorthand $a(\mathbf{{p}})\equiv a(p \!=\!(|\mathbf{{p}}|,\mathbf{{p}}))$ for creation and annihilation operators with $(d-1)$-dimensional vector arguments.}
\be
\label{eq:energy_detector_scalar_modes}
	\cE(\nhatbf)\propto\int_0^\infty dE\, E^{d-2} a^\dagger(E \nhatbf)\, a(E \nhatbf)\,.
\ee
For later purposes, it is much more convenient to write the energy detector operator in a covariant form which highlights the fact that an energy detector (capturing massless radiation) in a QFT is given by a null integral of the stress tensor $T_{\mu\nu}$ of the QFT along future null infinity $\scri^+$~\cite{Hofman:2008ar,Belitsky:2013bja}. Following the discussion in appendix B of \cite{Gonzo:2020xza}, we can identify a detector by its set of coordinates $x^\mu = (t, r \nhatbf)$. This can be covariantized by choosing a basis of null vectors $z^\mu = (1,\nhatbf)$, $\bar{z}^\mu = (1,-\nhatbf)$, together with a basis of transverse spacelike vectors $m^\mu_A$ which satisfy,
\begin{align}
\label{eq:covariant_basis_def}
\hspace{-.5cm}
z^2 = \bar{z}^2 = m_A {\cdot} z = m_A {\cdot} \bar{z} = 0\,, 
\quad
z\cdot \bar{z} = 2\,,
\quad
m_A{\cdot} m_B = -\delta_{AB}\,,
\quad
\forall A,B\in \{1,d{-}2\}\,.
\hspace{-.4cm}
\end{align}
The completeness relation is 
\begin{align}
\eta_{\mu \nu} = \frac{z_\mu \bar{z}_\nu + \bar{z}_\mu z_\nu  }{z\cdot \bar{z}} + \frac{(m_A)_\mu (m_A)_\nu}{m_A \cdot m_A} \, .
\end{align}
In this basis, a generic point in spacetime is written as $x^\mu = \frac{\bar{\alpha}}{4} z^\mu + \frac{\alpha}{4} \bar{z}^\mu + x_A m^\mu_A$, with
\begin{align}
\frac{\alpha}{4} = \frac{t-r}{2} = \frac{x\cdot z}{z\cdot \bar{z}}\,, 
\quad
\frac{\bar{\alpha}}{4} = \frac{t+r}{2} = \frac{x\cdot \bar{z}}{z\cdot \bar{z}}\,, 
\quad
x_A = \frac{x\cdot m_A}{m_A\cdot m_A}\,,
\end{align}
and in particular, the detector is located at $x^\mu = (t,r\nhatbf) = \frac{\bar{\alpha}}{4} z^\mu + \frac{\alpha}{4} \bar{z}^\mu$. With this, we can write the energy detectors as\footnote{If there are massive excitations that one wants to capture by the detector, one needs to be more careful in the order of limits \cite{Belitsky:2013bja}. }
\be
\label{eq:energy_detector_definition}
\cE(z)  = 2 \int_{-\infty}^\infty d\a \, T_{\a\a}(\a,z),
\ee 
where the stress tensor is evaluated at $\scri^+$,   
\be
T_{\a\a}(\a,z) = \lim_{L\to \infty} L^{d-2} \, 
(\partial_\a x^\mu) (\partial_\a x^\nu) T_{\mu\nu}(x+Lz),
\ee 
and the null coordinate $\a = 2x\cdot z$ is twice the retarded time at $\scri^+$. These definitions are now covariant, and we can recover~\eqref{eq:Energy_detector_definition_in_terms_of_stress_tensor} by picking the null vector $z^\mu$ to point in the direction of $\nhatbf$,
\be
\cE(z = (1,\nhatbf)) = \cE(\nhatbf)\, .
\ee
Expectation values of energy detector operators,
\begin{align}
\langle \Psi | \cE(\nhatbf_1)\cE(\nhatbf_2)\cdots \cE(\nhatbf_k) |\Psi \rangle\,,
\end{align}
are referred to as energy correlators (or for more general detectors, detector correlators). They were originally introduced in the context of states produced in $e^+e^-$ collisions in \cite{Basham:1978zq,Basham:1979gh,Basham:1977iq,Basham:1978bw}. 
As we will explain in detail in \Sec{sec:eec_def}, their calculation can be  related to the calculation of weighted cross-sections, which can be efficiently computed using scattering amplitudes.  However, combining perturbative techniques for the computation of detector correlators with a modern understanding of light-ray operators allows for a unified language in which both established and more exotic detector operators can be studied, which is why we have presented them in this more general context.

\subsubsection{Light-ray operator review}

%
\begin{figure}[h!]
\centering
\begin{tikzpicture}
\draw [gray] (0,-0.5) -- (0,4.5);
\draw [gray] (4,-0.5) -- (4,4.5);
\draw [gray,->] (0,3.1) -- (0,3.3);
\draw [gray,->] (0,3.18) -- (0,3.38);
\draw [gray,->] (4,3.1) -- (4,3.3);
\draw [gray,->] (4,3.18) -- (4,3.38);
\draw [dashed,gray] (0,2) -- (2,4) -- (4,2) -- (2,0) -- (0,2);
\draw [dashed,gray] (1.5,-0.5) -- (2,0) -- (2.5,-0.5);
\draw [dashed,gray] (1.5,4.5) -- (2,4) -- (2.5,4.5);
\draw [->,thick,blue] (0,2) -- (1,3);
\draw [thick,blue] (1,3) -- (2,4);
\draw[black,fill=blue] (0,2) circle (2pt);
\node [right] at (0.5,2.4) {$\cD$};
\node [above] at (0.8,2.95) {$\mathscr{I}^+$};
\node [below] at (0.8,0.95) {$\mathscr{I}^-$};
\node [left] at (0,2) {$i^0$};
\node [above] at (2.07,4.1) {$i^+$};
\node [below] at (2.07,-0.05) {$i^-$};
\end{tikzpicture}
\caption{\label{fig:detector_frame}
An illustration of the detector frame, where a detector $\cD$ lies along future null infinity $\mathscr{I}^+$ \cite{Caron-Huot:2022eqs}. In a CFT, placing the detector at infinity can be achieved by the choice of a suitable conformal frame, hence the terminology. However, as one moves from CFT to generic QFT or to gravity, this is not a choice of frame anymore, but a physical choice of what observable to study. In such general settings, light-ray operators placed at $\scri^+$ versus other null cones will result in observables not simply related by spacetime symmetry. In a CFT, the detector $\cD$ transforms like a primary operator at spatial infinity $i^0$ (the blue point) if it is invariant under Minkowski translation generators, and this statement is exact in perturbation theory. More generally, translation invariant detectors remain so under perturbation theory in arbitrary Lorentz-invariant theories, providing a useful simplification to the computation of observables. Similar to perturbative QFT, we propose to start with detectors placed at $\scri^+$ in perturbative quantum gravity and track them as interactions are turned on.}
\end{figure}
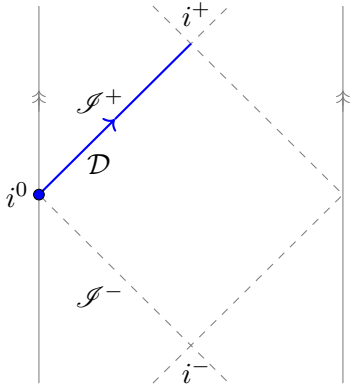
%

The energy detector \eqref{eq:energy_detector_definition} is but one example of a much more general class of operators referred to as light-ray operators \cite{Kravchuk:2018htv}. Light-ray operators are operators that are localized to the future null-cone of a point. Picking that point to be spacelike infinity is equivalent to picking the null cone $\scri^+$, as depicted in figure~\ref{fig:detector_frame}. This choice of anchoring the light-ray operator at infinity gives it the interpretation of a detector, as in~\eqref{eq:energy_detector_definition}. We will refer to this choice as the ``detector frame''. Since the language of light-ray operators will be crucial to generalize from the energy detector to more general detector operators, we now provide a review of the important properties of light-ray operators.

In order to discuss light-ray operators, it will be very useful to introduce null polarization vectors $z^\mu\in\R^{d-1,1}$ to keep track of Lorentz transformation properties of operators. Consider an operator $\cO_{\mu_1\cdots\mu_J}(x)$ in a spin $J$ traceless-symmetric representation of the Lorentz group. In index-free notation \cite{Costa:2011dw,Kravchuk:2018htv}, an equivalent object to study is 
\be \label{eq:definition contraction with null polarization}
\cO(x,z) \equiv z^{\mu_1} \cdots z^{\mu_J} \cO_{\mu_1\cdots\mu_J}(x)\,.
\ee 
The polarization vector is null, future-pointing and projective,
\be
z^2=0,\qquad z^0>0, \qquad z\sim \l z \quad (\l>0),
\ee 
which amounts to a choice of direction on the future null cone of $x$. 

The operator $\cO_{\mu_1\cdots\mu_J}(x)$ can be recovered from $\cO(x,z)$ by the action of certain differential operators in $z$. The upside of this definition is that Lorentz spin has simply become the homogeneity degree of the operator in $z$,
\be
\cO(x,\l z) = \l^J \cO(x,\l z).
\ee 
Readers may find this very similar to the familiar relationship between scaling dimension $\De$ and transformation properties under dilatations in the embedding space formalism \cite{Costa:2011mg}. Indeed, the null polarization $z$ could be thought of as the $d$ dimensional embedding space coordinates for the $d-2$ dimensional transverse space of the null cone anchored on $x$. Let us very briefly review the embedding space formalism in its typical incarnation. Embedding space linearizes the action of the conformal group; an operator $\cO$ of dimension $\De$ can be uplifted to embedding space as $\cO(X)$, where $X\in \R^{d,2}$ is on the projective null cone, 
\be
X^2=0,\qquad X^0>0, \qquad X\sim \rho X \quad (\rho>0),
\ee 
The downlift in Minkowski space $\cO(x)$ can be recovered by going to the patch given by $X=(X^+,X^-,X^\mu) = (1,x^2,x^\mu)$. 
In embedding space, the operator $\cO(X)$ has homogeneity determined by the scaling dimension $\De$,
\be
\cO(\rho X) = \rho^{-\De} \cO(X).
\ee 
Of course, the scaling dimension is typically some continuous real number. 

Analogously, the definition~\eqref{eq:definition contraction with null polarization} allows for a discussion of representations of the Lorentz group with continuous (even complex) spin. This is indeed the case for light-ray operator representations.\footnote{The Lorentz group $SO(d-1,1)$ has a noncompact generator corresponding to the $SO(1,1)$ boost symmetry, which indeed allows for representations with complex weights $J_L$.} 
We will denote a generic light-ray operator as $\mathbb{O}(x,z)$, or as $\mathbb{O}(X,Z)$ in embedding space (for further details, see~\cite{Kravchuk:2018htv}).\footnote{There may be additional indices (or polarizations) for operators in more complicated representations of the Lorentz group with multiple rows in their respective Young tableaux, as explained in~\cite{Chang:2020qpj}. These correspond to nontrivial representations $\l$ of the Euclidean rotation group $SO(d-2)$ of the celestial sphere. We will stick to the simple case of $\l$ being the trivial representation in our discussion here. We will sometimes abuse terminology and call such representations ``traceless-symmetric'' in analogy with their finite dimensional $(J\in\Z_{+})$ counterparts.} In order to distinguish the quantum numbers of light-ray operators from their local counterparts, we will denote the dilatation and Lorentz boost eigenvalues by $(\De_L,J_L)$, so
\be
\mathbb{O}(\r X,\l Z) = \r^{-\De_L}\l^{J_L} \mathbb{O}( X, Z)\, .
\ee
For the majority of our purposes in this present article, it will be sufficient to keep track of the transformation property
\be
\mathbb{O}(x,\l z) = \l^{J_L} \mathbb{O}( x, z)\, .
\ee
The simplest class of light-ray operators is the one defined as a null integral of local operators, like the energy detector. It is useful to grade operators---detector or local---by symmetry properties. In order to do so, it is advantageous to introduce the so-called \emph{light transform}~\cite{Kravchuk:2018htv}. The light transform $\wL[\cO](x,z)$ is a particular integral of $\cO$ along the null direction $z$ starting from the point $x$. We will not need its precise definition here, just that it turns a local operator $\cO_{\De,J}$ of scaling dimension $\De$ and Lorentz spin $J$ into the light-ray operator $\wL[\cO_{\De,J}]$ of definite scaling dimension $\De_L$ and Lorentz spin $J_L$, where
\be
\De_L = 1-J\, , \qquad J_L = 1-\De\, .
\ee
In fact, the light transform is a conformally-invariant integral transform that maps conformal primaries to conformal primaries. From the point of view of detector operators in the detector frame, the condition for a detector to be primary is practically useful as well. Since the detector is inserted at spatial infinity, the condition to be primary is that it is annihilated by the translation generators of the Minkowski patch, $[P^\mu,\cD]=0$, which act as null translations along $\scri^+$. Therefore, primary detectors are also those that preserve momentum, i.e. diagonal when acting on momentum eigenstates of the theory.

The energy detector is obtained by light transforming the stress tensor $T(x,z)$ (with quantum numbers $\Delta =d, J=2$) and going to the detector frame 
\be
\cE(\nhatbf) = 2\wL[T](\infty,z)\big|_{z=(1,\nhatbf)}\, .
\ee 
The light transform implements a generalization of the null integral in~\eqref{eq:energy_detector_definition} to integrals of arbitrary local operators anchored along arbitrary null cones. In the detector frame, the polarization vector $z$ encodes the angular position of the detector. Note that the scaling dimension of $\wL[T](x,z)$ is $\De_L=-1$, and its Lorentz spin is $J_L=1-d$. Placing the operator at infinity, $x=\infty$, flips the dilatation eigenvalue to $\De_L=1$, giving the energy detector units of energy, as expected. Charge detectors $\cQ$ measuring some charge $Q$ can similarly be obtained by light transforming the corresponding current $J_\mu$.\footnote{Depending on the theory, this detector may or may not be well defined.} 

One might be interested in defining more general detector operators corresponding to more general measurements, such as energy-to-some-power measurements $E^n$, or charge weighted energy measurements $E\times Q$. These detectors are strictly speaking only well-defined in free theory, and require renormalization once interactions are turned on. However, to even begin to discuss them we need to go outside the class of simple light-ray operators that are null integrals of local operators.\footnote{There appears to be no canonical distinction between light-ray operators supported on null lines versus null cones, and they are continuously connected as one varies $J_L$ in a trajectory.} In fact, even the above discussed examples of generalized detectors are but a small set of all possible detectors one might consider. This leads us to discuss more general light-ray operators and Regge trajectories.

\subsubsection{The space of detector operators and Regge trajectories}
\label{sec:cft regge trajectories}

In light of the above discussion of sample detectors in QFT, it is a natural question to ask about the space of \emph{all} allowed asymptotic measurements in a given QFT. One of the insights of Ref.~\cite{Caron-Huot:2022eqs} was the proposal that \emph{any} light-ray operator could be thought of as an asymptotic detector operator $\cD$ by simply placing it at $\scri^+$. An interesting and tractable case is that of CFT. In CFT, light-ray operators are intimately connected to the spectrum of local operators, where they provide an analytic continuation of local operators~\cite{Caron-Huot:2017vep,Kravchuk:2018htv}, forming analytic curves of Regge trajectories
\be
f_i(\De_L,J_L)=0 .
\ee
The rigidity of the spectrum of light-ray operators means that not any asymptotic measurement is possible in a given QFT. Rather, the spectrum dictates what the allowed asymptotic measurements are, just as the spectrum of local operators dictates what the allowed local excitations or measurements are.

\begin{figure}[ht!]
	\begin{center}
		\includegraphics[scale=.8]{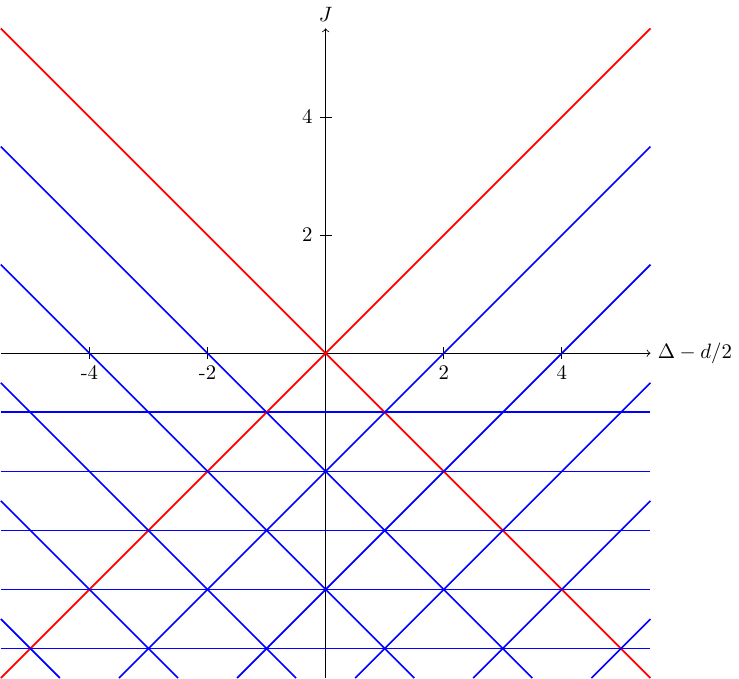}
		\caption{The perturbative structure of the ($\mathbb{Z}_2$-even, parity-even, traceless-symmetric) Regge trajectories in Wilson-Fisher theory from~\cite{Caron-Huot:2022eqs}. The red lines are the leading ``twist-two'' Regge trajectory of $E^{J-1}$ detectors and their shadows. The blue diagonal lines are the next higher twist trajectories and their shadows. The horizontal blue lines at $J=-1,-2,\cdots$ are the more exotic ``Reggeized scalar'' trajectories. The quantum numbers of the corresponding detector operators placed at $\scri^+$ are given by $(\De_L,J_L) = (1-J,1-\De)$. Once interactions are turned on, the trajectories acquire anomalous dimensions and are no longer straight lines. Furthermore, intersections are resolved and the trajectories combine into higher-degree complex curves. 
		\label{fig:perturbativelines}
		}
	\end{center}
\end{figure}

A convenient visualization of the spectrum of light-ray operators is the Chew-Frautschi plot. As an example, we reproduce the plot of the spectrum of detector operators in perturbative Wilson-Fisher theory from~\cite{Caron-Huot:2022eqs} in Fig.~\ref{fig:perturbativelines}. The spectrum consists of several interesting components. The easiest to understand are the diagonal $45^\circ$ trajectories of constant \emph{twist}, $\tau \equiv \De-J= 2\De_\f+2n$, with integer $n\geq0$, which are the analytic continuations of operators of the schematic type ``$\cO_{\Delta,J}(x)=[\f \partial^{2n}\partial^{\mu_1}\cdots \partial^{\mu_J}\f](x)$''. 
The corresponding light-ray operators fill out the diagonal trajectories 
\be
\label{eq:diagonal 45 degree trajectories}
\De_L - J_L =  2\De_\f + 2n\,.
\ee 
For $n=0$, these are the familiar ``twist-2'' 
trajectories, where the nomenclature is borrowed from $d=4$, where the twist $\tau = 2\De_\f=d-2 \stackrel{d=4}{\longrightarrow}2$ is two. 

Once we have a trajectory of detectors of constant twist, we can immediately obtain a second, ``shadow'' trajectory by a simple integral transform, $\wS_J$,\footnote{The subscript $J$ serves to distinguish the spin-shadow $\wS_J$ from the more familiar (dimension) shadow $\wS_\De$ which is an analogous integral over $x$ and implements the $\De\leftrightarrow d-\De$ shadow symmetry of local operators in a CFT leading to the $-45^\circ$ trajectories in Fig.~\ref{fig:perturbativelines}. In fact, the left-right symmetry of the Chew-Frautschi plot can be understood as a consequence of $\wS_J$ acting on light-ray operators, or as $\wS_\De$ acting on local operators.} called the ``spin shadow''~\cite{Kravchuk:2018htv}:
\be 
\label{eq:spin_shadow_def}
	\wS_J[\mathbb{O}](x,z)=\int D^{d-2}z' (2z\.z')^{2-d-J_L}\mathbb{O}(x,z').
\ee 
This integral over the future null cone is equivalent to the usual shadow transform~\cite{Ferrara:1972uq,Simmons-Duffin:2012juh} on the Euclidean celestial sphere $S^{d-2}$, and the measure is $D^{d-2}z=2 d^dz\, \de(z^2)\theta(z^0)/\mathrm{vol}(\R)$. The spin-shadow transform $\wS_J$ and therefore the existence of shadow trajectories is a consequence of Lorentz symmetry alone. Given a primary light-ray operator $\mathbb{O}$ with quantum numbers $(\De_L,J_L)$, $\wS_J[\mathbb{O}]$ is a primary operator with quantum numbers $(\De_L,2-d-J_L)$, implementing the left-right symmetry of the Chew-Frautschi plot, see e.g.~Fig.~\ref{fig:perturbativelines}. For the diagonal $45^\circ$ trajectories~\eqref{eq:diagonal 45 degree trajectories}, the shadow trajectories are given by the $-45^\circ$ curves
\be
   \De_L = (2-d-J_L) + 2\De_\f + 2n\ .
\ee
For detector operators in the detector frame, the spin-shadow transform corresponds to a Lorentz-covariant angular smearing of the detector on the celestial sphere, and is a true symmetry of detector operators in any Lorentz-invariant theory. 

Finally, there are the more exotic horizontal trajectories of Reggeized scalars, at $J=-1,-2,\cdots$. To give a feel for these trajectories, we review a schematic way to construct the trajectory at $J=-1$ by the following procedure. In free massless scalar theory, take two scalar detectors $\wL[\f^2](\infty,z_1)$ and $\wL[\f^2](\infty,z_2)$ inserted at generic positions on the celestial sphere. These have dimensions $\De_{L1}=\De_{L2} = 1$ and $J_{L1}=J_{L2}=-1$. In perturbation theory, we can take a product of these operators and normal order them to get rid of any singular behavior and define a physical detector
\be
\label{eq:horizontal scalars construction composite}
    :\wL[\f^2](\infty,z_1) \wL[\f^2](\infty,z_2):\, .
\ee
Since they are placed at the same point $x=\infty$, this object has dimension $\De_L = \De_{L1}+\De_{L2} = 2$, corresponding to $J=-1$. Furthermore, it is in a product representation $J_{L1}\times J_{L2}$ of the Lorentz group. We can diagonalize by projecting to irreps of the Lorentz group via a suitable kernel (which essentially is a Euclidean three-point function on the $d-2$-dimensional celestial sphere), obtaining
\be
\label{eq:J=1 trajectory in irreps}
\cH_{J_L}(z) = \int D^{d-2}z_1 D^{d-2}z_2\, \cK_{J_L}(z_1,z_2;z) :\wL[\f^2](z_1) \wL[\f^2](z_2):\, .
\ee
This operator has support at all $J_L$, filling out an entire horizontal line in the Chew-Frautschi plot at $J=1$. The unabridged story in Wilson-Fisher is more complicated, as one can construct the horizontal trajectories from products of many constituent detectors, and one needs to analyze which ones appear in correlation functions with other operators and gets renormalized. We refer to \cite{Caron-Huot:2022eqs} for further details. The construction is analogous to the construction of the BFKL trajectories of ``reggeized gluons'' in gauge theories~\cite{Kuraev:1977fs,Balitsky:1978ic,Mueller:1994jq,Balitsky:1995ub,Caron-Huot:2013fea}.

As a concrete example, let us take a closer look at the simple yet interesting generalization of the energy detector to ``energy-to-some-power'' detectors, $\cE_J$, measuring $E^{J-1}$. These operators are nothing but the leading Regge trajectory. In a free massless scalar theory, we can construct them as generalizations of the energy operator in \Eq{eq:energy_detector_scalar_modes} as
\be
\label{eq:EJnon-cov}
	\cE_J(\nhatbf )\propto\int_0^\infty dE\, E^{J+d-4} 
        a^\dagger(E \nhatbf)\, a(E\nhatbf),
\ee
using the mode expansion of the field $\f$ in terms of creation and annihilation operators, \Eq{eq:scalar_mode_expansion}. The resulting detector acts on asymptotic on-shell states $|X\> = |k_1,\cdots, k_n\>$ with $k_i = E_i(1,\khat_i)$ as
\be
\cE_J(\nhatbf) |X\> 
=\sum_{i\in X}  E_i^{J-1}\delta^{d-2}(\Omega_{\nhatbf} - \Omega_{\khat_i}) |X\>\, .
\ee
We can upgrade this to the covariant expression\footnote{This can be shown using the identity $\de(z^2) = \frac{1}{2z^0} \de(z^0-|\mathbf{z}|)$ which gives $\frac{\de^d(k-\b z)}{\de(z^2)}= \frac{2z^0}{\b} \de^{d-1}(\mathbf{k}-\b \mathbf{z})$.}
\be
\label{eq:E_J_detector_QFT_action_on_state}
\cE_J(z) |X\> 
= \sum_{i\in X} \int_0^\infty d\b \b^{J+d-2} \frac{\delta^{d}(k_i-\b z)}{2\de(z^2)} |X\>\, .
\ee
Strictly speaking, when $J\ne 2$ these detectors only make sense in free theory, and require renormalization when interactions are turned on. However, they are well-understood to form the leading Regge trajectory in an interacting CFT \cite{Caron-Huot:2022eqs}. It is more prudent to label them $\cD_{J_L}$, where $J_L$ is the Lorentz boost quantum number. In perturbation theory, Lorentz symmetry is preserved so $J_L$ is exact, while dilatations are broken and the energy dimension $J-1$ will get corrected. As always, we should make use of symmetries and organize our observables accordingly. In the free scalar theory, the $\cD_{J_L}$ detectors can equivalently be constructed as \cite{Caron-Huot:2022eqs}
\be 
\label{eq:definition of free scalar twist two trajectory}
\cD_{J_L}(z) = \frac{1}{C_{J_L}} \int_{-\infty}^\infty d\a_1 d\a_2 \, \psi_{J_L}(\a_1,\a_2) :\f(\a_1,z)\f(\a_2,z):\,.
\ee
We have made use of the scalar field at infinity
\be
\f(\a,z) = \lim_{L\to\infty} L^{\De_\f} \f(x+L z).
\ee
The ``wavefunction'' $\psi_{J_L}$ is fixed by symmetries if one requires a primary\footnote{Recall that since the detectors are inserted at infinity, primary operators are those annihilated by the momenta generators in the Minkowski patch, $[P^\mu,\cD]=0$. Momenta generators act as translations on $\scri^+$, so translation invariant detectors are primary, and thus a translation invariant kernel results in a primary detector. Since momenta generators are exact in perturbation theory, such detectors remain primary even when interactions are turned on. Said another way, detectors that carry zero momentum continue to carry zero momentum when interactions are turned on. More details can be found in~\cite{Caron-Huot:2022eqs}.} detector with definite Lorentz spin $J_L$,
\be
\psi_{J_L}(\a_1,\a_2) 
=  |\a_1-\a_2|^{2(\De_\f-1)+J_L}\, .
\ee 
The constant $C_{J_L}$ is a convenient normalization factor, 
\be
C_{J_L} &= 2^{J_L+d-1} \pi \sin\p{\pi \tfrac{J_L+2\De_\f}{2}} \G(2\De_\f+J_L-1),
\label{eq:constantfortwisttwo}
\ee
chosen to make the action on states as simple as possible and consistent with the $E^{J-1}$ detector interpretation. The expression~\eqref{eq:definition of free scalar twist two trajectory} is analytic in the Lorentz spin $J_L\in \C$, and defines a complex curve of light-ray operators labeled by $J_L$. This curve satisfies the familiar ``twist-2'' relation
\be
\De_L = 2\De_\f + J_L\, ,
\ee as it should as the twist-2 Regge trajectory. 
The relation between $\cD_{J_L}(z)$ and $\cE_{J}(\nhatbf)$ is 
\be
\cE_J(\nhatbf) = \cD_{J_L=3-d-J}(z)\Big|_{z=(1,\nhatbf)}\,.
\ee
In particular, the energy detector ($J=2$) is given by 
\be
\cE(z) = \cD_{J_L=1-d}(z) \,.
\ee
We can obtain the higher-twist Regge trajectories of detectors similarly, by using multiple asymptotic $\phi$ fields (or their derivatives). For example,
\be
\label{eq:multi phi detectors}
\cD_\psi(z) = \int_{-\infty}^\infty d\a_i \, \psi_{J_L}(\a_1,\cdots,\a_m) : \f(\a_1,z) \cdots \f(\a_m,z): \, ,
\ee 
are primary detectors satisfying $\De_L = J_L + m \De_\f \,,$ for any 1d translation-invariant and homogeneous kernel $\psi_{J_L}$. For even $m$, such detectors fall within the other $45^\circ$ lines to the right of the leading trajectory in the Chew-Frautschi plot Fig.~\ref{fig:perturbativelines}, which are highly degenerate in free theory. For odd $m$, these operators are $\Z_2$-odd, and fill out the analogous Regge trajectories in the corresponding Chew-Frautschi plot for that symmetry sector. The detectors with derivatives of $\f$ can be constructed 
via weight-shifting operators~\cite{Karateev:2017jgd}. For example, detectors corresponding to the analytic continuation of the $[\f \partial^{2n}\partial^{\mu_1}\cdots \partial^{\mu_J}\f]$ operators are given by~\cite{Kravchuk:2018htv}
\be
\label{eq:derivative higher twist trajectories}
\hspace{-.5cm}
\cD_{J_L}^{(n)}(z) = \int_{-\infty}^\infty d\a_1 d\a_2 \, \psi_{J_L}(\a_1,\a_2) : (D_{M_1}\cdots D_{M_n}\f)(\a_1,z)(D^{M_1}\cdots D^{M_n}\f)(\a_2,z):\,,
\hspace{-.4cm}
\ee 
where $D_M$ is the Thomas/Todorov differential operator which increases dimension by 1 and carries a vector embedding space index $M$.\footnote{We slightly abuse notation in the formula~\eqref{eq:derivative higher twist trajectories}; the derivatives $D_{M_1}\cdots D_{M_n}\f(X,Z)$ are evaluated in embedding space, then the expression is placed at $\scri^+$. 
} These operators have quantum numbers $\De_L = J_L + 2\De_\f+2n$, matching the trajectory~\eqref{eq:diagonal 45 degree trajectories}. Together, operators of the type~\eqref{eq:multi phi detectors} and~\eqref{eq:derivative higher twist trajectories} comprise the higher-twist trajectories. For recent studies of higher-twist light-ray operators, see \cite{Henriksson:2023cnh,Homrich:2024nwc,Ekhammar:2024neh}.

As discussed abstractly around \Eq{eq:spin_shadow_def}, for each primary detector $\cD$, there is a corresponding shadow detector $\widetilde \cD$ given by the smearing on the celestial sphere~\eqref{eq:spin_shadow_def}
\be
\widetilde{\cD}_{J_L} (z) = \wS_J[\cD_{2-d-J_L}](z)\,,
\ee
leading to the $-45^\circ$ lines in the Chew-Frautschi plot in Fig.~\ref{fig:perturbativelines}.

Once one moves away from free theory by turning on interactions, $E^{J-1}$ is not a meaningful quantum number that can be measured since interactions preserve energy but not powers of energy. This manifests as IR divergences in perturbative calculations of detector observables which need to be renormalized. The renormalized detectors match the interacting Regge trajectories with corresponding anomalous dimensions,
\be
\De_L = 2\De_\f + J_L + \g(J_L)\,.
\ee
The Regge trajectories of the interacting theory are now interpreted as the allowed detector operators, now explicitly constructed through perturbation theory.

In analogy with local operators, detector operators also exhibit an OPE \cite{Hofman:2008ar,Kologlu:2019mfz}, and interesting symmetry algebras \cite{Cordova:2018ygx,Korchemsky:2021htm}. For conciseness, we will not review these features here, but note that both should have interesting analogues in flat-space gravity that warrant further explorations.

In summary, we re-emphasize that light-ray operators provide a natural language to unify asymptotic measurements and the inherent operator structure of QFTs. We now turn to an exploration of detectors operators in perturbative quantum gravity in asymptotically flat spacetime.

\subsection{Initiating the investigation of detector operators in quantum gravity}
\label{subsec:GR_detectors}

In gravitational theories, there are no local gauge (diffeomorphism) invariant operators, as spacetime itself is fluctuating and $x$ in $\cO(x)$ no longer makes sense. The exception is operators placed at asymptotic infinity, where diffeomorphisms act trivially (except for large diffeomorphisms, which are physical). For this reason one typically studies scattering amplitudes of asymptotic states, which have been extensively explored and found to exhibit many remarkable properties. However, expectation values of light-ray operators inserted at $\scri^+$ are also physical observables. Furthermore, their operator definition makes them particularly appealing for better understanding the structure of quantum gravity. This warrants the study and exploration of general detector operators in gravitational theories. Ultimately, we are interested in answering the following questions: 
\begin{enumerate}
	\item What is the spectrum of asymptotic detector operators in quantum gravity? Do horizontal trajectories exist, and if so, do they encode information about the Regge limit?
	\item What is the dynamics of asymptotic states and detector operators in quantum gravity? In particular, what is the detector OPE in quantum gravity?
    \item What is the algebra of detector operators in quantum gravity and how does it relate to asymptotic symmetries?
    \item What is the mathematical structure of multi-point correlators of detector operators in perturbative (super-)  gravity?
\end{enumerate}	
As emphasized in the introduction, these are all questions associated with the infrared, and can therefore be answered without an understanding of a particular UV completion of quantum gravity. A first step in the exploration of detector operators in gravity was taken in \cite{Gonzo:2020xza}, which constructed the light-ray operators associated with the charges of the Lorentz group, and studied their symmetry algebra. Ref.~\cite{Gonzo:2020xza} also computed multi-point correlators of detector operators in the classical limit, where they factorize.

In this paper we extend the studies of \cite{Gonzo:2020xza}, both by broadening the space of detectors, and by calculating the quantum corrections to multi-point correlators of detector operators. In particular, we are going to study energy and ``energy-to-some-power'' detectors in asymptotically-flat Einstein gravity, viewed as an effective field theory. The generalization to ``energy-to-some-power'' detectors is important, since these are not conserved charges of the theory, and will ultimately allow us to study the renormalization of detector operators in perturbative quantum gravity. Mirroring our discussion of detectors in QFT (see Sec.~\ref{sec:QFT_detectors}), we will briefly outline the definitions and key properties of such operators in the gravitational context. In section~\ref{sec:eec_5pt}, we compute the two-point correlations of energy detectors (for generic detector positions) to leading nontrivial order in the gravitational coupling $G_N$.  We leave the important task of extracting important dynamical data such as detector OPE coefficients and detector renormalization to future work.

Before continuing, we note that the gravitational detector framework described below can also be applied to more general cases of interest, e.g.~when a CFT (or QFT) is coupled to gravity. For simplicity, this paper focuses on a minimalistic matter sector (see \Eq{eq:grav_action}), utilizing it solely to prepare a simple initial state to serve as a graviton source for subsequent asymptotic detections. We hope to explore the more general setup in future work.

Energy detectors in four-dimensional asymptotically-flat gravitational theories were introduced in \cite{Gonzo:2020xza}. Here, we will present a slightly generalized formulation more closely mimicking the QFT case we reviewed above. We will first consider the energy detector before generalizing to the $E^{J-1}$ and other trajectories. The parallel considerations in the coordinates of~\cite{Gonzo:2020xza} can be found in appendix~\ref{app:detectors_in_Bondi_gauge}.

We consider linearized gravity with metric $\mathfrak{g}_{\mu\nu} = \eta_{\mu\nu} + \kappa\, h_{\mu\nu}$, where $\kappa = \sqrt{32 \pi G_N}$. The key ingredient is the asymptotic graviton field, defined as, 
\be
\label{eq:covariant definition of the asymptotic shear}
h_{\mu\nu}(\a,z) =  \lim_{L\to \infty} L^{\De_h} \, h_{\mu\nu}(x+Lz)\, 
\ee 
where $\De_h=\frac{d-2}{2}$, and as above, $\a = 2x\cdot z$ is the retarded detector time and $z^\mu$ is a future pointing null-vector. A similar construction was used in Ref.~\cite{Strominger:2014pwa} when discussing gravitational memory. Discussions in the asymptotic symmetry literature often involve the choice of Bondi coordinates (reviewed in appendix~\ref{app:detectors_in_Bondi_gauge}). In order to translate our expressions to these coordinates, one identifies the radial coordinate $r=L$, the retarded time $v=\a/2$, and the celestial coordinates $z^\mu=(1,\nhatbf)$. We can write the graviton energy detector in terms of the asymptotic metric field in \Eq{eq:covariant definition of the asymptotic shear} as
\be 
\label{eq:GR_energy_detector_definition_covariant}
\cE_h(z) = 2 \int_{-\infty}^\infty d\a : (\partial_\a h_{\mu\nu}(\a,z)) (\partial_\a h^{\mu\nu}(\a,z)) :\, .
\ee 
We have introduced normal ordering since we would like to interpret $\cE_h(z)$ as an operator in perturbation theory. As noted by Ref.~\cite{Gonzo:2020xza}, the integrand in~\eqref{eq:GR_energy_detector_definition_covariant} agrees with the $\a\a$ component of the effective stress tensor for gravitational waves \cite{Maggiore:2007ulw}
\begin{align}
T^{\rm eff, GW}_{\mu\nu} = \langle \partial_\mu h^{\alpha \beta} \partial_\nu h_{\alpha \beta} \rangle \,,
\end{align}
where $\langle\cdot\rangle$ denotes the Isaacson averaging prescription \cite{Isaacson:1968hbi,Isaacson:1968zza} over the short-wavelength graviton modes. We can check the definition~\eqref{eq:GR_energy_detector_definition_covariant} by utilizing the mode expansion of the graviton field,
\be
h_{\mu\nu} (x) = \int
\hat d^d p \, \hat \delta^+(p^2)\sum_{s} \left[ \varepsilon^{s,*}_{\mu\nu}(p) a_s(p) e^{-ip\cdot x}+\varepsilon^{s}_{\mu\nu}(p) a_s^\dagger(p) e^{ip\cdot x}\right]\, .
\ee 
Here, $\varepsilon^{s}_{\mu\nu}$ are polarization tensors, and $s$ runs over the $\frac{d (d-3)}{2}$ physical polarizations of a graviton. We use the normalization
\begin{align}
[a_{s_1}(p),a_{s_2}^\dagger(p')]= 2p^0 \hat{\delta}^{d-1}(\bp-\bp') \delta_{s_1 s_2}\,.
\end{align}
Evaluating $h_{\mu\nu}(\a,z) $ in terms of the mode expansion using the stationary phase approximation yields 
\be
\hspace{-.5cm}
h_{\mu\nu}(\a,z)=\!\! \int_0^\infty 
\!\!
\frac{d\b\, \b^{\De_h-1} }{(2\pi)^{d/2}} 
\sum_{s} \left[ie^{-i\frac{d\pi}{4}} \varepsilon^{s,*}_{\mu\nu}(\b z) a_s(\b z) e^{-i\frac{\b \a}{2}}- i e^{i\frac{d\pi}{4}} \varepsilon^{s}_{\mu\nu}(\b z) a_s^\dagger(\b z) e^{i\frac{\b \a}{2}}\right] ,
\hspace{-.4cm}
\ee
which can be inserted into \Eq{eq:GR_energy_detector_definition_covariant} to give
\be
\cE_h(z) \propto 
\int_0^\infty d\b \b^{d-2} \sum_s a^\dagger_s(\b z) a_s(\b z) \, .
\ee 
This indeed has the right form for an energy detector. The energy detector acts on asymptotic graviton states $|X\> = | \{p_1,\s_1\},\cdots,\{p_n,\s_n\}\>$ as in \Eq{eq:energy_detector_on_asymptotic_states} (see also the discussion around \Eq{eq:E_J_detector_QFT_action_on_state}), namely
\begin{align}
\label{eq:graviton_energy_detector_on_asymptotic_states}
    \cE_h(z) \ket{X} = 
    \sum_{i\in X} \int_0^\infty d\b \b^{d} \frac{\delta^{d}(k_i-\b z)}{2\de(z^2)} |X\>\,.
\end{align}
In the presence of matter fields, the total energy detector $\cE$ is simply the sum of the graviton detector $\cE_h$ and the matter one $\cE_{\mathrm{matter}}$ constructed from the stress tensor as~\eqref{eq:energy_detector_definition}. 

Ref.~\cite{Gonzo:2020xza} used an equivalent definition of energy detectors in gravity to compute the two-point function of energy flux operators in the classical limit. As commented in Ref.~\cite{Gonzo:2020xza}, the energy flux operator and its correlators are BMS invariant observables, which is equivalent to their infrared safety in perturbation theory. Ref.~\cite{Gonzo:2020xza} also constructed the detectors associated with other asymptotic charges of the Lorentz group, and found that their algebra was the BMS group familiar from the study of asymptotic symmetries. In this paper we will focus on detectors that measure energy, and powers of the energy, but it would be interesting to perform a detailed study of these other detector operators as well. From an amplitudes perspective, one expects certain complications associated to zero energy gravitons, see e.g.~Refs.~\cite{Elkhidir:2024izo,Biswas:2024ept} for a discussion in the classical context. 

We are now in a position to construct the $E^{J-1}$ trajectory of graviton detectors. Taking inspiration from the scalar QFT case, it is straightforward to write 
\be
\label{eq:graviton E^J detectors}
\cD_{J_L}(z) = \frac{1}{C_{J_L}} \int_{-\infty}^\infty d\a_1 d\a_2 \, \psi_{J_L}(\a_1,\a_2) : h_{\mu\nu}(\a_1,z)  h^{\mu\nu}(\a_2,z): \, .
\ee 
The wavefunction is 
\be
\psi_{J_L}(\a_1,\a_2) =  |\a_1-\a_2|^{2(\De_h-1)+J_L}\, ,
\ee
and the coefficient $C_{J_L}$ is the same as in the scalar case, c.f.~\Eq{eq:constantfortwisttwo},
\be
C_{J_L} &= 2^{J_L+d-1} \pi \sin\p{\pi \tfrac{J_L+2\De_h}{2}} \G(2\De_h+J_L-1)\,,
\ee
since $\De_h=\De_\f=(d-2)/2$ at tree level. With this choice, we have 
\be
\cD_{J_L}(z) = \frac{
1
}{(2\pi)^d} \int_0^\infty d\b \, \b^{-J_L-1} \sum_s :a_s^\dagger(\b z) a_s(\b z): \, .
\ee
Indeed, one can check that the action of this detector on asymptotic graviton states is
\be
\cD_{J_L}(z) |X\> 
= \sum_{i\in X} \int_0^\infty d\b \b^{1-J_L 
} \frac{\delta^{d}(k_i-\b z)}{2\de(z^2)} |X\>\,.
\ee
This construction establishes the leading trajectory of $E^{J-1}$ detectors $\cD_{J_L}$ in gravity. In addition, we automatically have the shadow trajectory $\widetilde{\cD}_{J_L} = \wS_J[\cD]_{J_L}$ by virtue of the spin-shadow transform $\wS_J$ given in~\eqref{eq:spin_shadow_def}. As discussed above, this symmetry is a consequence of Lorentz symmetry alone, and persists in gravity. We believe that this leading trajectory of light-ray operators will play an important role in understanding the structure of renormalization of light-ray operators in gravity. It would also be interesting to study the action of BMS generators on these more general $E^{J-1}$ detectors, as well as their symmetry algebra. More generally, one might expect that these more general detectors (or even all detectors) would organize under representations of the asymptotic BMS algebra, which would be interesting to work out.

One might guess that the true definition of detectors~\eqref{eq:graviton E^J detectors} and their further generalizations involve gravitational Wilson lines stretched between the insertions of the asymptotic metric fields. This is the case for light-ray operators built out of gauge fields in gauge theories~\cite{Balitsky:1987bk}. Typically, the Wilson line can be set to $1$ by going to light-cone gauge. We expect an analogous story to hold in gravity as well, although we will avoid spelling it out. In perturbation theory, the nontrivial presence of such a Wilson line necessarily enters at higher order, so we expect that it will not modify our calculations below regardless.

\begin{figure}[tb]
	\begin{center}
		\includegraphics[scale=.8]{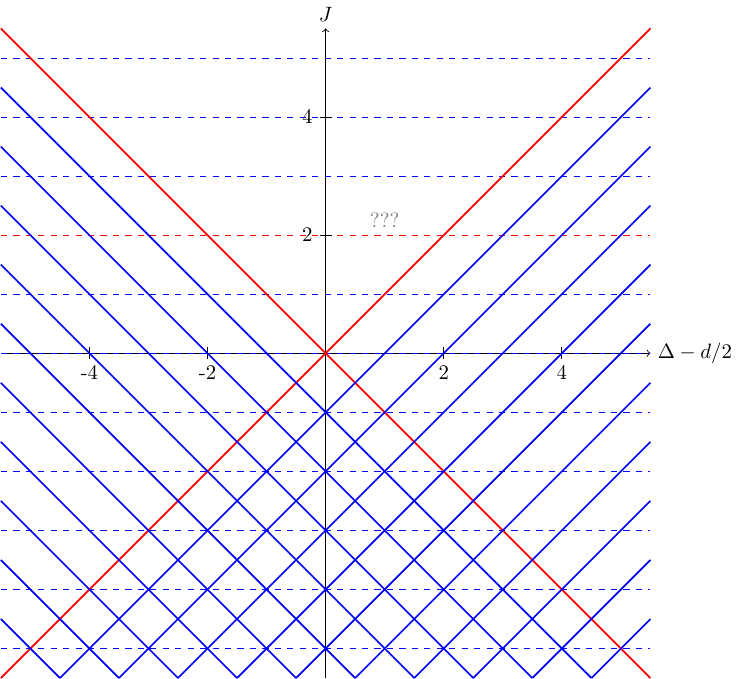}
		\caption{A guess for part of the perturbative structure of the (parity-even, traceless-symmetric) Regge trajectories of translation-invariant detectors in a gravitational theory. The red lines are the leading ``twist-two'' Regge trajectory of $E^{J-1}$ detectors and their shadows. The blue diagonal lines are the next higher twist trajectories and their shadows. Note that unlike Wilson-Fisher theory, even and odd twist trajectories appear in the same plot, as there is no $\Z_2$ symmetry that distinguishes them. One might guess the existence of the dashed horizontal lines at integer $J$, with the dashed red line at $J=2$ the putative ``reggeized graviton" trajectory, and the others the subleading or composite trajectories. 
        Their existence is unclear since although reggeized gravitons appear in Regge scattering, it is not guaranteed for free that they can be placed at infinity and interpreted as detectors. Once gravitational interactions are turned on, the trajectories are expected to mix vertically and the vertical (scaling dimension $\De_L =1-J$) direction stops being a sensible quantum number.  
		\label{fig:GRperturbativelines}
		}
	\end{center}
\end{figure}

In addition to the leading trajectory of $E^{J-1}$ detectors, we expect that one can construct more general detectors similar to the case of free scalar theory, by using multiple asymptotic graviton fields,
\be
\cD_\psi(z) = \int d\a_i \, \psi_{J_L}(\a_1,\cdots,\a_m) : h(\a_1,z) \cdots h(\a_m,z): \, .
\ee 
Here, $\psi$ is any 1d translation invariant kernel of homogeneity $J_L$. We have left the contractions of indices of the asymptotic graviton field implicit, there are many choices of those contractions. Before interactions, these detectors have quantum numbers $J_L$ and $\De_L= J_L + m \De_h$, where $\De_h=\frac{d-2}{2}$. Similar to Eq.~\eqref{eq:derivative higher twist trajectories}, there are also graviton detectors with contracted derivatives between the asymptotic graviton fields, analytically continuing operators of the type $\wL[h(x,z) \partial^{2n_1} (z\cdot\partial)^{\ell_1}h(x,z) \cdots h(x,z)\partial^{2n_k} (z\cdot\partial)^{\ell_k}h(x,z)]$, and even those with transverse spin. Therefore, we expect that part of the perturbative structure of detectors in gravity is similar to that of a free field theory, with multiple diagonal trajectories of constant twist, and their corresponding shadows. One difference one might expect is that unlike scalar field theory where operators with even and odd number of $\f$ fields---i.e. even and odd twist---appear in different $\Z_2$ sectors, operators with even or odd number of gravitons likely appear in the same sector, as there are cubic interactions between gravitons. 
Once interactions are turned on, this structure will change drastically, especially since the interaction strength $G_N$ is dimensionful. With a scale in the problem, dilatations are broken, and scaling dimension stops being a good quantum number. We expect that light-ray operators of different perturbative scaling dimensions start talking to each other, and the Chew-Frautschi plot starts collapsing in the vertical direction.

Furthermore, one can ask about the potential presence of horizontal trajectories in the spectrum of detectors in flat space gravity. One might expect that in analogy with reggeized scalars and gluons in scalar and gauge theories, there might be trajectories of reggeized gravitons. However, since we are not in a CFT, the relation between light-ray operators exchanged in Regge scattering and detector operators might be broken.\footnote{For recent progress on an operator based approach to Regge scattering in quantum gravity, see \cite{Rothstein:2024nlq}. It would be interesting to try and relate the operators they use to describe forward scattering to detector operators in gravity.} This is because we cannot simply put any light-ray operator in a theory with scale and conformally-transform it to infinity. Instead, we should think carefully in the language of detector operators, starting from the detector frame, what kind of singularities might be present and could give rise to horizontal trajectories. Regardless, we expect that similar processes to Regge scattering will be present in detector correlators, and give rise to interesting structures, likely to horizontal trajectories. We hope to comment on these in the future. We summarize our findings and speculations in a guess Chew-Frautschi plot in Figure~\ref{fig:GRperturbativelines}.

%
\section{Correlation functions of detector operators from scattering amplitudes}
\label{sec:eec_def}
%

To begin our exploration of detector operators and their correlators in perturbative quantum gravity, we must first develop efficient techniques for their calculation. In QFT, two primary approaches have been used. First, they can be directly obtained by performing the light transform of correlation functions of local operators. This has been highly successful in the case of $\mathcal{N}=4$ SYM, enabling the calculation of the two-point energy correlator to next-to-next-to-leading order \cite{Belitsky:2013xxa,Belitsky:2013bja,Henn:2019gkr}. Knowledge of correlators of local operators to high loop orders \cite{Bourjaily:2016evz,He:2024cej} has also been used to obtain the integrand for multi-point correlators \cite{He:2024hbb} up to 11 point.   Alternatively, they can be computed as phase space integrals over squared form factors of local operators. This approach has been used for the two-point correlator in QCD \cite{Dixon:2018qgp,Gao:2020vyx}, and for higher point correlators in both QCD and $\mathcal{N}=4$ SYM \cite{Chen:2019bpb,Yan:2022cye,Yang:2022tgm,Chicherin:2024ifn}. In the case of gravity, neither of these options are available, due to a lack of gauge invariant local operators. Instead we must compute energy correlators in states specified by asymptotic in-states, which in perturbation theory entails the use of scattering amplitudes.  This approach is also used to compute energy correlators at hadron colliders such as the LHC \cite{Gao:2019ojf,Lee:2022ige,Chen:2023zlx}.

In this section, we will discuss how to compute correlators of detector operators perturbatively using scattering amplitudes. We begin in \Sec{sec:calc_abs} with a general discussion of the formalism for calculating correlators of detector operators, and then show how this can be concretely performed in terms of scattering amplitudes in \Sec{sec:calc_amps}. Part of our goal is to connect the formal language of detector operators to concrete calculations that can be performed using the wealth of known amplitudes in perturbative (super-) gravity. Throughout this section, we mainly focus on $E^{J-1}$ detectors for concreteness, but other detectors should work analogously with the vertex factors appropriately replaced.

\subsection{Review: correlation functions of detector operators}\label{sec:calc_abs}

Let us review some salient features of the formalism of \cite{Caron-Huot:2022eqs} for computing correlation functions of detector operators in a weak coupling expansion. We are interested in an observable of the kind
\be
\label{eq:detector_matrix_element_def}
     \frac{1}{N} \<\Psi| \cD_1(z_1) \cdots \cD_n(z_n)|\Psi\>\, ,
\ee 
where $|\Psi \>$ is some state of interest, $\cD_i$ are detector operators, and $N$ is a normalization factor that depends on the particular initial state $|\Psi \>$ of interest. In situations where the initial state is generated by the action of a local operator on the vacuum (e.g. in the Hoffman-Maldacena setup \cite{Hofman:2008ar}), it is customary to divide out by the norm of that state $N=\<\Psi|\Psi\>$. If we are interested in states generated by asymptotic scattering particles, it is more convenient to normalize the observable by the incoming particle flux which leads to the interpretation of the observables as weighted cross sections that are commonly defined in collider settings. Generically, we can think of the detectors as light-ray operators inserted at $\scri^+$, $\cD (z) = \mathbb{O}(\infty,z)$. We refer to the frame where light-ray operators are inserted along $\scri^+$ (anchored at spatial infinity) as the ``detector frame''. (See previous sections for details.)  In this work, we primarily consider correlation functions of energy detectors $\cE(z)$ or ``energy-to-some-power'' detectors $\cD_{J_L}(z)$.\footnote{Recall that the energy detector is recovered as $\cE = \cD_{J_L=1-d}$.} 

\begin{figure}
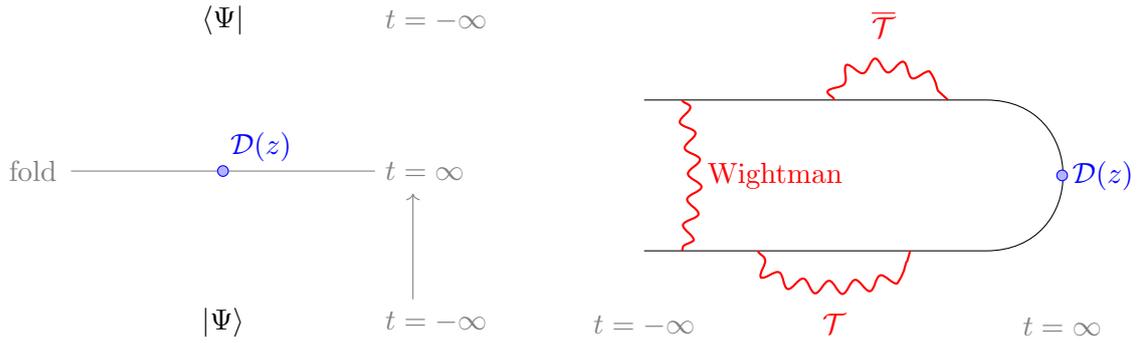

\centering
\begin{subfigure}[t]{0.5\textwidth}
    \schwingerKeldyshA
\end{subfigure}
\begin{subfigure}[t]{0.45\textwidth}
    \schwingerKeldyshB
\end{subfigure}
\caption{
\label{fig:detector_in-in_diagrams}
A depiction of the diagrammatic setup for computing correlation functions of detector operators in the in-in formalism. On the left, we present the template for diagrams used in the computation of detector observables. A state is prepared in the far past, evolved to the future where the detector (blue) lies along the time fold at $t=\infty$, then time evolves backwards to compute the overlap with the in state. On the right, we present a cartoon summary of the Schwinger-Keldysh contour and the associated Feynman rules (red). One can view the two pictures as sideways views of each other.
}
\end{figure}

To compute an ``in-in''-type observable such as~\eqref{eq:detector_matrix_element_def}, one needs to use a Schwinger-Keldysh contour, as laid out in \cite{Caron-Huot:2022eqs}. We briefly review the Feynman rules associated with computing such an observable, and refer to \cite{Caron-Huot:2022eqs} for further details. The computation involves a time fold, where the ket state $|\Psi\>$ is prepared with usual Feynman rules along the first part of the contour from $t=-\infty$ to $t=\infty$, the detectors sit at the time fold at $t=\infty$, and then time runs backwards to $t=-\infty$ (with conjugated Feynman rules) to compute the overlap with the bra state $\<\Psi|$. Interactions and propagators below the time fold are the usual time-ordered ones ($\cT$), interactions and propagators above the fold are conjugated and anti-time-ordered ($\overline\cT$), and propagators crossing the cut have Wightman ordering (on-shell). The in-in setup and the associated Feynman rules are summarized pictorially in figure~\ref{fig:detector_in-in_diagrams}.

\begin{figure}[ht!]
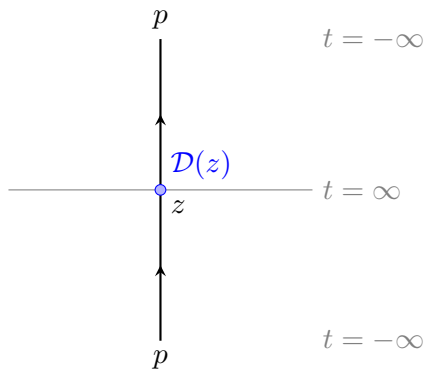

\centering
    \detectorVertexFig
\caption{\label{fig:tree_level_vertex}
The $\cD$-detector tree-level vertex in the in-in formalism.
}
\end{figure}

Perturbatively, each detector insertion at $\scri^+$ corresponds to a vertex factor in Feynman diagrams where a massless particle passes through the detector and has the corresponding quantity (such as its energy, or charge) recorded. Diagrammatically, it is given by figure~\ref{fig:tree_level_vertex}. The detector vertex can be computed at tree level, and receives corrections at loop order. It's renormalization in weakly-coupled QFT was discussed in~\cite{Caron-Huot:2022eqs}. For ``energy-to-some-power'' detectors measuring $E^{J-1}$, the detector vertex is universally given by the distribution
\be
\label{eq:detector_vertex_gen_J}
V_{J_L}(z;p) = \int_0^\infty d\b \b^{-J_L-1} \delta^{d}(p-\b z)\, ,
\ee 
up to additional factors propagating polarization (Lorentz) and color indices. For detectors~\eqref{eq:definition of free scalar twist two trajectory} in a theory of massless scalars, the detector vertex in perturbation theory is
\be
\<0|\f(-q) \cD_{J_L}(z) \f(p)|0\>=\hat \delta^d(p-q) V_{J_L}(z;p)\, .
\ee 
The detector vertex is zero unless the incoming momenta $p$ is aligned with the detector position $z$, and the integral extracts the energy $\b$ of the quanta and weighs it to a power determined by $J_L$. Note that the detector vertex is not ``amputated'' and includes the propagators of the fields entering and exiting the detector.

In gravity, we have the detectors~\eqref{eq:graviton E^J detectors} constructed out of the asymptotic graviton field. In that case, the perturbative vertex for the detection of gravitons is 
\be \label{eq:graviton detector vertex}
\<0| h_{\mu\nu}(-q) \cD_{J_L}(z) h_{\rho\s}(p)|0\>
= \hat{\delta}^d(p-q) \, \Pi_{\mu\nu\rho\s}(p) \, V_{J_L}(z;p)\, .
\ee
The calculation of this detector vertex is presented in \App{app:graviton detector vertex}. The physical-state projector $\Pi$ naturally arises in the vertex, making sure only physical polarizations are detected, and the universal vertex factor $V_{J_L}$ weighs by the corresponding power of energy. For now, the detailed form of $\Pi$ is irrelevant, but will be discussed in section~\ref{sec:amplitude_review} when we consider sewing and squaring amplitudes (see \Eq{StateSumGravity}). Essentially $\Pi$  implements that only \emph{physical} graviton states are registered by the detector.

\subsection{Calculation from scattering amplitudes}\label{sec:calc_amps}

In many cases of interest, asymptotic single particle states are eigenstates of the action of the detector operator. This is true in particular for the ANE operator, or its generalized $E^{J-1}$ detectors. In this case correlators of detector operators can be computed as ``weighted cross-sections".  While this is well known, we review it here to connect it to the discussion in \Sec{sec:calc_abs}. We also wish to emphasize the importance of the availability of scattering amplitudes in perturbative gravity for the calculation of energy correlators, and to give a simple formula which can be applied to calculate energy correlators in perturbative gravity.

As an example, let us consider an $n$-point correlation function of energy-to-some-power detectors 
\be
 \mathbb{E}_n=\frac{1}{N}\<\Psi| \cD_{J_{L1}}(z_1) \cdots \cD_{J_{Ln}}(z_n)|\Psi\>
\ee 
for some initial state $|\Psi\>$ and a normalization factor $N$ discussed below \Eq{eq:detector_matrix_element_def}. For convenience, let us pick $|\Psi\>$ to be a state of definite total momentum $P$. Using the in-in formalism, the perturbative computation of the detector correlator is organized in terms of Feynman diagrams.  The states in the asymptotic future, where the measurement takes place, are related to the in-states via the S-matrix
\be
\ket{\Psi}_{{\rm out}} = S \ket{\Psi}_{{\rm in}} = (1+i\, T) \ket{\Psi}_{{\rm in}}\,.
\ee
We end up with a perturbative expansion of \Eq{eq:detector_matrix_element_def} that can be interpreted as follows: We generate an initial state in the asymptotic past, time evolve it to the asymptotic future where we measure the expectation value of the detector operators. Ignoring contact terms where multiple detectors are on top of each other, i.e. requiring all $z_i$ to be distinct, we need at least $n$ particles crossing the time fold at $\scri^+$ and hitting the $n$ detectors to get a non-zero answer. We also have processes where $m>n$ particles are produced in the final state. In perturbation theory, each additional particle in the final state requires additional powers of the small coupling constant $g^2$ (or $G_N$ in gravity) so that the desired observable can be computed systematically in a small coupling expansion. Beyond the leading order, higher-loop processes with fewer particles in the final state can interfere with lower loop processes with additional particles in the final state. Schematically we have,
\begin{align}
\hspace{-.5cm}
\begin{split}
\label{eq:detector_matrix_element_2}
 \mathbb{E}_n= \frac{1}{N}\sumint \left[\prod^n_{i=1} \hat{d}^dq_i\right] d{\rm LIPS}_X & \
 _{\rm in}\<\Psi|q_1,\dots, q_n;X\>_{\rm out} 
 \,\times\, \\
 &\hspace{-4.2cm} _{\rm out}\<q_1,\dots,q_n;X| \cD_{J_{L1}}(z_1) {\cdots} \cD_{J_{Ln}}(z_n) |q_1,\dots, q_n;X\>_{\rm out}  
 \,\times\,
 _{\rm out} \<q_1,\dots, q_n;X|\Psi\>_{\rm in}\,.
 \end{split}
 \hspace{-.5cm}
\end{align} 
The $|q_i\>$-states are ``out'' states with $q_i^2=0$, representing massless particles that travel to $\scri^+$, and $X$ schematically denotes any other (massive or massless) excitations that can be produced in the final state, including no excitation at all. The phase space integration \begin{footnotesize}$\sum$\end{footnotesize}$ \kern -.3cm \int$ includes a sum over final state polarizations or colors. 

In our explicit discussion in \Sec{sec:eec_5pt}, we will take \emph{all} outgoing states to be gravitons, but generalizing to other particle species is straightforward. In writing \Eq{eq:detector_matrix_element_2}, we have used the fact that each detector vertex is diagonal in momentum space so that we can identify the momenta across the timefold at $\scri^+$. Even though we still write the $X$-dependence in the states entering the detector matrix element, it is indeed independent on $X$. For each of the $n$ particles hitting the $n$ detectors, we are instructed to insert a detector vertex factor, and for each particle contained in $X$ that crosses the cut without hitting a detector, we are instructed to insert a Wightman propagator 
\be
W(q) = \hat{\delta}^+(q^2), \qquad \hat{\delta}^+(q^2) = \theta(q^0) \hat{\delta}(q^2-m^2)\,,
\ee 
where the Heavyside-theta function $\theta(q^0)$ ensures the propagation of positive energy quanta and $m$ is the (possibly zero) mass of that particle. Altogether, we have 
\be
\label{eq:observable_def_ini}
\mathbb{E}_n = \frac{1}{N} \sum_X
 \int   \left[\prod_{a=1}^n   \hat{d}^dq_a\, V_{J_{La}}(q_a,z_a)\right] 
        \left[\prod_{b\in X}  \hat{d}^dq_b\, W(q_b) \right] \,\, 
        |\<q_1,\dots, q_n;X|\Psi\>|^2 \,,
\ee 
where we suppress the explicit ``in'' and ``out'' subscripts on the states for brevity and we define the squared matrix elements to include the sum over final state polarizations or colors. Note that this definition conveniently absorbs the explicit physical-state projector of the graviton energy detector vertex~\eqref{eq:graviton detector vertex} into the squared, polarization summed matrix elements. As mentioned before, there are different physical scenarios on how to prepare the initial state $|\Psi\>$. In cases where the state is obtained by the action of a local operator $\cO(x)$ on the vacuum \cite{Hofman:2008ar}
\begin{align}
 \ket{\Psi} = |\cO(E) \> \equiv \int d^d x \ e^{-i \, Ex^0 - x^2_E/\s^2}\, \cO(x) \ket{0}\,,
\end{align}
where the euclidean norm $x^2_E$ ensures the normalizability of the state where the initial excitation is localized in coordinate space to a region of size $\cO(\s)$. In this case, it is natural to choose the two-point function of $\mathcal{O}$ as normalization of the observable, i.e. $N=\bra{\Psi}\Psi\rangle \sim \langle 0| \overline{\mathcal{O}}(E) \mathcal{O}(E) | 0 \rangle $. In the limit of $\s E \gg 1$, the state becomes a momentum eigenstate with time-like momentum $P^\mu = (E,\mathbf{0})$. In this limit, the matrix elements in \Eq{eq:observable_def_ini} become nothing but the \emph{form factors} $\mathcal{F}_{\cO}$ of the operator $\cO(x)$ defined as the overlap of the off-shell state created by $\cO$ from the vacuum $\ket{0}$ with an on-shell final state $\bra{q_1,\dots,q_n;X}$
\begin{align}
    \mathcal{F}_{\mathcal{O}}
        & = \<q_1,\dots, q_n;X|\Psi\> 
        = \int d^d x \ e^{-i \, P\cdot x}\, \<q_1,\dots, q_n;X| \mathcal{O}(x) \ket{0}
        \\
        & = \hat{\delta}^{d}(q_1{+}\dots{+}q_n{+}Q_X - P) \ \<q_1,\dots, q_n;X| \mathcal{O}(0) \ket{0}
        \,
\end{align}
where $Q_X=q_{n+1}+\dots+q_m$ is the total momentum of the non-observed outgoing states.

\medskip

In a collider setting, and for later discussions in this work, we consider $|\Psi\>$ to be a collection of $k$ scattering ``in'' states,
\begin{align}
|\Psi\>=|p_1,\dots,p_k\>\,, 
\end{align}
with total momentum $P=\sum\limits^k_{i=1} p_i$. In this case, the overlap $_{\rm out}\<q_1,\dots q_n;X|p_1,\dots,p_k\>_{\rm in}$ is nothing but the $k\to n+|X|$ scattering amplitude
\be
\label{eq:scat_amp_def}
 \<q_1,\dots, q_n;X|p_1,\dots,p_k\> = \hat{\delta}^{d}(q_1{+}\dots{+}q_n + Q_X - P)\, \cM_{k\to n+|X|} (p_i,q_j)\,, 
\ee 
for the transition of $k$ particles in the asymptotic past to $n+|X|$ final state particles in the asymptotic future. In \Eq{eq:scat_amp_def}, we have explicitly exposed the momentum-conserving delta function of the scattering process. In order to obtain a non-zero result for $n$ detectors located at locations $z_i$ on $\scri^+$, we require at least $n$ light-like states with momenta $q_i$ in the final state, whereas $X$ can be empty, or a collection of massless and massive states. 

As the observables in \Eq{eq:observable_def_ini} can be interpreted as weighted cross sections, it is natural to adjust the normalization factor $N$ so that we are only left with the usual flux factor in standard cross-section calculations, see e.g.~\cite{Schwartz:QFTbook2013}. Standard considerations take care of the naive $\left[\hat{\delta}^{d}(x)\right]^2$ terms coming from the squared matrix elements which can be dealt with by formally taking the large but finite time and volume limit for the scattering process. Putting everything together, we can perturbatively compute the observable in \Eq{eq:detector_matrix_element_2} for an initial state given by $k$ incoming scattering particles in terms of a weighted integral over the modulus square of $k\to n+|X|$ scattering amplitudes 
\begin{align}
\label{eq:observable_scattering_initial_state_def}
\begin{split}
\mathbb{E}_n=\frac{1}{N_{k}} \sum_{X}
 \int   \left[\prod_{a=1}^n   \hat{d}^dq_a\, V_{J_{La}}(q_a,z_a)\right] &
        \left[\prod_{b\in X}  \hat{d}^dq_b\, W(q_b) \right] \,\, \times 
        \\
        & \hspace{-2cm} \hat{\delta}^{d}(q_1{+}\dots{+}q_n + Q_X - P)\, \big|\cM_{k\to n+|X|} (p_i,q_j)\big|^2 \,,
\end{split}
\end{align}
where $V_{J_{La}}$ is the general detector vertex defined in \Eq{eq:detector_vertex_gen_J}, $Q_X=q_{n+1}+\dots+q_m$ is the total momentum of the un-observed outgoing states in $X$, and $P = p_1 + \dots + p_k$ is the total momentum of the incoming states. In \Eq{eq:observable_scattering_initial_state_def} all volume and time factors have canceled and we are left with one overall momentum preserving delta function and a remaining normalization factor $N_{k}$ that only depends on the number of initial states, $k$. Diagrammatically, the final-state sum over $X$ in  \Eq{eq:observable_scattering_initial_state_def} can be represented as follows
\begin{align}
\label{eq:event_shape_expansion}
\mathbb{E}_n 
= \vcenter{\hbox{\genDetectorNone}} \hspace{-.0cm}+ \vcenter{\hbox{\genDetectorOne}} +\quad  \cdots\,,
\end{align}
where the subscript in $\mathbb{E}_n$ denotes the number of detector insertions and the terms in the sum correspond to additional (non-observed) states in $X$ (including massive particles) that can cross the cut. The blue dots represent the detector vertex, and the ellipses denote additional detectors not drawn explicitly. Again, the summation over final state polarization or color degrees of freedom is understood. In perturbation theory, the $k\to m$ amplitude $\mathcal{M}_{k\to m}$ (and its complex conjugate $\bar{\mathcal{M}}_{k\to m}$) has the usual weak coupling expansion in powers of $g$,
\begin{align}
\hspace{-.4cm}
\mathcal{M}_{k\to m} = \hspace{-.3cm} \vcenter{\hbox{\scalebox{0.8}{\ampGen}}} \hspace{-.3cm} = g^{k+m-2} 
\left[
    \vcenter{\hbox{\scalebox{0.8}{\ampGenOrder{0}}}} 
    \hspace{-.3cm} + g^2 \hspace{-.2cm} \vcenter{\hbox{\scalebox{0.8}{\ampGenOrder{1}}}}\hspace{-.4cm} + \cdots
\right]\,,
\hspace{-.4cm} 
\end{align}
where we have pulled out the explicit factors of the coupling constant, and we denote the expansion order of the amplitude by a superscript. We are finally in the position to perturbatively expand the detector correlator in powers of $g$
\begin{align}
\label{eq:event_shape_g_expansion}
\mathbb{E}_n  
& =  g^{2(k+n-2)}\!\! \left[ 
    \overbrace{\vcenter{\hbox{\scalebox{0.8}{\genDetectorNonePert{0}{0}}}}}^{\widetilde{\mathbb{E}}^{(0)}_n } 
    \!\!\!\! + g^2 \! \left(\!\!
            \overbrace{
             \underbrace{\vcenter{\hbox{\scalebox{0.8}{\genDetectorNonePert{1}{0}}}}
            \!+\!\vcenter{\hbox{\scalebox{0.8}{\genDetectorNonePert{0}{1}}}}}_{\mathbb{E}^{(1)}_{n,{\rm contact}} }  
            \!+\!
            \underbrace{\vcenter{\hbox{\scalebox{0.8}{\genDetectorOnePert{0}{0}}}}}_{\mathbb{E}^{(1)}_{n}}
            }^{\widetilde{\mathbb{E}}^{(1)}_n}
        \!\!\!\right)\!
    {+} \cdots    
\right]\,,
\end{align}
where the terms in the coupling expansion are denoted by a tilde, $\widetilde{\mathbb{E}}^{(i)}_n$, to streamline our subsequent discussion of structurally distinct contributions. Let us first look at the leading term in the perturbative expansion of the $n$-point correlator in \Eq{eq:event_shape_g_expansion} and consider the $n$ positions of the detectors $z_i$ to be distinct.\footnote{There are additional contact terms when the detectors are allowed to lie on top of each other. For simplicity, we ignore these contributions in the present discussion.} In that case, $X$ in \Eq{eq:observable_scattering_initial_state_def} is empty, $Q_X=0$, and all amplitudes are at tree-level
\begin{align}
\label{eq:observable_scattering_initial_state_LO}
\begin{split}
\widetilde{\mathbb{E}}^{(0)}_n=\frac{1}{N_{k}}
 \int   \left[\prod_{a=1}^n   \hat{d}^dq_a\, V_{J_{La}}(q_a,z_a)\right]
         \hat{\delta}^{d}(q_1{+}\dots{+}q_n - P)\, \big|\cM^{(0)}_{k\to n} (p_i,q_j)\big|^2 \,.
\end{split}
\end{align}
Each detector vertex $V_{J_{La}}$ forces one of the momenta $q_a$ to lie along the (fixed) null detector position $z_a$, $ V_{J_{La}} \sim \delta^{d}(q_a - \b_a z_a)$, thereby localizing all of the $q_a$ integrals. In addition, the overall momentum conserving delta function can be used to localize energy integrals over the $\beta_a$ from the detector vertex. When the number of detectors $n$ is smaller than the spacetime dimension $d$, the observable has distributional support only. We refer to such terms as ``contact terms'', as they only have support on a restricted kinematic configuration. This can be understood from the fact that the absence of additional final states excludes kinematic ``recoil'' of the particles that hit the detectors. When the number of detectors becomes larger than the dimension of space-time $n>d$, we are left with some nontrivial energy integrals. 

For certain detectors, there can be additional diagrams with (spacelike) propagators between detectors along the cut. These diagrams are known to give rise to ``detector cross-talk'', which arises physically from zero-modes. These contributions are zero for energy detectors, but are nonzero for ``scalar flow'' detectors in some theories~\cite{Belitsky:2013bja, Belitsky:2013xxa}. It would be interesting to explore this issue in the context of gravity, where it should be related to the choice of BMS frame. For a recent discussion from the amplitudes perspective, see \cite{Elkhidir:2024izo}. Although it will not contribute for energy detectors, one expects that detector cross-talk would be important for angular-momentum correlators. Since we focus on energy detectors, we omit these diagrams for simplicity in our discussion. They are once again relevant for $E^{J-1}$ detectors for small enough $J$, but we will leave their computation and consequences for future work. 

As is evident in \Eq{eq:event_shape_g_expansion}, at next-to-leading order in the perturbative expansion, there are two structurally distinct contributions. The first comes from a perturbative correction to one of the $k{\to}n$ matrix elements, without additional unobserved particles in the final state. The second contribution involves leading order higher-point matrix elements with one additional unobserved particle in the final state. Explicitly, we have 
\begin{align}
 \widetilde{\mathbb{E}}^{(1)}_n = \mathbb{E}^{(1)}_{n,{\rm contact}} + \mathbb{E}^{(1)}_n\,.
\end{align}
Crucially, the latter contribution has support on finite angle kinematics irrespective of the number of detectors vis a vis the spacetime dimension. In what follows, we will direct our attention to these finite angle contributions, where the leading term is given by the last image in \Eq{eq:event_shape_g_expansion}.

In summary, we have provided a concrete recipe for computing energy correlators, or more generally correlators of different detector operators, in perturbative quantum gravity. We hope that this will enable the exploration of detector correlators in perturbative quantum gravity using the wealth of scattering amplitudes which have been computed in the past decade. For completeness, we spell out in more detail the perturbative calculation of the $1$, and $2$-point correlators, leaving the discussion of general $n$ to appendix \ref{app:multi-pt-kinematics}.

\subsubsection{The 1-point correlator}

For concreteness, let us elaborate on the computation of the 1-point correlator
\be
 \mathbb{E}_1=\frac{1}{N}\<\Psi| \cD_{J_{L}}(z) |\Psi\>\,.
\ee 
Diagrammatically, the perturbative expansion of the 1-point correlator is obtained by restricting the general form in \Eq{eq:event_shape_g_expansion} to a single $n=1$ detector.
%
%
The first term in the perturbative expansion, $\widetilde{\mathbb{E}}^{(0)}_1$, is supported on restricted kinematics where the detected particle needs to have null momentum pointing in the direction of the detector position $q_1\propto z$. At the same time, momentum conservation implies $q_1=P$ so that this term only has support at $P\propto z$. In terms of the underlying amplitude, we have
\be
\label{eq:EC_LO}
\widetilde{\mathbb{E}}^{(0)}_1(p_i;z)
=
\vcenter{\hbox{\oneDetectorNoneLO}}
&= \frac{1}{N_k}\int
    \hat{d}^d q_1 \, 
    V_{J_{L}}(q_1,z) \,  \big|\cM^{(0)}_{k\to 1}(p_i, q_1)\big|^2 \, \hat{\delta}^{d}(q_1-P) 
    \nn \\[-35pt]
&= \frac{1}{N_k } 
    \big|\cM^{(0)}_{k\to 1}(p_i, q_1 {=} P)\big|^2 V_{J_L}(P,z)\, .
\ee 
This term is therefore distributional by virtue of the distributional nature of the detector vertex \Eq{eq:detector_vertex_gen_J}.

At next-to-leading order, the nontrivial generalization of \Eq{eq:EC_LO} involves an additional unobserved particle in the final state
\be
\mathbb{E}^{(1)}_1(p_i;z) 
=
\!\!\! \vcenter{\hbox{\oneDetectorOneLO}} \!\!\!
&= \frac{1}{N_k}\int \!\! \prod^2_{a=1}
    \hat{d}^d q_a \,
    V_{J_{L}}(q_1,z) \,   \hat{\delta}^+(q_2^2) 
    \big|\cM^{(0)}_{k\to 2}(p_i, q_1, q_2)\big|^2 \, 
    \hat{\delta}^{d}(q_1{+}q_2{-}P) 
	\nn    \\[-45pt]
&= \frac{1}{N_k\, (2\pi)^{d}}
    \int_0^\infty d\b \b^{-J_{L}-1} \hat{\delta}^+((P{-}\b z)^2)  \big|\cM^{(0)}_{k\to 2}(p_i,\beta z,P{-}\b z)\big|^2\,,
\ee 
where we have assumed that particle $2$ is massless, and we have made use of the detector vertex delta function, together with momentum conservation to localize the final state momenta to $q_1 {=} \beta z, \, q_2 {=}P{-}\b z$. There are two different behaviors, depending on whether $P$ is timelike or null. First, let us consider $P$ to be timelike. In that case, it is useful to rescale the $\b$ variable, $\b = \frac{P^2}{2P\cdot z} \a$. We can go to the center of mass frame with $P=(\sqrt{s},\mathbf{0})$, and $z=(1,\nhatbf)$ to resolve the $\theta$ function within $\hat{\de}^+$, and obtain
\be
\hat{\delta}^+((P{-}\b z)^2) = \frac{1}{P^2} \theta(2-\a) \hat \de(1-\a)\, .
\ee  
Therefore, the energy integral is totally fixed and we have
\be
\mathbb{E}^{(1)}_1(p_i;z) &=\frac{1}{N_k\, (2\pi)^{d{-}1}} 
\frac{1}{P^2} \left[\frac{2P\cdot z}{P^2}\right]^{J_L} 
\!\!
\left|
\cM^{(0)}_{k\to 2}\left(p_i, q_1 {=} \frac{P^2}{2P\cdot z} z, q_2 {=}P{-}\frac{P^2}{2P\cdot z} z \right)
\right|^2 \!\!\! \!.
\ee
When $P$ is null and future pointing, we can write $P=\gamma z'$, with $z'^2=0, \gamma>0$, and the integral becomes
\be
\mathbb{E}^{(1)}_1(p_i;z)  &{=} \frac{1}{N_k\, (2\pi)^{d}}
    \!\!\int_0^\infty\!\!\! d\b \b^{-J_{L}-1} 
    \theta(\g{-}\b) \hat\delta(2 \g \b\, z{\cdot} z')  \big|\cM^{(0)}_{k\to 2}(p_i, q_1 {=} \beta z, q_2 {=}\g z'{-}\b z)\big|^2\,.
\ee 
For $\beta,\gamma>0$, the delta function can only be satisfied when $z\cdot z'=0$. 

The two terms discussed so far, $\widetilde{\mathbb{E}}^{(0)}_1$ and $\mathbb{E}^{(1)}_1$, involved relatively simple operations on the underlying squared amplitudes, and did not require any nontrivial loop integration. The next-to-next-to-leading order term $\mathbb{E}^{(2)}_1$ is far more complicated and requires ``loop'' integration:
\be \label{eq:loop correction to one point event shapes}
\mathbb{E}^{(2)}_1(p_i;z)
=
\!\!\! \vcenter{\hbox{\oneDetectorTwoLO}} \!\!\!
&= \frac{1}{N_k} \!\!\int\!\! 
    \prod^3_{a=1}\hat{d}^d q_a \, 
    V_{J_{L}}(q_1,z)   
    \hat\delta^+(q_2^2)  \hat\delta^+(q_3^2) 
     \\[-45pt]
     &\qquad\qquad\times \big|\cM^{(0)}_{k\to 3}(p_i, q_1, q_2,q_3)\big|^2 \,\hat{\delta}^{d}(q_1{+}q_2{+}q_3{-}P) 
    \nn \\
&= \frac{1}{N_k\, (2\pi)^{d}}
   \int_0^\infty d\b \b^{-J_{L}-1} \int \hat d^d q_2 \hat \delta^+(q_2^2) \hat \delta^+((P{-}\b z -q_2)^2) 
    \nn  \\ &\qquad \qquad
    \times \big|\cM^{(0)}_{k\to 3}(p_i, q_1 {=} \beta z, q_2, q_3 {=} P{-}\b z {-} q_2)\big|^2\, . \nn
\ee 
The integral over $q_2$ is effectively a two-particle phase space integral, or a cut one-loop integral when viewed through the lens of reverse unitarity \cite{Anastasiou:2002yz,Anastasiou:2002qz,Anastasiou:2003yy,Anastasiou:2015yha} and depends on the form of the underlying amplitude $|\cM^{(0)}_{k\to3}|^2$. Such integrals are common in collider physics cross section calculations. Notably, these phase-space integrals can develop divergences in limits where the final state particles become collinear or soft. In the context of detector correlators, such an integral was evaluated in the simple case of Wilson-Fisher theory in \cite{Caron-Huot:2022eqs}, and the resulting divergences led to the renormalization of detector operators $\cD_{J_L}$ and detector anomalous dimensions for general $J_L$, as alluded to in section~\ref{sec:cft regge trajectories}. In QCD they have been performed (at one higher loop order), and are the standard way to compute the time-like DGLAP anomalous dimensions \cite{Mitov:2006ic,Mitov:2006wy}. It would be very interesting to compute loop corrections such as \eqref{eq:loop correction to one point event shapes} and study the renormalization and mixing of detector operators in gravity, or more simply in QFTs with a scale. We leave the investigation of such issues beyond the scope of our current paper and hope to address them in the future.

\subsubsection{The 2-point correlator}
\label{subsubsec:2pt_correlator}

As our primary example, let's compute the two point correlator in the state created by some number $k$ of incoming scattering states,
\be
\mathbb{E}_2(p_i;z_1,z_2) = \frac{1}{N_k} \< p_1,\dots,p_k | \cD_{J_{L1}}(z_1) \cD_{J_{L2}}(z_2) | p_1, \dots p_k\>\, ,
\ee 
as an integral of the underlying scattering amplitude squared. Later, we will consider the initial state created by $k=2$ massive scalar particles that evolves into $2+|X|$  outgoing gravitons in general relativity, but the derivation below works for any $k$ and is theory agnostic, so we will keep things general for now. 

Perturbatively, the 2-point correlator is given by restricting the general formula in \Eq{eq:event_shape_g_expansion} to two ($n=2$) detectors.
%
%
The leading contribution $\widetilde{\mathbb{E}}^{(0)}_2$ has support only on restricted kinematics, and we will not discuss it further here. Instead, we direct our attention to the leading term with support at generic angles, $\mathbb{E}^{(1)}_2$, obtained by squaring and sewing the $k\to 3$ amplitude and performing the weighted energy integral
\begin{align}
\mathbb{E}^{(1)}_2(p_i;z_1,z_2) 
= 
\!\!\!\! \vcenter{\hbox{\twoDetectorOneLO}}\!\!\!
&= \frac{1}{N_k}\int
    \prod^3_{a=1}\hat{d}^d q_a \,
    V_{J_{L1}}(q_1,z_1) \, V_{J_{L2}}(q_2,z_2)\,  \hat{\delta}^+(q_3^2) 
     \\[-45pt] &\qquad \qquad
    \times  \big|\cM^{(0)}_{k\to 3}(p_i, q_1, q_2,q_3)\big|^2 \, \hat{\delta}^{d}(q_1+q_2+q_3-P) 
   \nn \\
&= \frac{1}{N_k\, (2\pi)^{2d}}
    \!\! \int_0^\infty \!\!\! d\b_1 d\b_2 \b_1^{-J_{L1}-1}\b_2^{-J_{L2}-1} 
    \hat{\delta}^+((P{-}\b_1 z_1 {-}\b_2z_2)^2) 
    \nn \\ &\qquad \qquad
    \times \big|\cM^{(0)}_{k\to 3}(p_i, q_1 {=} \beta_1 z_1, q_2 {=} \beta_2 z_2, q_3 {=} P{-}\b_1z_1 {-} \b_2 z_2)\big|^2\, . \nn
\end{align} 
We have used the delta functions in the detector vertices to write $q_1 =\beta_1 z_1$ and $q_2=\beta_2 z_2$ and momentum conservation to eliminate the $q_3$ integral. Once again, we made use of the definition $ P \equiv \sum_i p_i$. We are left with two integrals over the detector energies $\b_1$ and $\b_2$, one of which can be eliminated by virtue of the additional on-shell delta function. Effectively, we will be left with an integral over the energy fraction of the two detectors. It is convenient to define the following change of integration variables,
\be \label{eq:alpha definition}
\b_a = \frac{P\cdot P}{2 P\cdot z_a} \a_a\,,
\ee 
and to introduce the cross ratio
\be \label{eq:zeta definition 2 detector}
\zeta \equiv \zeta_{12} = \frac{(2 z_1\cdot z_2) (P\cdot P)}{(2 P\cdot z_1) (2 P\cdot z_2)}\,,
\ee
which takes values $0<\zeta < 1$ for physical scattering kinematics. We can furthermore go to the center of mass frame of the incoming states where $P^\mu=(P,\mathbf{0})$ and $z^\mu_i = (1,\nhatbf)$ to evaluate the theta function coming from the Wightman propagator. Here we explicitly take the additional particle with momentum $q_3$ to be massless, i.e. $q^2_3=0$ to find
\be
\mathbb{E}^{(1)}_2(p_i;z_1,z_2) &= 
\frac{1}{N_k\, (2\pi)^{2d-1}}
\frac{1}{P\cdot P}
\left[\frac{2P\cdot z_1}{P\cdot P}\right]^{J_{L1}}
\left[\frac{2P\cdot z_2}{P\cdot P}\right]^{J_{L2}} 
\\
& \hspace{-1.4cm}\times 
\int_0^\infty d\a_1 d\a_2 \, \a_1^{-J_{L1}-1} \, \a_2^{-J_{L2}-1} \,
\theta(2{-}\a_1{-}\a_2) \,
\delta\left(1{-}\a_1{-}\a_2{+}\a_1\a_2 \zeta \right) 
\big|\cM_{k\to3}(p_i, q_j^{**})|^2 \,.\nn
\ee 
The amplitude squared is evaluated at the following intermediate momentum configuration
\be
q_1^{**} &= \beta_1 z_1 = \frac{P\cdot P}{2P\cdot z_1} \a_1 z_1, \nn \\ 
q_2^{**} &= \beta_2 z_2 = \frac{P\cdot P}{2P\cdot z_2} \a_2 z_2,  \\ 
q_3^{**} &= P-q_1^{**}-q_2^{**} 
          = P - \frac{P\cdot P}{2P\cdot z_1} \a_1 z_1 
          - \frac{P\cdot P}{2P\cdot z_2} \a_2 z_2\, . \nn 
\ee 
We resolve the delta function and perform the $\a_2$ integral to fix
\be
\a_2 = \frac{1-\a_1}{1-\a_1 \zeta}\, .
\ee 
Since $2-\a_1> \a_2>0$, the theta function is nonzero for $\a_1<1$ which restricts the final rescaled ``energy-integration'' range
\begin{align}
\begin{split}
\label{eq:gen_2pt_detector_final}
\mathbb{E}^{(1)}_2(p_i;z_1,z_2) &= 
\frac{1}{N_k\, (2\pi)^{2d-1}}
\frac{1}{P\cdot P}
\left[\frac{2P\cdot z_1}{P\cdot P}\right]^{J_{L1}}
\left[\frac{2P\cdot z_2}{P\cdot P}\right]^{J_{L2}}   \\
&\quad \times 
\int_0^1 d\a_1 \, \a_1^{-J_{L1}-1}\left(1-\a_1\right)^{-J_{L2}-1}(1-\a_1 \zeta)^{J_{L2}} \big|\cM^{(0)}_{k\to3}(p_i, q_j^*)\big|^2\,.
\end{split}
\end{align}
The amplitude squared is evaluated at the ``detector configuration''
\be \label{eq:EEC_2to3_detector_config}
q_1^* &=  \frac{P\cdot P}{2P\cdot z_1} \a_1 z_1  \,, \nn \\ 
q_2^* &=  \frac{P\cdot P}{2P\cdot z_2} \frac{1-\a_1}{1-\a_1 \zeta} z_2 \,, \\ 
q_3^* &= P-q_1^*-q_2^* 
        = P - \frac{P\cdot P}{2P\cdot z_1} \a_1 z_1 
        - \frac{P\cdot P}{2P\cdot z_2} \frac{1-\a_1}{1-\a_1 \zeta} z_2\, . \nn
\ee 
\Eq{eq:gen_2pt_detector_final} is the final result for the 2-point correlator recoiling against a single massless particle. For a concrete theory under consideration, we can insert the desired amplitude, perform the final state sum, evaluate the resulting expression on the above detector configuration and perform the single energy integral. We will explicitly show results for a state generated by two massive scalar particles minimally coupled to general relativity in \Sec{sec:eec_5pt}. 

The calculation of higher-point correlators proceeds similarly, and is briefly discussed in appendix \ref{app:multi-pt-kinematics}.

\subsection{Detector correlators in the collinear limit}\label{sec:collinear}

The above expressions for general kinematics of the detectors is unwiedly due to the numerous kinematic variables, and the strong dependence on the state. In the context of gauge theories, an interesting kinematic limit that has been extensively studied is the multi-collinear limit. In this case the $N$-point correlator becomes a function of $2(N-2)$ cross-ratios. At lowest order, it can be expressed as a finite integral in $N-1$ dimensional projective space. It has been computed at both three- and four-points in $\mathcal{N}=4$ SYM \cite{Chen:2019bpb,Chicherin:2024ifn}, and its integrand has been obtained up to 11 points \cite{He:2024hbb}. Due to the particularly simple structure of gravity in the collinear limit, we believe that these multi-point correlators of detector operators also deserve exploration in perturbative (super-) gravity. We therefore briefly review the simplification of the integration in the collinear limit.

In the collinear limit we can take one cross ratio, for example $\zeta = \zeta_{12}$, as a scaling parameter, leaving a non-trivial dependence on the shape parameters $\zeta_{ab}/\zeta$. Taking the leading scaling behavior in $\zeta$  of the higher point correlator formula~\eqref{eq:gen_npt_detector_final} gives
\be 
\label{eq:collinear limit integral formula for n detectors from n+1 amplitude}
\mathbb{E}_{n,{\rm col.}}^{(1)}(p_i,z_a) 
&= \frac{1}{N_k(2\pi)^{2d-1} } \frac{1}{P\cdot P}\left[\frac{2P\cdot z_1}{P\cdot P}\right]^{\sum\limits^n_{a=1} J_{La}}  
\nn \\ & \times 
\int_0^\infty \prod^n_{a=1} d\a_a\,  \a_a^{-J_{La}-1} \, \delta(1-\sum^n_{a=1}\a_a ) 
\big|\cM^{(0),{\rm col.}}_{k\to n{+}1}(p_i, q_j^{c})\big|^2\, .
\ee 
Here $\cM^{(0),{\rm col.}}_{k\to n{+}1}$ is the collinear limit of the scattering amplitude evaluated for momenta $q^c_j$ specified by the shape parameters. Explicit parameterizations can be found in Refs.~\cite{Chen:2019bpb,Chicherin:2024ifn}. Compared to the integral for generic kinematics in~\Eq{eq:gen_npt_detector_final}, the collinear limit has simplified the constraint on the integration parameters, $\alpha_a$, reducing it to a simple $n-1$ dimensional projective integral. This makes the collinear limit of detector operators a particularly appealing target of study for the application of integration techniques for finite loop integrals \cite{Caron-Huot:2014lda,Henn:2022vqp,Gambuti:2023eqh}, particularly in Feynman parameter representation \cite{Artico:2023bzt,Artico:2023jrc,Britto:2023rig,Chen:2019mqc,Lee:2014tja,Chicherin:2024ifn,Chen:2024xwt}. We will systematically explore the structure of these multi-point correlators in the collinear limit of (super-)gravity in future work.

\newpage
\section{Computation of (squared) S-matrix elements}
\label{sec:amplitude_review}

Our calculation relies on the availability of perturbative scattering amplitudes in gravitational theories. Such amplitudes are notoriously difficult to compute using standard Feynman diagrammatic techniques.  In the past decades there has been tremendous progress in our understanding of scattering amplitudes in perturbative gravity driven by generalized unitarity \cite{Bern:1996fj,Bern:1994cg,Bern:1994zx}, and the color-kinematics duality \cite{Bern:2008qj, Bern:2010yg}  (see e.g.~the reviews \cite{Travaglini:2022uwo,Bern:2019prr,Elvang:2013cua}). We do not wish to give a comprehensive review of the multi-dimensional progress in the field of modern on-shell methods here, but instead, focus on the advances relevant for our present work and for immediate future directions in the calculation of detector observables in perturbative gravity.

As discussed in \Sec{sec:eec_def}, the calculation of detector correlators requires the calculation of state-summed squared amplitudes. These can be obtained either by directly squaring and performing the state sum of available amplitudes, or from unitarity cuts of multi-loop amplitudes. After a brief review of the color-kinematics duality, which is used to construct the scattering amplitudes themselves, we then discuss how to construct the squared amplitudes by directly performing the state sums in \Sec{sec:state_sum}, and through unitarity cuts in \Sec{sec:unitarity}.

\subsection{Color-kinematics duality and the double copy}

We start our discussion by reviewing several features of gravitational scattering amplitudes. Notoriously, starting from the Einstein-Hilbert action \Eq{eq:grav_action} and deriving the associated Feynman rules leads to an enormous proliferation of terms, especially at higher multiplicity due to diffeomorphism redundancies. In contrast, inspired by string theory, there were first hints that gravitational tree-level scattering amplitudes are the ``square'' of much simpler gauge theory ones \cite{Kawai:1985xq}. Subsequently, the ``squaring'' idea has been vastly generalized and extended to e.g.~loop level in the form of the color-kinematics duality and the associated double-copy construction~\cite{Bern:2008qj, Bern:2010yg} that have been the main tools for constructing higher-loop integrands in gravitational theories with applications to e.g. the study of UV properties of supergravity theories (see e.g. reviews \cite{Bern:2011qn,Bern:2023zkg}), or in classical gravitational-wave physics~\cite{Bern:2019nnu, Bern:2019crd, Bern:2021dqo, Bern:2021yeh}. (For a recent extension of the double copy to all loop orders, see \cite{Bern:2024vqs}.) In special circumstances it is also possible to extend the double copy beyond perturbation theory \cite{Monteiro:2014cda}. For a comprehensive review, see e.g.~\cite{Bern:2019prr}.  

The direct application of the double copy relies on the presence of gauge-theory amplitudes that exhibit color-kinematics duality. At higher loop orders, it becomes increasingly challenging to find explicit forms of gauge-theory integrands that meet this criterion~\cite{Bern:2015ooa, Mogull:2015adi, Bern:2012uc}. In contrast, tree-level amplitudes demonstrating color-kinematics duality are readily accessible (see e.g.~\cite{Edison:2020ehu}), allowing for the construction of gravitational tree-level amplitudes with virtually any number of external legs through the double copy, independent of four-dimensional spinor-helicity techniques. Since our discussion will focus on gravitational detector correlators at leading order, we will only need tree-level amplitudes. Additionally, the novel approach to the all-orders loop-level double copy \cite{Bern:2024vqs} fundamentally relies on the tree-level double copy, so we will only cover these aspects here.

Our discussion of the double copy begins with color-dressed tree-level amplitudes from gauge theory. We primarily focus on pure gluon and graviton amplitudes, later detailing how to obtain amplitudes involving massive scalars through dimensional reduction. While generalizations to include fermionic external states are known, they are not central to our discussion; interested readers can refer to sources such as~\cite{Bern:2019prr, Edison:2020ehu}. By appropriately multiplying and dividing by inverse propagators, we can generally reorganize the amplitudes as a sum over diagrams that consist solely of three-point vertices (cubic diagrams).
\begin{equation}
i {\cal A}^\tree_m(1,2,3,\ldots,m)\,= g^{m-2} \sum_{i}
                \frac{n_i c_i }{\prod_{\alpha_i} p^2_{\alpha_i}}\,,
\label{Anrep}
\end{equation}
where the summation label $i$ runs over $(2m-5)!!$ diagrams. The $c_i$ are color factors obtained by attaching a group-theory structure constant $f^{abc}$ to each vertex of the diagram and a $\delta^{ab}$ to each internal line. The kinematic numerators  $n_i$ depend on external momenta and polarization vectors.  The $p^2_{\alpha_i}$ in the denominator represent the inverse propagators for the $(m-3)$ internal lines $\alpha_i$ of diagram $i$. The gauge-theory coupling constant is $g$.

\begin{figure}[tb]
\centering
\begin{subfigure}[b]{0.3\textwidth}
\centering
\renewcommand\thesubfigure{(i)}
    $\vcenter{\hbox{\includegraphics{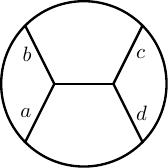}}}$
    \caption{}
    \label{subfig:4ptGluon_s_blob}
\end{subfigure}
\begin{subfigure}[b]{0.3\textwidth}
\centering
    \renewcommand\thesubfigure{(j)}
    $\vcenter{\hbox{\includegraphics{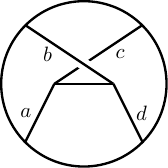}}}$
    \caption{}
    \label{subfig:4ptGluon_u_blob}
\end{subfigure}
\begin{subfigure}[b]{0.3\textwidth}
\centering
    \renewcommand\thesubfigure{(k)}
    $\vcenter{\hbox{\includegraphics{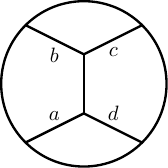}}}$
    \caption{}
    \label{subfig:4ptGluon_t_blob}
\end{subfigure}

\caption{
\label{GeneralJacobiFigure}
The color factors of the three indicated diagrams form a color Jacobi identity, see \Eq{jacobic}. The circle surrounding the exposed lines represents an embedding in a larger diagram. In a color-dual representation, the kinematic numerators of the diagrams satisfy the same relation as the color factors, see \Eq{jacobin}. 
}
\renewcommand\thesubfigure{(\arabic{subfigure})}
\end{figure}

The $f^{abc}$'s are completely antisymmetric and obey the Jacobi relation, 
\begin{align}
    f^{abe} f^{ecd} - f^{ace} f^{ebd} - f^{aed} f^{ebc}=0 \,.
\label{ColorJacobi}
\end{align}
Pictorially, the three terms in the Jacobi relation correspond to the color factors of the three sub-diagrams in Fig.~\ref{GeneralJacobiFigure}. If these three sub-diagrams are embedded in a larger diagram with the remaining parts identical, the Jacobi identity continues to hold, as it is independent of kinematics.  Labeling the three diagrams appearing in such a Jacobi triplet by $i,j,k$,  Eq.~\eqref{ColorJacobi} translates to 
\begin{equation}
    c_i - c_j - c_k = 0\, .
\label{jacobic}
\end{equation}
where the $c_{i}$, $c_{j}$ and  $c_{k}$ are the color factors of the larger diagrams. For every internal line of every diagram, there is one such Jacobi relation connecting it with two others.

The color Jacobi relations allow us to modify the numerator factors $n_i$ without modifying the amplitude. The amplitude is said to obey the duality between color and kinematics if the kinematic numerators $n_i$ are arranged to obey the same algebraic relations as the corresponding color factors $c_i$
\begin{align}
c_i \rightarrow -c_i \quad \Rightarrow \quad  n_i \rightarrow -n_i\,,  
\hskip 1.5 cm 
c_i =  c_j + c_k  \quad \Rightarrow \quad n_i =  n_j + n_k \, .
\label{jacobin}
\end{align}
In accordance with Fig.~\ref{GeneralJacobiFigure}, the kinematic Jacobi identity mandates that lines shared among the three participating graphs be labeled consistently. Additionally, it is crucial to avoid imposing any relations that apply solely to specific gauge groups or representations. 

The duality between color and kinematics has been established at tree level in gauge theories~\cite{Cachazo:2012uq, Chen:2011jxa, Feng:2010my, Stieberger:2009hq, Bjerrum-Bohr:2009ulz, Bern:2010yg}, with amplitudes demonstrating this duality available for all multiplicities~\cite{Edison:2020ehu}. One of the most striking features of amplitudes that exhibit color-kinematics duality is that a gravitational amplitude can be constructed by substituting the color factors of one gauge-theory amplitude with the color-dual kinematic numerators from another:
\begin{equation}
\label{cTon}
c_i\rightarrow n_i \, .
\end{equation}
Specifically, focusing on tree-level amplitudes, consider a tree-level $m$-point gauge-theory amplitude as given in Eq.~\eqref{Anrep}, alongside a second $m$-point amplitude from another gauge theory, which may have different field content or a different gauge group. The corresponding gravitational amplitude can then be expressed as follows:
\begin{eqnarray}
&& i {\cal M}^\tree_m(1,2,\ldots,m)\,=
 \Big(\frac{\kappa}{2}\Big)^{m-2}  \sum_{i} \frac{n_i\, \widetilde{n}_i}{ \prod_{\alpha_i} p^2_{\alpha_i}} \, ,
\label{squaring}
\end{eqnarray}
where the sum runs over the same set of $(2m-5)!!$ diagrams as in Eq.~\eqref{Anrep}, and the $\widetilde{n}_i$ are the kinematic numerators of the second gauge theory. The gravitational coupling is related to Newton's constant via $\kappa^2 = 32 \pi G_N$.

Up to this point, we have focused on pure gluon and pure graviton amplitudes. However, for this paper, we are interested in scattering amplitudes involving a massive scalar pair interacting with gravitons. In our discussion so far, we have remained completely agnostic about the spacetime dimension. Thus, we can begin with higher-dimensional massless gauge or gravity amplitudes and dimensionally reduce them to four dimensions with a massive scalar pair. For example, consider pure Yang-Mills amplitudes in $d=6$, and then select the polarization vectors of the two legs—let’s say legs 1 and 2—that we wish to convert into massive scalars.
\begin{align}
\varepsilon^A_1 = \varepsilon^A_2 = (0^\mu,0,1)
\end{align}
and their momenta as 
\begin{align}
 p^A_1 = (p^\mu_1,m,0)\,, \qquad \qquad 
 p^A_2 = (p^\mu_2,-m,0)\,,
\end{align}
where $A=0,\ldots,5$ is a six-dimensional Lorentz index and $\mu=0,\ldots,3$ is a four-dimensional one. From the four-dimensional perspective, particles 1 and 2 are massive scalars with mass $m$. The polarization vectors and momenta of the remaining particles live in the four-dimensional subspace specified by the first four entries 
\begin{align}
\varepsilon^A_i = (\varepsilon^\mu_i,0,0) \,, \qquad p^A_i = (p^\mu_i,0,0)\,, \qquad \text{for}\quad i=3,\ldots,m\,.
\end{align}
Under double-copy, the scalars remain in place in the gravitational theory as well.

\subsection{Simplified state-sums for squared amplitudes}\label{sec:state_sum}

Having discussed the construction of the tree-level amplitudes of massive scalars minimally coupled to gravity from the double copy, we are left with performing the summation over final state graviton polarizations in \Eq{eq:eec_LO_GR} which was left implicit in writing the matrix element square 
\begin{align}
\label{eq:sample_state_sum}
\big|\cM^{(0)}_{2\to3}(p_1,p_2; q_i^*)\big|^2 \equiv 
\sum_{{\rm states}} \cM^{(0)}_{2\to3}(p_1,p_2; q_i^*) \bar{\cM}^{(0)}_{2\to3}(p_1,p_2; q_i^*)
\end{align}
To evaluate the state sums in \Eq{eq:sample_state_sum} we use completeness relations for the states that cross the cut. The sum over graviton states is realized in terms of sums over the single-copy gluon states
\begin{align}
 \Pi^{\mu\nu}(\ell,q) =\!\! \sum_{{\rm states}} \!\!
  \varepsilon^\mu(\ell)\varepsilon^\nu(-\ell) = \eta^{\mu\nu} - \frac{q^\mu \ell^\nu + \ell^\mu q^\nu}{q\cdot\ell}\,,
  \label{StateSumGauge}
\end{align}
For each color, gluons have $d-2$ states in $d$ dimensions, so the double copy naturally yields $(d-2)^2 = \frac{1}{2}d(d-3) + \frac{1}{2}(d-2)(d-3) + 1$ states, corresponding to a graviton, an antisymmetric tensor, and a dilaton. To obtain Einstein's gravity, we need to remove the antisymmetric tensor and dilaton from the cuts, which is accomplished by using the graviton physical-state projector 
\begin{align}
  \Pi^{\mu\nu \alpha \beta}(\ell,q)  =\!\! \sum_{{\rm states}} \!\!
  \varepsilon^{\mu\nu}(\ell)\varepsilon^{\alpha \beta}(-\ell) 
   = \frac{1}{2} \left[
      \Pi^{\mu \alpha}\Pi^{\nu \beta}
    + \Pi^{\nu \alpha}\Pi^{\mu \beta}
    -\frac{2}{d-2}\Pi^{\mu \nu}\Pi^{\alpha \beta}
    \right].
    \label{StateSumGravity}
\end{align}
Both \Eqs{StateSumGauge}{StateSumGravity} utilize a null reference vector $q^\mu$ to define the physical polarizations. This null reference vector can be effectively eliminated by carefully selecting the form of the tree amplitudes. Specifically, $q$ always appears in conjunction with a linearized gauge transformation, $\varepsilon^\mu(\ell) \to \ell^\mu$, of a tree amplitude. While tree amplitudes are gauge invariant, this typically necessitates the transversality of the other polarizations. However, it is possible to arrange them so that they satisfy the generalized Ward identity \cite{Kosmopoulos:2020pcd}, meaning they vanish when replacing $\varepsilon^\mu(\ell)$ with $\ell^\mu$ without imposing additional constraints on the other polarizations. Employing such tree amplitudes ensures that all $q$-dependent terms vanish from the outset, thereby simplifying the projectors
\begin{align}
\label{StateSumsSimple}
\Pi^{\mu\nu}(\ell,q)\to \eta^{\mu\nu}\,, 
\qquad
\Pi^{\mu\nu \alpha \beta}(\ell,q)  \to
    \frac{1}{2} \left[
        \eta^{\mu \alpha}\eta^{\nu \beta}
    + \eta^{\nu \alpha}\eta^{\mu \beta}
    -\frac{2}{d-2}\eta^{\mu \nu}\eta^{\alpha \beta}
    \right]  \,;
\end{align}
see Ref.~\cite{Kosmopoulos:2020pcd} for further details.

\subsection{Generalized and reverse unitarity}\label{sec:unitarity}

There is yet another perspective that one could take when looking at the squared matrix elements, such as the ones in \Eq{eq:sample_state_sum}, appearing in the calculation of detector correlators. Based on the idea of reverse unitarity \cite{Anastasiou:2002yz,Anastasiou:2002qz,Anastasiou:2003yy,Anastasiou:2015yha}, which is well established in collider physics applications, the squared matrix elements can be interpreted as the \emph{unitarity cuts} of loop amplitudes in the forward limit. For example, the squared matrix element in \Eq{eq:sample_state_sum} can be interpreted 
\begin{align}
\hspace{-.4cm}
\text{Cut}\left[\cM^{(2)}_{2\to 2} (p_1,p_2 {\to} p_1,p_2)\right] 
= \!\!\vcenter{\hbox{\scalebox{.8}{\revUnitarityGR}}} \!\!
= \!\! \sum_{{\rm states}}\!\!\cM^{(0)}_{2\to3}(p_1,p_2; q_i) \bar{\cM}^{(0)}_{2\to3}(p_1,p_2; q_i),
\hspace{-.3cm}
\end{align}
as the unitarity cut of a two-loop amplitude of massive external scalars in forward scattering kinematics, where the $q_i$ are re-interpreted as on-shell loop momenta. Such a framework meshes well with the construction of higher loop amplitudes within the generalized unitarity method \cite{Bern:1994zx, Bern:1994cg, Bern:1995db, Bern:1997sc, Britto:2004nc}. Such a perspective was also successfully adopted to compute observables relevant for classical gravitational wave physics, see e.g.~\cite{Herrmann:2021lqe,Herrmann:2021tct,Herderschee:2023fxh}. For the purposes of this article, where the explicit examples only involve tree-level amplitudes, the full machinery of generalized unitarity is not necessary, so we mostly refer to review articles \cite{Bern:1996je, Bern:2011qt} and a recent streamlined approach \cite{Bern:2024vqs} for details. However, going beyond leading order calculations, generalized unitarity has a track record with a wide array of applications (see, e.g., Refs.~\cite{Bern:1998ug, Bern:1998sv, Bern:2007hh, Bern:2009kd, Drummond:2008bq, Berger:2008sj, Berger:2010zx, Abreu:2020xvt, Abreu:2021asb}) as an efficient way forward, which is why we mention it here. The key idea of the latest incarnations of the generalized unitarity framework is the notion that multi-loop \emph{integrands}, roughly the sum of Feynman diagrams before loop integration, are rational functions in the loop and external momenta. Being rational functions, such loop integrands are then completely determined by the knowledge of the location of the poles and their associated residues. Exploiting unitarity and analytic properties of scattering amplitudes of course has a long history dating back to the inception of quantum mechanics. The concept of the analytic S-matrix theory~\cite{Cutkosky:1960sp, Eden:1966dnq} became a fundamental tool for constraining particle interactions during the 1960s and the factorization properties of loop level integrands are essentially based on repeated use of the optical theorem in perturbation theory
\begin{align}
\label{eq:basic_unitarity_relation}
-i (T-T^\dagger) = T T^\dagger\,,
\end{align}
which allows to construct loop amplitudes from tree-level ones. For further details, see e.g. the brief summary in Ref.~\cite{Bern:2024vqs} and references therein. 

Although we will not pursue this approach further in this paper, we believe that this approach will be particularly convenient for the calculation of energy correlators in supergravity theories. In the case of $\mathcal{N}=8$ SUGRA, the four-point scattering amplitude is known to five-loops \cite{Bern:2008pv,Bern:2009kd,Bern:2010tq,Bern:2012uc}, which would enable the calculation of multi-loop or multi-point energy correlators.  Obtaining the integrand for the energy correlators directly from these loop amplitudes avoids the need to perform the state sum, which becomes particularly advantageous in theories with high degrees of supersymmetry. We leave an investigation of energy correlators in perturbative supergravity theories to a future paper.

%
\section{The EEC in general relativity}
\label{sec:eec_5pt}
%

In this section we explicitly compute the two-point energy correlator in Einstein-Hilbert gravity \Eq{eq:grav_action}, in a state of gravitons produced through the annihilation of two massive scalars. To our knowledge, this is the first calculation of a multi-point correlator in quantum gravity.

\subsection{Kinematics and calculation}

\begin{figure}[hb!]
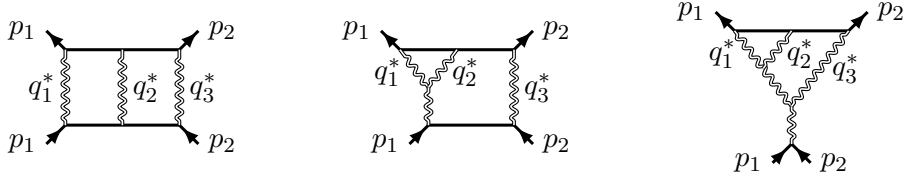

\centering
    $\vcenter{\hbox{\sampleDiagEECa}}$
\qquad
    $\vcenter{\hbox{\sampleDiagEECb}}$
\qquad
    $\vcenter{\hbox{\sampleDiagEECc}}$
\caption{\label{fig:amp_square_gr_sample_diag_5pt}Three sample diagrams contributing to the (final state polarization summed) squared amplitude relevant for the computation of the two-detector correlator in \Eq{eq:eec_LO_GR}. Solid lines represent massive scalars, and wiggly lines denote gravitons. The $q^*_i$ are defined in \Eq{eq:EEC_2to3_detector_config} and the unlabeled lines in the diagram are off-shell particles.}
\end{figure}

The observable of interest is
\be
\mathbb{E}_2 (p_1,p_2;z_1,z_2) &= \frac{1}{N_2}
\<p_1, p_2| \cD_{J_{L1}}(z_1) \cD_{J_{L2}}(z_2) |p_1,p_2\>\,.
\ee 
which can be expanded in perturbation theory. Adapting our general expression in \Eq{eq:event_shape_g_expansion} to the gravitational context and specializing to $k=2$ initial state particles and $n=2$ detectors, we have 
\be
\mathbb{E}_2 (p_1,p_2;z_1,z_2) &= 
  \kappa^4 \, \widetilde{\mathbb{E}}^{(0)}_2 
+ \kappa^6  \left(\mathbb{E}^{(1)}_{2,{\rm contact}} + \mathbb{E}^{(1)}_2\right) + \cO(\kappa^8)\,,
\ee
taking into account that $\kappa$ plays the role of the weak coupling constant $g$. As explained previously, the terms $\widetilde{\mathbb{E}}^{(0)}_2$ and $\mathbb{E}^{(1)}_{2,{\rm contact}}$ are localized to restricted kinematic support and can be described as ``contact terms''. Here, we will compute the leading term with support on generic detector angles, $\mathbb{E}^{(1)}_2$. The process can be depicted as follows:
\begin{align}
\label{eq:eec_LO_GR}
\mathbb{E}^{(1)}_2 (p_1,p_2;z_1,z_2) 
=
\!\!\!
\vcenter{\hbox{\twoDetectorOneGR}}\!\!\!
&= 
\frac{1}{N_2\, (2\pi)^{2d-1}}
\frac{1}{P\cdot P}
\!\left[\frac{2P\cdot z_1}{P\cdot P}\right]^{J_{L1}}\!\!
\left[\frac{2P\cdot z_2}{P\cdot P}\right]^{J_{L2}}  
\hspace{-1cm}
\\[-45pt]& \hspace{0cm}\times \!\!
\int_0^1\!\!\!\! d\a \, 
\a^{-J_{L1}-1}\left(1{-}\a\right)^{-J_{L2}-1}(1{-}\a \zeta)^{J_{L2}} \big|\cM^{(0)}_{2\to3}(p_1,p_2; q_i^*)\big|^2\,. \nn \\ \nn
\end{align}
The incoming two-particle state is comprised of two massive scalars with momenta $p_1$ and $p_2$ that both square to $m^2$. Gravitons are denoted by wavy lines, and the detector vertices (defined in \Eq{eq:detector_vertex_gen_J}) are represented by small blue dots with a wavy graviton line entering and exiting. Specializing \Eq{eq:gen_2pt_detector_final} to the case of two massive incoming states, the $q_j^*$ are given by the detector configuration~\Eq{eq:EEC_2to3_detector_config} and we have renamed the integration variable $\alpha_1\to \alpha$ for brevity. The flux normalization factor $N_2$ for two-particle states is defined as
\begin{align}
\label{eq:flux_factor_2_particles}
  N_{2}= 4\,  \sqrt{ (p_1 \cdot p_2)^2 - m^4}\, .
\end{align}  
Having reviewed the basics on how to obtain gravitational scattering amplitudes from the double copy and perform the final state sums for the squared matrix elements in \Sec{sec:amplitude_review}, we can proceed with our discussion of the leading finite-angle contribution to the two-point energy correlator in perturbative quantum gravity, defined in \Eq{eq:eec_LO_GR}. Concretely, this observable requires the tree-level scattering amplitude $\cM^{(0)}_{2\to 3}$ of two scalars to three gravitons, $\Phi\Phi\to hhh$ which can be written as a sum of 15 cubic diagrams, where the diagram numerators can be obtained by suitably squaring the gauge theory numerators from Ref.~\cite{Edison:2020ehu}, according to \Eq{squaring}. The squared matrix element can then be evaluated by performing the final state polarization sum for the gravitons with the help of the graviton projector in \Eq{StateSumGravity} (or its simplified version in \Eq{StateSumsSimple}). Naively, this leads us to evaluate $15^2=225$ cubic contributions, many of which are related by diagram symmetries. A representative selection of the relevant graphs is shown in \Fig{fig:amp_square_gr_sample_diag_5pt}. 

These diagrams are to be interpreted as follows: lines labeled by $q^*_i$ are the on-shell final states of the individual trees appearing in \Eq{eq:eec_LO_GR} which are sewn with the projector in  \Eq{StateSumGravity}. Schematically, an individual contribution looks as follows, e.g.
\begin{align}
\label{eq:sample_amp_square_contribution_5pt}
\vcenter{\hbox{\sampleDiagEECb}} = \frac{n(p_i\cdot q^*_j)}{(q^*_1+q^*_2)^2 \left[(p_1{-}q^*_1{-}q^*_2)^2-m^2\right]^2 \left[(p_1{-}q^*_1)^2-m^2\right]^{\phantom{2}}}
\end{align}
where each individual contribution (and therefore the final (polarization summed) squared matrix element) only depends on Lorentz dot products between the various momenta $p_1,p_2,q^*_1,q^*_2,q^*_3$. Using on-shell conditions and momentum conservation, one can reduce the number of independent variables to six (five scalar products and $m^2$).  Despite the fact that individual tree-amplitudes are obtained by squaring simple gauge-theory amplitudes, the graviton state sum destroys this simple factorized structure. In order to avoid writing a proliferation of terms, we left the polynomial numerator $n(p_i\cdot q^*_j)$ (of degree 6 in Lorentz dot products) implicit.

\medskip

With the polarization summed squared amplitude at hand, let us briefly discuss the kinematic variables relevant for the two-point energy correlator \Eq{eq:eec_LO_GR}. A detailed summary of all kinematic variables is provided in \App{app:kinematics}. Since the initial state is comprised of two massive scalars with momenta $p_1$ and $p_2$ ($p^2_1=p^2_2=m^2$), the initial state kinematics is slightly more involved compared to a single incoming momentum eigenstate relevant to scenarios where the initial state is generated by the action of a local operator. The key difference between the two setups is the residual symmetry of the source. For a single (timelike) source, one can go to the rest frame of the source and retain a $(d{-}1)$-dimensional spatial rotation symmetry, whereas the center-of-mass system of the two-particle initial state singles out a ``beam axis'' so that we are left with a $(d{-}2)$-dimensional symmetry of axial rotations.  In the scattering setup of interest here, it is convenient to define
\be
P=p_1+p_2, \qquad Q=p_1-p_2\,,
\ee 
that satisfy $P^2=s$, $P\cdot Q=0$, and $Q^2 = -(s-4m^2)$. From dimensional analysis, we can factor out the overall mass dependence, leaving us with a dimensionless ratio $x$
\begin{align}
    x^2 \equiv \frac{-Q^2}{P^2} = \frac{s-4m^2}{s}\,,
\end{align}
where, in physical scattering kinematics for which $s>4m^2$ and $Q^2<0$, we can choose the branch $0\leq x\leq 1$. The lower bound $x=0$ corresponds to the threshold limit $s=4m^2$, whereas $x=1$ can be interpreted either as the massless limit $m=0$, or the ultra-relativistic limit $s\gg m^2$.  
It is convenient to extract the mass dependence of the mod square of the amplitude \Eq{eq:eec_LO_GR} 
\be
\label{eq:amp_square_mass_scaled}
\left|\cM^{(0)}_{2\to3}\left(p_i, q_j^*\right) \right|^2 = s^2 \, \left|\cM^{(0)}_{2\to3}\left(\frac{p_i}{\sqrt{s}}, \frac{q_j^*}{\sqrt{s}}\right) \right|^2
\ee
which follows from standard dimension counting of an $n$-point amplitude $\cM^{(0)}_n \sim s^{n-4}$. We can slightly re-write \Eq{eq:eec_LO_GR} 
\be
\mathbb{E}^{(1)}_{2}(p_1,p_2;z_1,z_2) \equiv 
\frac{P\cdot P}{N_2\, (2\pi)^{2d-1}} 
\left[\frac{2P\cdot z_1}{P\cdot P}\right]^{J_{L1}}
\left[\frac{2P\cdot z_2}{P\cdot P}\right]^{J_{L2}}
\cG^{(1)}_{2}(\zeta,\chi_1,\chi_2,x) \, ,
\ee 
by introducing the dimensionless integral
\be
\label{eq:cG_LO_GR} 
\cG_{2}^{(1)}(\zeta,\chi_1,\chi_2,x) &= 
\!\!\int_0^1\!\! d\a \, 
    \a^{-J_{L1}-1} 
    \left({1{-}\a}\right)^{-J_{L2}-1} 
    ({1{-}\a\, \zeta})^{J_{L2}} 
    \left|\cM^{(0)}_{2\to3}\left(\frac{p_i}{\sqrt{s}}, \frac{q_j^*}{\sqrt{s}}\right) \right|^2
\,,
\ee
where the amplitude is evaluated on the ``detector configuration'' defined in \Eq{eq:EEC_2to3_detector_config}. The fact that $\cG_{2}^{(1)}(\zeta,\chi_1,\chi_2,x)$ depends on four dimensionless variables can be understood as follows. Upon scaling out the mass dependence from the squared matrix element in \Eq{eq:amp_square_mass_scaled}, the result depends on five independent Lorentz invariants. Performing the $\alpha$ integral eliminates one further variable, leaving us with the four remaining ones. 

In order to describe the kinematics associated with the detectors, we have already defined the cross ratio $\zeta$ in \Eq{eq:zeta definition 2 detector}, but require two additional cross ratios $\chi_1, \chi_2$
\be
\label{eq:chi_def_gr_eec}
    \chi_a &= -\frac{Q\cdot z_a}{P\cdot z_a} \,, \qquad \text{where} \quad a=1,2\,,
\ee 
to characterize the location of the detector with respect to the ``beam axis" of the incoming particles. For physical scattering kinematics, the cross ratios are constrained to the range $-x \leq \chi_a \leq x$. 

\noindent
For the EEC in $d=4$, we set $J_{L1}{=}J_{L2} = 1{-}d=-3$ leaving the final expression to compute 
\be 
\label{eq:EEC_final_G}
\cG^{(1)}_{\text{EEC}}(\zeta,\chi_1,\chi_2,x) 
&= \! \int_0^1 \!\! d\a\, \a^{2}\left({1-\a}\right)^{2}({1-\a\, \zeta})^{-3} 
\left|\cM^{(0)}_{2\to3}\left(\frac{p_i}{\sqrt{s}}, \frac{q_j^*}{\sqrt{s}}\right) \right|^2\,.
\ee
As explained above, the polarization summed squared matrix element is written as a sum over 225 contributions of the form in \Eq{eq:sample_amp_square_contribution_5pt}. Each of the contributions can be written in terms of the ``detector variables'' $\alpha,x,\chi_1,\chi_2, \zeta$ by inserting the momenta as written in \Eq{eq:EEC_2to3_detector_config} together with the cross-ratio definitions \Eqs{eq:zeta definition 2 detector}{eq:chi_def_gr_eec} into the basis of kinematic invariants for the $2\to3$ scattering process:
\begin{align}
&\frac{m^2}{s} = \frac{(1{-}x^2)}{4}\,,
&&\frac{(p_1 \cdot p_2)}{s} = \frac{(1{+}x^2)}{4}\,,
&&&\frac{(q^*_1\cdot q^*_2)}{s} = \frac{\alpha\, \zeta \, (1{-}\alpha)}{2(1{-}\alpha\, \zeta)}\,,
\\
&\frac{(p_1 \cdot q^*_1)}{s} = \frac{\alpha (1{-}\chi_1)}{4} \,,
&&\frac{(p_2 \cdot q^*_1)}{s} = \frac{\alpha (1{+}\chi_1)}{4} \,,
&&&\frac{(p_1 \cdot q^*_2)}{s} =  \frac{(1-\alpha)(1-\chi_2)}{4(1-\alpha\, \zeta)}\,.
\end{align}
At the end of the day, we are left with the one-fold integral over $\alpha$ over a rational function in $\alpha$. The denominators in the squared matrix elements have $\alpha$-dependent singularities at the following locations
\begin{align}
\left\{
\frac{1}{1+ \alpha^2 \zeta - 2 \alpha\, \zeta \pm 
    \big[(1-\alpha)\chi_2+\alpha(1-\alpha\zeta)\chi_1\big]} 
, 
\frac{1}{1{-}\alpha\, \zeta} 
\right\}
\,,
\end{align}
all other potential singularities, e.g. at $\alpha =0$ or $\alpha=1$ disappeared in the sum over contributions. It is advantageous to group the alpha integrals as follows
\begin{align}
\cI[a_1,a_2,a_3,a_4] = \int^1_0 d\alpha\, \alpha^{a_1}\, (1-\alpha \zeta)^{a_2} [\text{poly}_1(\alpha)]^{a_3}[\text{poly}_2(\alpha)]^{a_3}\,,
\end{align}
where $\text{poly}_{1,2}(\alpha)= 1+ \alpha^2 \zeta - 2 \alpha\, \zeta \pm \big[(1-\alpha)\chi_2+\alpha(1-\alpha\zeta)\chi_1\big]$. In the next step, we can simplify the integrals further by partial fractioning the integrand with respect to $\alpha$ and recollecting the result. At the end of this procedure, we are left with 21 finite, elementary integrals that can be evaluated in e.g. \texttt{Mathematica}.

\subsection{Results for the full-angle EEC}

We obtain a result for the leading quantum correction to the two-point energy correlator in Einstein gravity,  which can be written compactly as
\begin{align}
\label{eq:eec_fi_representation}
\cG^{(1)}_{\text{EEC}}(\zeta,\chi_1,\chi_2,x) = r^{(0)}(\zeta,\chi_1,\chi_2,x) + \sum^7_{i=1} r^{(i)}(\zeta,\chi_1,\chi_2,x) \times f^{(i)}(\zeta,\chi_1,\chi_2)\,,
\end{align}
where we introduce the transcendental functions
\begin{align}
\begin{split}
f^{(1)}(\zeta,\chi_1,\chi_2) & =\frac{
    \text{arctan}\left[\frac{\chi_2-\chi_1 + 2 \zeta \, \chi_1}{\sqrt{\Delta }}\right]
    -
    \text{arctan}\left[\frac{\chi_2-\chi_1 - 2 \zeta}{\sqrt{\Delta }}\right]}{\sqrt{\Delta }} \,, 
\\
f^{(2)}(\zeta,\chi_1,\chi_2) & = \frac{
    \text{arctan}\left[\frac{\chi_2-\chi_1 + 2 \zeta}{\sqrt{\Delta }}\right]
    -
    \text{arctan}\left[\frac{\chi_2-\chi_1 - 2 \zeta}{\sqrt{\Delta }}\right]}{\sqrt{\Delta }}\,,
\\
f^{(3)}(\zeta,\chi_1,\chi_2) & = \log \left(1-\chi_1\right)\,, \qquad \qquad
f^{(4)}(\zeta,\chi_1,\chi_2)  = \log \left(1+\chi_1\right)\,,
\\
f^{(5)}(\zeta,\chi_1,\chi_2) & = \log \left(1+\chi_2\right)\,, \qquad \qquad 
f^{(6)}(\zeta,\chi_1,\chi_2)  = \log \left(1-\chi_2\right)\,,
\\
f^{(7)}(\zeta,\chi_1,\chi_2) & = \log \left(1-\zeta\right)\,.
\end{split}
\end{align}
The discriminant $\Delta$ is given by 
\begin{align}
\Delta =4 \zeta (1 - \zeta) + 2 \chi_1 \chi_2 (1 - 2 \zeta) - \chi^2_1 - \chi^2_2  \,.
\end{align}
The identical coefficients $r^{(3)}=r^{(6)}$, as well as $r^{(4)}=r^{(5)}$ of the respective logarithms are related by a $\chi_1\leftrightarrow \chi_2$ interchange, so that there are only five independent transcendental functions left. The final result is given in the ancillary file \texttt{eec\textunderscore curlG\textunderscore to\textunderscore functions.m}.

As expected, the result is expressed in terms of weight-1 functions, and takes a form quite similar to its gauge theory counterpart \cite{Basham:1978zq,Basham:1979gh,Basham:1977iq,Basham:1978bw}. However, particularly in its kinematic limits, we will see that it behaves quite differently than in gauge theory, due to the different infrared structure of gravity.

In order to get a better feeling for the physics, it is beneficial to go to the center-of-mass frame, where we can re-parameterize the cross ratios in terms of CoM angles (see appendix \ref{app:kinematics} and \Fig{fig:com_angles} for details)
\begin{align}
\label{eq:eec_xrats_to_com_angles}
\hspace{-.2cm}
  \zeta = \frac{1}{2}\left[1{-}\cos \theta\right]\,, \quad 
  \chi_1 = x \cos \psi \,, \quad
  \chi_2 = x \left[\cos \theta \cos \psi {-} \cos\phi\sin\theta\sin\psi \right]\,,
\hspace{-.2cm}
\end{align}
where $\psi$ is the scattering angle of detector 1 w.r.t.~the beam axis, $\theta$ is the opening angle between the two detectors, and $\phi$ is an azimuthal rotation angle of detector 2 around detector 1. 


%
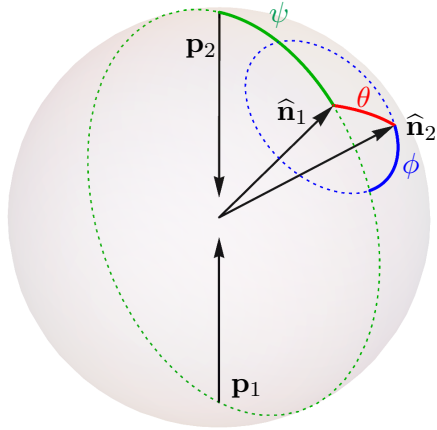
\begin{figure}
\begin{center}
\begin{overpic}[scale=0.166,unit=1mm]{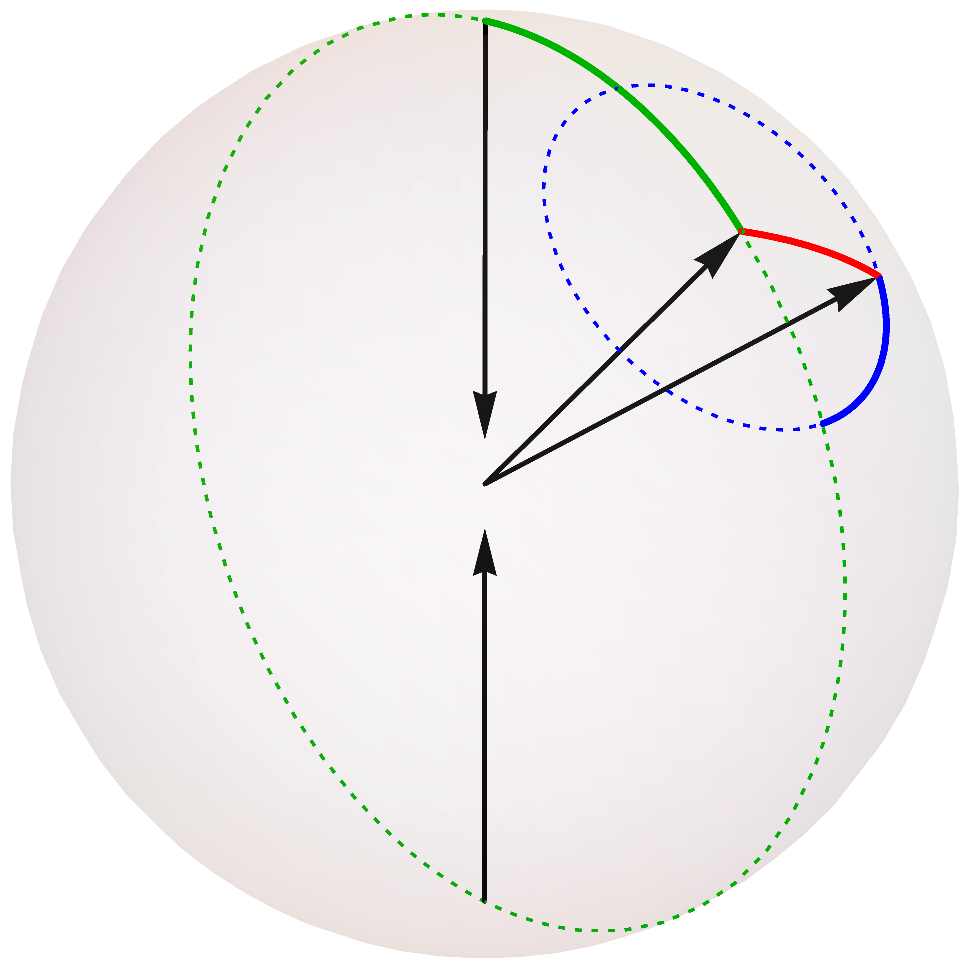}
    \put(35,54){{\color{ForestGreen} $\psi$}}
    \put(52.5,34){{\color{blue} $\phi$}}
    \put(46.5,42.8){{\color{red} $\theta$}}
    \put(30,5){$\mathbf{p}_1$}
    \put(24,50){$\mathbf{p}_2$}
    \put(53,39){$\nhatbf_2$}
    \put(36,41){$\nhatbf_1$}
\end{overpic}
\end{center}
\caption{Kinematics of a two-point energy correlator in the process $\Phi\Phi\to \mathrm{gravitons}$. The picture shows the spatial geometry in the center of mass frame; time is suppressed. Particles $\Phi$ propagate along the $z$-axis, and we measure energy flux in the directions $\nhatbf_1,\nhatbf_2$. $\psi$ (solid green arc) is the angle between one of the detectors $\nhatbf_1$ and the beam direction, $\theta$ (solid red arc) is the angle between detectors, and $\f$ (solid blue arc) parametrizes the angle of $\nhatbf_2$ around $\nhatbf_1$ on the celestial sphere.  Figure inspired by \cite{Chang:2020qpj}.}
\label{fig:com_angles}
\end{figure}
%

\begin{figure}[ht!]
\centering
\begin{subfigure}[]{.48\textwidth}
\centering
\includegraphics[width=\linewidth]{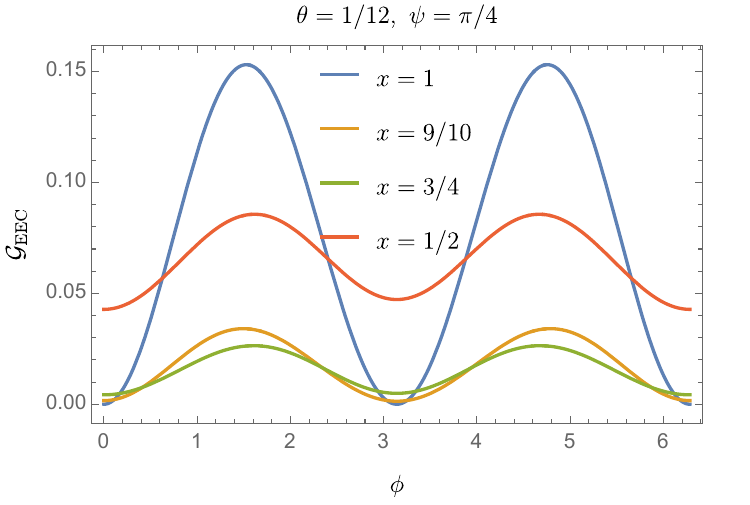}
\end{subfigure}
\quad
\begin{subfigure}[]{.48\textwidth}
\centering
\includegraphics[width=\linewidth]{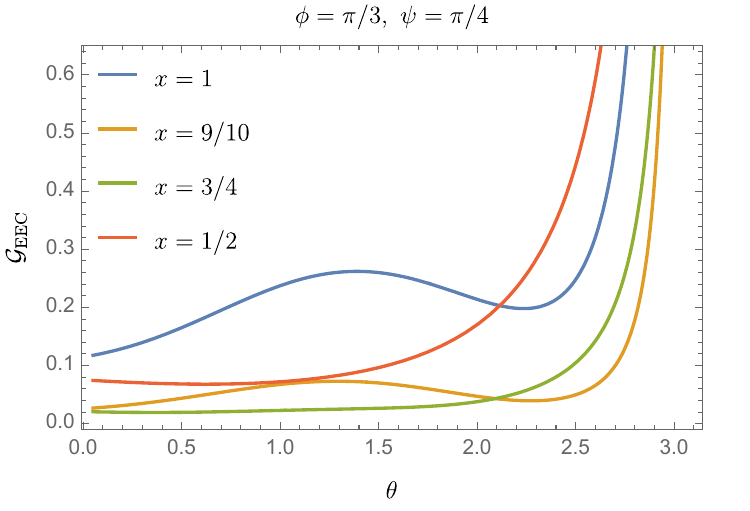}
\end{subfigure}
\begin{subfigure}[]{.48\textwidth}
\centering
\includegraphics[width=\linewidth]{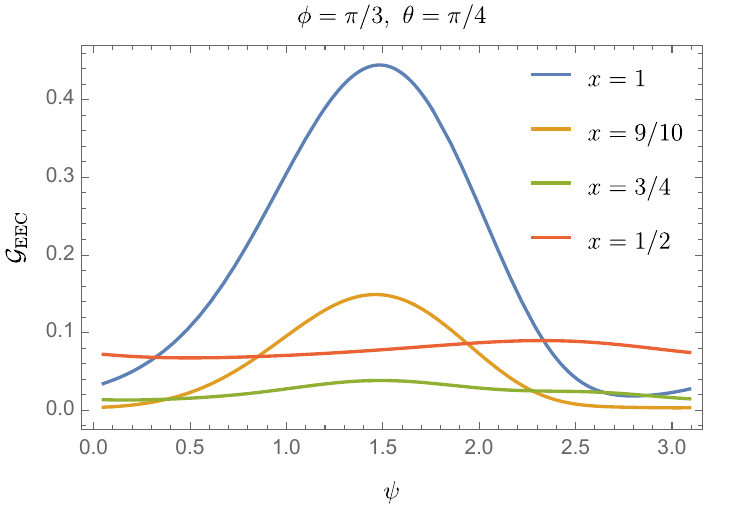}
\end{subfigure}
\quad
\begin{subfigure}[]{.48\textwidth}
\centering
\includegraphics[width=\linewidth]{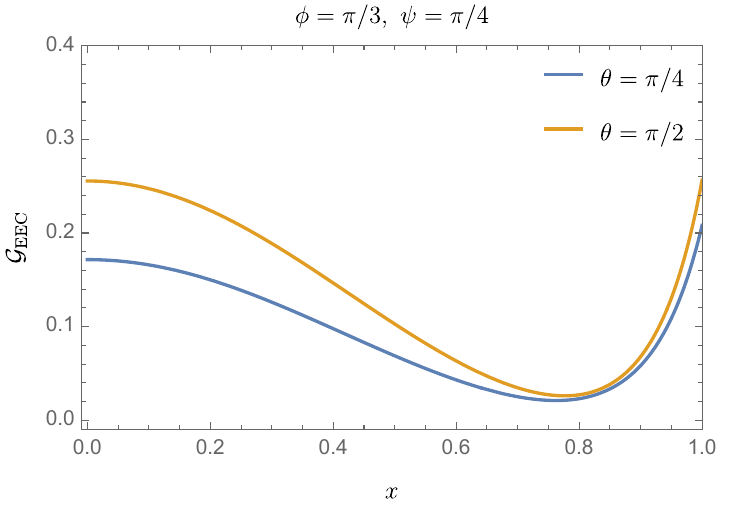}
\end{subfigure}
\begin{subfigure}[]{.48\textwidth}
\centering
\includegraphics[width=\linewidth]{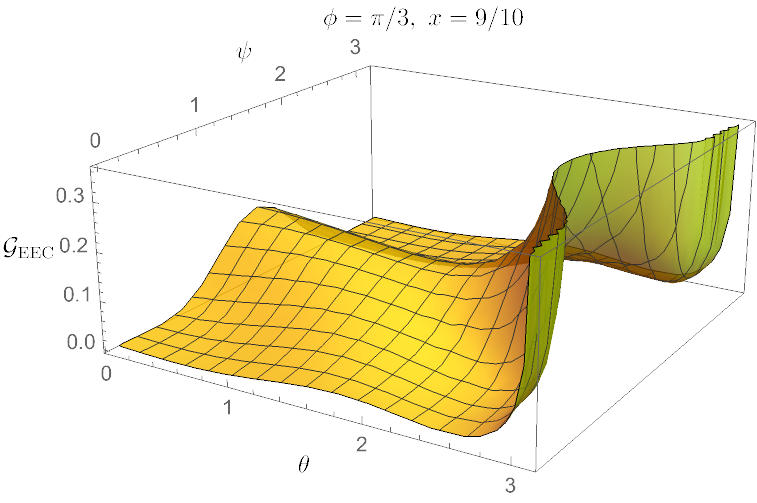}
\end{subfigure}
\ 
\begin{subfigure}[]{.48\textwidth}
\centering
\includegraphics[width=1.1\linewidth]{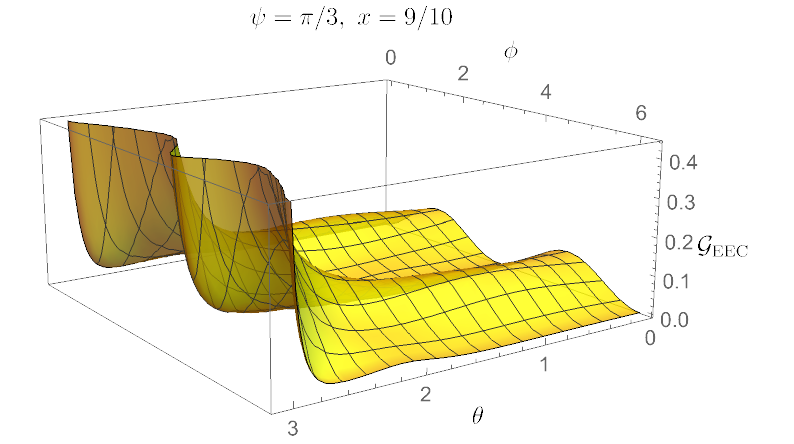}
\end{subfigure}
\caption{\label{fig:eec_plots} Plot of one-dimensional slices of $\cG_{{\rm EEC}}$, defined in \Eq{eq:eec_fi_representation}, in terms of center-of-mass angular variables \Eq{eq:eec_xrats_to_com_angles}. $\phi$ denotes the azimuthal rotation angle of detector 2 around detector 1, $\theta$ is the detector-detector opening angle, and $\psi$ denotes the angle of detector 1 from the beam axis. }
\end{figure}

With the final result of the gravitational EEC in the CoM angular variables at hand, we can plot the result in full kinematics for several interesting kinematic slices shown in \Fig{fig:eec_plots}. Here we immediately see a difference between the energy correlator in gravity and in gauge theory. In gauge theory, the perturbative energy correlator has a collinear singularity as $\theta\to 0$, and a soft singularity as $\theta\to \pi$. Due to the milder singularity structure in the collinear limit in gravity, the energy correlator exhibits a finite limit in the $\theta\to 0$ limit.  In the $\theta\to \pi$ limit, the gravity result exhibits a soft singularity. 
%
%
%
%
%

\subsection{Kinematic limits}

We can now investigate analytically the kinematic limits of the EEC. We first consider the collinear, $\theta\to 0$, limit, where find the following finite result
\begin{align}
\label{eq:eec_col_from_expansion_of_full_result}
\cG^{(1)}_{\text{EEC}}(\theta\to 0 ,\psi,\phi,x)=
g_0(\psi,x) + g_2(\psi,x) \cos (2\phi) + g_4(\psi,x) \cos (4\phi) + \cO(\theta)\,,
\end{align}
where we define the coefficient functions
\begin{align}
\begin{split}
g_0(\psi,x) & = \left(1-x^2\right)^2 
            \left(
                \frac{11 y^2}{36}
                +\frac{14 y}{9}
                +\frac{2839}{840}
                +\frac{1037}{252 y}
                +\frac{361}{105 y^2}
                +\frac{9}{5 y^3}
                +\frac{1}{2 y^4}
            \right)
\\
g_2(\psi,x)& = \left(1-x^2\right)^2 
            \left(
                -\frac{11 y^2}{36}
                -\frac{14 y}{9}
                -\frac{1003}{315}
                -\frac{459}{140 y}
                -\frac{92}{45 y^2}
                -\frac{7}{10 y^3}
            \right)
\\
g_4(\psi,x) & = \left(1-x^2\right)^2 \frac{(1+y)^2}{1260 y^2}
\end{split}
\end{align}
and the convenient parameter $y = \frac{(x^2 \cos ^2\psi -1)}{(1-x^2)}$. Note the peculiar sign pattern where each term comes with a `$+$' for $g_0(\psi,x)$ and each term comes with a `$-$' for $g_2(\psi,x)$. In the massless limit ($x=1$), $\cG^{(1)}_{\text{EEC}}$ drastically simplifies, and we find
\begin{align}
\cG^{(1)}_{\text{EEC}}(\theta\to 0 ,\psi,\phi,x=1)=
\frac{11}{18} \sin^4(\psi ) \sin^2(\phi ) + \cO(\theta)\,.
\end{align}
The azimuthal dependence is clearly seen in \Fig{fig:eec_plots}. Interestingly, in \Fig{fig:eec_plots} we only see a $2\phi$ periodicity, whereas in \Eq{eq:eec_col_from_expansion_of_full_result}, there is a $\cos(4\phi)$ in the analytic result. The fact that the  $\cos(4\phi)$ dependence is invisible in our plots is due to its small coefficient, which is numerically smaller by a factor of $10^3$ throughout the physical kinematic range. This azimuthal dependence arises from light-ray operators with transverse spin \cite{Chang:2020qpj}. The structure of transverse spin operators has been studied in detail in perturbative gauge theories \cite{Chen:2021gdk,Chen:2020adz,Chen:2022jhb,Chang:2022ryc}, and it would be interesting to extend this to the case of gravity.

Note that the behavior of the energy correlator in the collinear limit is quite distinct in gravity as compared to gauge theory. In gauge theory it exhibits a $1/\theta$ divergence associated with a collinear singularity, while in gravity it is completely regular. The amplitude level factorization properties of gravity amplitudes have been well studied. Instead of a pole, as in gauge theory, one has a phase singularity \cite{Bern:1998zm}. It would be interesting to understand the all orders structure of the energy correlator in the collinear limit using either an extension of the light-ray OPE, or the development of factorization theorems using the soft-collinear effective theory of gravity \cite{Beneke:2022pue,Beneke:2022ehj,Beneke:2021aip,Okui:2017all,Beneke:2012xa}, building on those in gauge theory \cite{Dixon:2019uzg}. 

Another interesting limit to study is in the back-to-back limit, where $\theta \to \pi$. This limit has been extensively studied in gauge theories \cite{Moult:2018jzp,Gao:2019ojf,Korchemsky:2019nzm}, where it is associated with soft physics. In gravity, we also find an expected soft divergence as we approach $\theta\to\pi$:
\begin{align}
\cG^{(1)}_{\text{EEC}}(\theta\to \pi ,\psi,\phi,x) \sim \frac{b(\psi,\phi,x)}{(\theta-\pi)^2} + \cO\left(\frac{1}{(\theta-\pi)}\right)\,,
\end{align}
where $b(\psi,\phi,x)$ is a transcendental function of the form
\begin{align}
b(\psi,\phi,x) = b_0 
                + b_1\, \text{arctanh}\left(x \cos \psi \right)
                + b_2\, \frac{\text{arctan}\left(\frac{x \cos \phi \sin\psi}{\sqrt{\widetilde{\Delta}(\psi,\phi,x)}} \right)}{\sqrt{\widetilde{\Delta}(\psi,\phi,x)}}\,,
\end{align}
$\widetilde{\Delta}(\psi,\phi,x)=1-x^2\left(\cos^2 \psi + \cos^2\phi \sin^2\psi\right)$, and the coefficient functions $b_i$ are provided in the ancillary file \texttt{eec\textunderscore curlG\textunderscore CoM\textunderscore angles\textunderscore back\textunderscore to\textunderscore back\textunderscore limit\textunderscore LO.m}.
In the massless limit ($x=1$), the result dramatically simplifies
\begin{align}
b(\psi,\phi,x=1)= \frac{s_{\psi }^4 \left(2 c_{\psi } \, \text{arctanh}\left(c_{\psi }\right) s_{\phi }^2+s_{2 \phi } \, \text{arctan}(\cot \phi )\right)}{c_{\phi }^2+\cot^2\psi }\,.
\end{align}
Here we introduced the shorthand notation $c_a \equiv \cos\alpha$ and $s_\alpha \equiv \sin \alpha$. We can further average over the azimuthal $\phi$ dependence to obtain a result of the scattering angle $\psi$ only:
\begin{align}
\begin{split}
\langle b(\psi,x{=}1)\rangle_\phi \equiv & \frac{1}{2\pi} \int^{2\pi}_0 d\phi\ b(\psi,\phi,x{=}1)
\\
=&  2\, s^4_\psi \left[
    \log \left[\frac{2}{1+|c_\psi|}\right]
    - c_\psi \,\left(\frac{|c_\psi|-1}{|c_\psi|}\right) \text{arctanh}\left(c_\psi\right) 
\right]\,,
\end{split}
\end{align}
where we also have the expression of the `arctanh' in terms of logarithms:
\begin{align}
\text{arctanh}\left(c_\psi\right) = \frac{1}{2} \log\left[\frac{1+c_\psi}{1-c_\psi}\right]\,.
\end{align}
Although the energy correlator exhibits a divergence in the back-to-back limit in quantum gravity, it is also different in structure than in a gauge theory. In a gauge theory the back-to-back limit has a double logarithmic structure governed by the cusp anomalous dimension \cite{Moult:2018jzp,Gao:2019ojf,Korchemsky:2019nzm}. In gravity it is only single logarithmic. This seems to suggest that the all orders structure in this limit should be describable in terms of a factorization theorem analogous to that in \cite{Moult:2018jzp} for gauge theories, but with only rapidity renormalization group evolution, due to the lack of a standard collinear singularity in gravity. We leave the study of this to future work.

\subsection{Generalized detector-detector correlators}
\label{sec:ddc collinear limit}

We can also compute correlators of more general detectors, by keeping the energy weights $J_{L1}$ and $J_{L2}$ generic.  This provides a concrete generalization of the EEC to $E^{J_1-1}\times E^{J_2-1}$ detectors. We will dub these more general detector correlators ``DDC''. For simplicity, we compute these only in the collinear limit, since they are primarily of interest for developing an understanding of the light-ray OPE in gravity.

Using~\eqref{eq:collinear limit integral formula for n detectors from n+1 amplitude}, or its dimensionless version
\be
\label{eq:ddc from collinear amplitude}
\cG_{J_{L1},J_{L2},\text{col.}}^{(1)}(\psi,\phi,x) &= \int_0^1 d\a_1 \a_1^{-J_{L1}-1}\left({1-\a_1}\right)^{-J_{L2}-1} \left|\cM_{2\to3}^{(0),\text{col.}}\left(\frac{p_i}{\sqrt{s}}, \frac{q_j^c}{\sqrt{s}}\right) \right|^2
\,,
\ee 
it is straightforward to compute the $\cD_{J_{L1}}(z_1)\cD_{J_{L2}}(z_2)$ two-point correlator in the collinear limit. Note that here we have chosen to parametrize the cross ratios using the CoM frame angles $\psi,\phi$, as these remain untouched by the collinear limit.  We obtain
\be
\label{eq:ddc collinear limit}
\cG_{J_{L1},J_{L2},\text{collinear}}^{(1)}(\psi,\phi,x)
&= \frac{\Gamma \left(2-J_{{L1}}\right) \Gamma \left(2-J_{{L2}}\right)}{\Gamma \left(4-J_{{L1}}-J_{{L2}}\right)} (1-x^2)^2 \frac{(1+y)^2}{2y^2} \cos 4 \phi 
\nn \\ & \quad + (1-x^2)^2 h_2(y) \cos 2\f+(1-x^2)^2 h_0(y),
\ee 
where the polynomials $h_2(y)$ and $h_0(y)$ are given in appendix~\ref{app:collinear ddc functions} and the ancillary file \texttt{ddc\textunderscore collinear.m}. Setting $J_{L1}=J_{L2}=-3$, we recover the EEC in the collinear limit, which indeed matches~\eqref{eq:eec_col_from_expansion_of_full_result}. We believe that this result will provide interesting data for understanding the light-ray OPE in perturbative quantum gravity, but we leave its exploration to future work.

\section{Conclusions}
\label{sec:concl}

In this paper we have initiated the investigation of detector operators and their correlation functions in perturbative quantum gravity. We provided a review of recent developments in the study of detector operators in field theory and pointed out how they can be generalized to the context of perturbative quantum gravity. We then introduced a family of detector operators, ``energy-to-some-power" detectors, in perturbative quantum gravity, which we believe will play an important role in understanding the renormalization properties of detector operators in CFTs coupled to gravity.

After introducing this family of detector operators, we began an exploration of their correlation functions through concrete perturbative calculations. We described in detail how these calculations can be efficiently performed using modern scattering amplitude technology, e.g. generalized unitarity and the double copy, which will be useful for a broader set of future calculations. We then performed an explicit calculation of the energy-energy correlator in Einstein gravity, in a state comprised of two massive scalars that annihilate into gravitons. Our analytic result takes a simple form and is expressed in terms of logarithms accompanied by algebraic factors. Starting from the result in generic kinematics, we are able to analyze interesting kinematic limits. In the collinear limit of the two detectors, our result is finite, due to the absence of collinear singularities in gravity compared to gauge theory. In the back-to-back limit, our result exhibits a soft singularity. Compared to gauge theory, the all orders behavior of these limits is much less understood, and it would be interesting to explore them in more detail. We also computed the generalized $E^{J_1-1} \times E^{J_2-1}$ detector two-point function in the collinear limit. These are the first results for a correlator of detector operators in quantum gravity, and will provide valuable data for the further study of light-ray operators in quantum gravity. 

This paper represents only a preliminary investigation of the properties of detector operators and their correlators in quantum gravity, leaving open many interesting future directions, including:

\paragraph{OPE of detector operators in gravity}~\\
Our perturbative calculation of the two-point correlator provides valuable data for the development of the light-ray OPE in quantum gravity. It would be particularly interesting to decompose our result into celestial blocks~\cite{Kologlu:2019mfz,Chang:2020qpj}, and extract important dynamical data such as detector OPE coefficients. (See Ref.~\cite{Chen:2021gdk} for an analogous computation in QCD.) This will also provide insight into the broader space of detector operators in quantum gravity, as it gives access to all the operators appearing in the $\mathcal{E} \mathcal{E}$ OPE. More ambitiously, it might lead to a general understanding of the light-ray OPE in CFTs coupled to gravity.

\paragraph{The space of detector operators in (quantum) gravity}~\\
It would be interesting to explore the space of detector operators in both classical and quantum gravity. Many of the classical observables considered for characterizing radiation in binary inspirals or hyperbolic encounters, such as momentum deflections, radiated energy \cite{Kosower:2018adc,Herrmann:2021lqe,Herrmann:2021tct}, waveforms \cite{Cristofoli:2021vyo,Herderschee:2023fxh,Bini:2024rsy}, and memory effects \cite{Elkhidir:2024izo} should be or have been expressed as detector operators. This perspective could help to clarify subtleties in their definitions, or aid in their calculations. For example, the gravitational waveform is naturally given by the following detector operator,
\be
\wL_\omega[h](\infty,z)=\int_{-\infty}^\infty d\a\, e^{i\a \omega}\, h(\a,z)\, .
\ee 
Such operators fall under a class of generalized detectors introduced by~\cite{Korchemsky:2021okt}. This is suggestive of an even broader class of detector operators in gravity of practical and theoretical interest even beyond the classical limit.

\paragraph{Renormalization of detectors in quantum gravity}~\\
In this paper we introduced a broad class of ``energy-to-some-power" detectors in perturbative quantum gravity. Our motivation for introducing these operators is that they do not correspond to conserved charges, allowing us to study their renormalization. Due to the absence of collinear singularities in gravity, and the presence of a dimensionful coupling $G_N$, we expect this renormalization to take a very different structure in gravity as compared with gauge theories. In particular, we do not expect a standard renormalization of the detectors, but rather a mixing between shadow trajectories and with horizontal trajectories (poles in $J$ or rapidity renormalization). This could provide interesting insights into the Regge limit in gravity, and deserves exploration.

\paragraph{Perturbative structure of multi-point correlators in supergravity}~\\
The exploration of multi-point correlators in gauge theory, particularly $\mathcal{N}=4$ SYM has revealed remarkable simplicity in the perturbative structure of physical observables. However, $\mathcal{N}=8$ supergravity is famously claimed to be the ``simplest quantum field theory" \cite{Arkani-Hamed:2008owk}, due to the remarkably simple structure of its scattering amplitudes. How does this imprint into the structure of energy correlators? Can the calculation of energy correlators in perturbative supergravity enable calculations beyond what has been achieved in gauge theories, especially in light of the available amplitude data through five loops \cite{Bern:2018jmv}?

\paragraph{Asymptotic symmetries and their interplay with detector operators}~\\
Asymptotic symmetries have an already explored interplay with certain light-ray operators, see e.g.~\cite{Gonzo:2020xza}. What is the relation between asymptotic symmetries, their generators, and more general detector operators? Does the space of detector operators inherit a representation of asymptotic symmetries?

\paragraph{Detectors in nontrivial spacetimes and nonperturbative effects in quantum gravity}~\\
It would be very interesting to ask about detector correlators in nontrivial spacetimes. For example, one could consider detector measurements in the state of an evaporating black hole. The detectors would capture outgoing Hawking radiation. Multiple detectors could be used to probe correlations in the Hawking radiation. One could envision unpacking the fine-grained structure of Hawking radiation via detector measurements. Many recent developments have seen the computation of various black hole observables, such as the black hole index in supergravity~\cite{Sen:2012kpz,Iliesiu:2021are,H:2023qko,Boruch:2023gfn,Chowdhury:2024ngg}; can Lorentzian analogues with detector insertions be computed? More generally, one could try to push beyond perturbation theory and ask what are the contributions from nonperturbative gravitational effects to detector observables. In the spirit of recent developments in low-dimensional quantum gravity in AdS, one can envision detector observables inserted at different copies of asymptotic boundaries connected via spacetime wormholes.

\paragraph{Conformal collider bounds in gravitational theories}~\\
Due to the positivity of the ANEC operator, energy correlators famously allow one to derive the so called ``conformal collider bounds" \cite{Hofman:2008ar,Cordova:2017zej} on CFT data. What can be derived using similar considerations in gravity? In QFT the ANEC is closely related to causality \cite{Hartman:2016lgu}. In gravity could it be used to derive the CEMZ bounds \cite{Camanho:2014apa}?

\paragraph{Understanding color-kinematics duality beyond scattering amplitudes}~\\
It is of interest to understand how to extend the color-kinematics duality beyond scattering amplitudes, in particular to off-shell observables. There has been progress both in the study of classical solutions (see \cite{Bern:2019prr} for a review) and form factors \cite{Boels:2017ftb,Yang:2016ear,Boels:2012ew,Boels:2017skl}. In field theory, light-ray operators and the light transform provide a link between correlation functions of local operators and asymptotic observables. As such, they may provide an interesting observable for understanding the color-kinematics duality for correlation functions. 

\paragraph{Detector operators in flat space from the flat space limit of AdS/CFT}~\\
The flat space limit of the AdS/CFT correspondence has been much studied recently~\cite{Li:2021snj,Alday:2024yyj,Kraus:2024gso}. High energy collisions in AdS probe only a small Minkowski patch embedded in the bulk of AdS. How do detector measurements in the flat space patch embed inside the AdS spacetime, and what (limit of an) observable do they correspond to on the holographic dual CFT?

\vspace{0.5cm}

The language of detector operators is particularly well suited to exploring the structure of perturbative quantum gravity. By introducing these observables into the study of quantum gravity, we hope to both learn more about detector operators, and use them to learn more about the structure of quantum gravity.  We hope to build on the first steps presented in this paper, and address some of these bigger questions in future work.

\acknowledgments
We thank Zvi Bern, Thomas Hartman, Petr Kravchuk, Gregory Korchemsky, Yue-Zhou Li, Julio Parra-Martinez, Miguel Paulos, David Simmons-Duffin, Walter Goldberger, Riccardo Gonzo, Mark Gonzalez, Michael Saavedra for useful discussions. I.M. is supported by the DOE Early Career Award DE-SC0025581.
E.H. is supported in part by the U.S. Department of Energy (DOE) under award number DE-SC0009937, and by the Mani L. Bhaumik Institute for Theoretical Physics. M.K. is supported by the Yale Mossman Prize Postdoctoral Fellowship.

\newpage
\appendix 

%
\section{Notation, conventions, and coordinate systems}
\label{app:coordinate_systems}
%

Our metric conventions are mostly-minus, $\{+,-,\dots,-\}$.  We use $x,z$ with indices $\mu,\nu$ to refer to $d$ dimensional vectors. Unless otherwise noted, $d$ dimensional index contractions are performed with the flat metric $\eta_{\mu\nu}$. We also use $\mathbf{x},\mathbf{n}$ to refer to $d-1$ dimensional spatial vectors, and $\vec n^\perp, \vec y^\perp$ to refer to $d-2$ dimensional transverse vectors.

In this appendix, we review several coordinate systems used for calculations involving detector operators. As nicely explained in e.g.~\cite{Firat:2023lbp}, instead of the time and radial coordinates used to write the energy detector in \Eq{eq:energy_detector_t_r_coords}, it is convenient to adapt a more suitable coordinate parametrization. As we imagine the state formed either by a local excitation or a scattering event centered around $t=0$ with the characteristic length scale $\sigma$, in theories with massless particles, such states spread asymptotically at the speed of light. This implies that after some time $t$, the excitation will be localized around some radius $r \sim t \pm \mathcal{O}(\sigma)$. From here, we see that the main contribution to the flux integrals in \Eq{eq:energy_detector_t_r_coords} comes from the integration region where $t\sim r\pm \cO(\sigma)$ approaches infinity.  Therefore, \emph{radial light-cone coordinates} $r^{\pm} = t\pm r,$
\begin{align}
\label{eq:radial_LC_coords}
x^\mu = 
        \begin{pmatrix}
              x^ 0 \\
              r \, \vec{n}^{\perp\phantom{d1}} \\
              r \, n^{d-1}
        \end{pmatrix}
        =
         \begin{pmatrix}
           \frac12 (r^++r^-) \phantom{n^{d-1}} \\
           \frac12 (r^+-r^-) \vec{n}^{\perp\phantom{d1}} \\
           \frac12 (r^+-r^-) n^{d-1}
        \end{pmatrix}\,,
\end{align}
provide a more suitable parametrization of the relevant physics. For later convenience, we split the unit vector $\nhatbf$ into sub-components $\nhatbf = (\vec{n}^\perp, n^{d-1})$, $\nhatbf^2 =1 \Rightarrow 1 = (\vec{n}^\perp)^2 + (n^{d-1})^2$.  For any value of $r^+$, the state is localized around a finite value of $r^- \sim \cO(\sigma)$. In these coordinates, the Minkowski metric is given by:
\begin{align}
\label{eq:metric_radial_LC_coords}
 ds^2 =  \eta_{\mu\nu} dx^\mu dx^\nu = dr^- dr^+ - \frac{1}{4}(r^+-r^-)^2 d \Omega^2_{d-2}\,,
\end{align}
where $d\Omega^2_{d-2}$ denotes the square of the line element on the $(d-2)$-dimensional celestial sphere. Note that $r^\pm$ are null coordinates that satisfy $\eta^{\mu\nu} (\grad_\mu r^{\pm}) (\grad_\nu r^\pm) =0$. In conformal theories, one can show that not only the \emph{total} charge, but also the time integrated flux is independent of the choice of surface if it approaches future null infinity $\scri^-$. In the radial light-cone coordinates, the charge and energy detector read
\begin{align}
    \mathcal{Q}(\nhatbf) & =  \frac{1}{2} \int^{\infty}_{-\infty} dr^- \
    \lim_{r^+\to \infty} \left(\frac{r^+}{2}\right)^{d-2}
    J^{\,r^+}(r^+, r^-, \nhatbf)\,,
    \\
    \mathcal{E}(\nhatbf) &= \frac{1}{2} \int^{\infty}_{-\infty} dr^- \ 
    \lim_{r^+\to \infty} \left(\frac{r^+}{2}\right)^{d-2} 
    T^{\,r^+ r^+}(r^+, r^-, \nhatbf)\,.
\end{align}
It is important to stress that for external non-localized states, like momentum eigenstates, we need to take the large $r^+$ limit first and then perform the $r^-$ integral. The opposite is usually done in CFT for localized states, for which all notions are equivalent \cite{Hofman:2008ar,Belitsky:2013xxa}.
Despite their more convenient description of the asymptotic region $r^+\to \infty$, $r^-,\nhatbf$, finite, the radial light-cone coordinates lead to the nontrivial Minkioski metric in \Eq{eq:metric_radial_LC_coords}. 

\medskip

If one is allowed to perform a conformal transformation $x^\mu \to y^{\mu}(x)$, one can set up coordinates that retain the advantageous of radial light-cone coordinates, but keep the metric simple. Starting from \emph{ordinary light-cone coordinates} defined via $x^{\pm} = x^0 \pm x^{d-1}$ and $\vec{x}^\perp= (x^1,\ldots,x^{d-2})$, (with $t \equiv x^0$) where the Minkowski metric reads
\begin{align}
\label{eq:light_cone_var_metric}
 ds^2 = dx^- dx^+ - d\vec{x}^\perp \cdot d\vec{x}^\perp\,,
\end{align}
one can perform the conformal mapping \cite{Hofman:2008ar} 
\begin{align}
 y^+ = -\frac{1}{x^+}\,, \quad
 y^- = x^- - \frac{|\vec{x}^\perp|^2}{x^+}\,, \quad
 \vec{y}^\perp = \frac{\vec{x}^\perp}{x^+}\,,
\end{align}
to the $y$-light-cone variables that are defined in analogy to $x$. With this conformal transformation, the metric in \Eq{eq:light_cone_var_metric} changes by a conformal factor 
\begin{align}
 ds^2 = \frac{1}{(y^+)^2} \left(dy^- dy^+  - d \vec{y}^\perp \cdot d\vec{y}^\perp \right)
\end{align}
and $\scri^+$ gets mapped to the null plane $y^+=0$. Parametrizing $x^\mu$ via the radial light-cone coordinates introduced in \Eq{eq:radial_LC_coords} maps the surface where $r^+=\infty$ (keeping $r^-, \hat{n}$ fixed) to the plane $y^+=0$. On this plane,
\begin{align}
\begin{split}
 x^+ & = x^0 + x^{d-1} = \frac12 (r^++r^-) +  \frac12 (r^+-r^-) n^{d-1} \stackrel{r^+ \to \infty}{\longrightarrow} \frac{r^+}{2} (1 + n^{d-1})
 \\
 \vec{x}^\perp &= \frac12 (r^+-r^-) \vec{n}^\perp \stackrel{r^+ \to \infty}{\longrightarrow} \frac{1}{2} r^+ \vec{n}^\perp
 \end{split}
\end{align}
one can show that $y^-$ and $\vec{y}^\perp$ are finite functions of $r^-$ and $\hat{n}$
\begin{align}
y^- \stackrel{r^+ \to \infty}{\longrightarrow} \frac{2 r^-}{1+n^{d-1}}\,, 
\quad
\vec{y}^\perp \stackrel{r^+ \to \infty}{\longrightarrow} \frac{\vec{n}^\perp}{1+n^{d-1}}\,,
\end{align}
where we used the fact that $\hat{n}$ is normalized $\hat{n}^2 = 1 = \left(\vec{n}^\perp\right)^2 + \left(\vec{n}^{d-1}\right)^2 $. Furthermore, the null line defined by $r^+= \infty,\, \hat{n}=\text{fixed}$ maps to the null line defined by $y^+=0, \, \vec{y}^\perp=\text{fixed}$. In the limit where $t\approx r$, i.e. $y^+ \to 0^-$, the various coordinate systems are related by
\begin{align}
  t & = \frac{1}{2} \left(\frac{1 + |\vec{y}^\perp|^2}{-y^+} + y^-\right)\,, \quad 
  && r^{\phantom{-}} = \frac{1}{2} \left(\frac{1 + |\vec{y}^\perp|^2}{-y^+} - \left[\frac{1-|\vec{y}^\perp|^2}{1+|\vec{y}^\perp|^2}\right] y^-\right)\,,
  \\
  r^+ & = t+ r = \frac{1 + |\vec{y}^\perp|^2}{-y^+} - \frac{|\vec{y}^\perp|^2}{1+|\vec{y}^\perp|^2} y^-\,, \quad 
  && r^-= t-r = \frac{y^-}{1+|\vec{y}^\perp|^2} \,,
\end{align}
and the perpendicular directions are related via $\vec{n}^\perp = \left(\frac{2}{1+|\vec{y}^\perp|^2}\right) \vec{y}^\perp$, so that the spherical measure becomes
\begin{align}
 d \Omega_{d-2} = \left(\frac{2}{1+|\vec{y}^\perp|^2}\right)^{d-2} d \vec{y}^\perp\,.
\end{align}
In the conformal coordinate system, the energy flux operators are related to light-ray operators by an integral over the null line
\begin{align}
 \mathcal{Q}(\vec{y}^\perp) = \int^{\infty}_{-\infty} dy^-\, J_{-}(y^+=0,y^-,\vec{y}^\perp)\,, \quad
 \mathcal{E}(\vec{y}^\perp) = 2 \int^{\infty}_{-\infty} dy^-\, T_{--}(y^+=0,y^-,\vec{y}^\perp)\,.
\end{align}
The relation to the original flux operators is obtained by taking into account the transformation of the measure and the currents
\begin{align}
\label{eq:detector_relation_coord_systems}
\mathcal{Q}(\hat{n}) = \left(\frac{1+|\vec{y}^\perp|^2}{2}\right)^{d-2}  \mathcal{Q}(\vec{y}^\perp)\,,
\quad
\mathcal{E}(\hat{n}) = \left(\frac{1+|\vec{y}^\perp|^2}{2}\right)^{d-1} \mathcal{E}(\vec{y}^\perp)\,.
\end{align}
A detailed discussion of these coordinate systems can be found in Appendix A of \cite{Firat:2023lbp}.


Note that Ref.~\cite{Gonzo:2020xza} starts from light-cone coordinates 
\begin{align}
\label{eq:light_cone_metric_PG}
ds^2 = dx^- dx^+ -dx^1 dx^1 - dx^2 dx^2\,,
\end{align}
where $x^\pm = x^0 \pm x^3$ and subsequently changes variables to \emph{flat null coordinates}
\begin{align}
\label{eq:flat_null_metric_PG}
ds^2 = du\, dr  - r^2 dz d\bar{z}\,.
\end{align}
In these coordinates, the null boundaries are located at $r\to \infty$ for $(u,z,\bar{z})$ held fixed, and $z,\bar{z}$ are stereographic coordinates on the celestial sphere. The two sets of coordinates in \Eqs{eq:light_cone_metric_PG}{eq:flat_null_metric_PG} are related via the transformation
\begin{align}
 x^+ = r \,, \quad
 x^- = u + r\, z\bar{z} \,, \quad
 x^1 = \frac{1}{2} r (z + \bar{z})\,,\quad
 x^2 = -\frac{1}{2} i\, r (z-\bar{z})\,.
\end{align}
Converting back to the standard coordinates, gives the following relations:
\begin{align}
x^\mu = \left(
 \frac{r}{2}\left(1+ z\bar z + \frac{u}{r}\right) , 
 \frac{1}{2} r (z + \bar{z}),
-\frac{1}{2} i\, r (z-\bar{z}),
 \frac{r}{2}\left(1- z\bar z - \frac{u}{r}\right)
 \right)\,.
\end{align}
In the gravitational context, Ref.~\cite{Gonzo:2020xza} makes use of yet another set of coordinates which are tailored for the asymptotic large $r$ expansion near $\scri^+$. In $d=4$, the \emph{Bondi coordinates} are given by 
\begin{align}
\label{eq:bondi_coords}
    x^\mu = \left( 
                v+r ,
                r \frac{w + \bar w}{1 + w \bar w}, 
                - i \, r \frac{w - \bar w}{1 + w \bar w} ,
                r \frac{1- w \bar w}{1 + w \bar w} 
            \right)\,,
\end{align}
where the retarded time is $v=t-r$. The Minkowski metric in these coordinates is
\begin{align}
 ds^2 = dv^2 + 2 dv\, dr - 2 r^2 \gamma_{w \bar w} dw d\bar w\,,
\end{align}
where the metric on the celestial sphere is given by
\begin{align}
 \gamma_{w \bar w} = \frac{2}{(1 + w \bar w)^2}\,.
\end{align}
\emph{Asymptotically flat geometries} that admit a large $r$ asymptic expansions
\begin{align}
\label{eq:bondi_metric}
 ds^2 & =   dv^2 + 2 dv\, dr - 2 r^2 \gamma_{w \bar w} dw d\bar w\  +  \\
      \Bigg\{& -\frac{2 m_B}{r} dv^2 - r C_{w w} dw^2 
        -D^w C_{ww} dv dw 
       -\frac{1}{r} \left[\frac{4}{3} N_w - \frac{1}{4} \partial_w \left(C_{ww}C^{w w}\right)\right]dvdw  + \left(w \leftrightarrow \bar w \right) + \cdots
       \Bigg\} \nn
\end{align}
beyond the leading order Minkowski spacetime are called Christodoulou-Klainerman spaces. Here one introduces  the Bondi mass aspect $m_B(v,w,\bar w)$, the shear tensor $C_{\alpha\beta}(v,w,\bar w)$, and the Bondi angular momentum aspect $(N_w(v,w,\bar w),\, N_{\bar{w}}(v,w,\bar w))$. $D^w$ is the covariant derivative on the celestial sphere. 

%
\section{Detectors in Bondi gauge}
\label{app:detectors_in_Bondi_gauge}
%

For a review of different coordinates used for taking limits in asymptotically flat spacetimes, see \App{app:coordinate_systems}. To construct the detector operators, it is convenient to use Bondi coordinates (see \Eq{eq:bondi_coords}),
\be
x^\mu = (v+r, r \nhatbf)
\ee 
The coordinate $v$ is the retarded time, and $\nhatbf$ is a $d-1$ dimensional unit vector parametrizing the celestial sphere. In $d=4$ in Bondi coordinates, one uses Poincar\'e coordinates $w,\bar w$ on the celestial $S^2$,
\be
\nhatbf = \left(\frac{w+\bar{w}}{1+w \bar{w}},- i \frac{\left(w-\bar{w}\right)}{1+w \bar{w}},\frac{1-w \bar{w}}{1+w \bar{w}}\right)\, .
\ee
The essential object is the shear tensor (see \Eq{eq:bondi_metric}), which in Bondi coordinates is related to the leading $1/r$ term of the gravitational field via
\be
C_{ww}(v,w,\bar w) = - \kappa \lim_{r\to \infty} \frac{1}{r} h_{ww} (r,v,w,\bar w)\, .
\ee
The graviton energy detector in $d=4$ then takes the form
\begin{align} \label{eq:GR_energy_detector_definition_Gonzo}
\cE_h(\nhatbf) & = \frac{1}{16\pi G_N} \int_{-\infty}^\infty dv (\partial_v C_{ww}) (\partial_v C^{ww})
              & = 2 \int^\infty_{-\infty} dv \lim_{r\to \infty} \frac{1}{r^2} (\gamma^{w \bar w})^2 \left(\partial_v h_{w w}\right)\left(\partial_v h_{\bar w \bar w}\right)\,.
\end{align}
%
%
This agrees with our expression Eq.~\eqref{eq:GR_energy_detector_definition_covariant} in Bondi gauge. 
The energy detector acts on asymptotic graviton states $|X\> = | \{p_1,\s_1\},\cdots,\{p_n,\s_n\}\>$ as in \Eq{eq:energy_detector_on_asymptotic_states}, namely
\begin{align}
    \cE(\nhatbf) \ket{X} = \sum^m_{i=1} E_i\, \delta^{d-2}(\Omega_{\khat_i} - \Omega_{\nhatbf}) \ket{X}\,.
\end{align}
Ref.~\cite{Gonzo:2020xza} used this definition to compute the two-point function of energy flux operators in the classical limit. 


It is straightforward to generalize the definition~\eqref{eq:GR_energy_detector_definition_Gonzo} to the Regge trajectory of $E^{J-1}$ detectors, as follows: 
\be\label{eq:graviton_E^J_detectors_bondi}
\cD_{J_L}(z) \propto 
\frac{1}{16 \pi G_N} \int_{-\infty}^\infty d\a_1 d\a_2 \, \psi_{J_L}(\a_1,\a_2) : C_{ww}(\a_1,z) C^{ww}(\a_2,z): \, .
\ee 
Once again, this agrees with the definition Eq.~\eqref{eq:graviton E^J detectors} in Bondi gauge.
%

%
\section{Graviton detector vertex}
\label{app:graviton detector vertex}
%

In this appendix we compute the detector vertex
\be
\<0|h_{\mu\nu}(-q) \cD_{J_L}(z) h_{\rho\s}(p)|0\>\, .
\ee
The analogous detector vertex for scalar field theory was computed in~\cite{Caron-Huot:2022eqs}, we proceed similarly. Plugging in for the detector~\eqref{eq:graviton E^J detectors} and Wick contracting, we obtain 
\be
\frac{1}{C_{J_L}} \int_{-\infty}^\infty d\a_1 d\a_2 \, \psi_{J_L}(\a_1,\a_2)\<0|h_{\mu\nu}(-q) h_{\a\b}(\a_1,z) |0\> \<0| h^{\a\b}(\a_2,z) h_{\rho\s}(p) |0\> + (\a_1\leftrightarrow \a_2).
\ee 
Noting that 
\be
h_{\mu\nu}(p) = \hat \delta(p^2) \sum_{s} \left[ \theta(-p^0) \varepsilon^{s,*}_{\mu\nu}(p) a_s(p) +\theta(p^0) \varepsilon^{s}_{\mu\nu}(p) a_s^\dagger(p) \right],
\ee
we can compute
\be
\< 0 |  h_{\mu\nu}(\a,z) h_{\rho\s}(p) |0\> = \frac{ie^{-i\frac{d\pi}{4}}}{(2\pi)^{d/2}} \Pi_{\mu\nu\rho\s}(p) \, \int_0^\infty d\b \b^{\De_h-1}  e^{-i\frac{\b \a}{2}} \delta^d(p-\b z).
\ee
With this, we can compute the detector vertex by evaluating the integrals over $\a_1$ and $\a_2$, obtaining~\eqref{eq:graviton detector vertex}.

%
\section{Detector kinematics}
\label{app:kinematics}
%

In this appendix we provide a review of the different coordinates used to parameterize our detectors. We frequently use null vectors $z^\mu$ to parametrize the angular position of detectors. These null vectors are naturally polarization vectors for underlying light-ray operators. They satisfy the following conditions
\be
z^2=0, \qquad z\sim \lambda z \qquad (\l>0).
\ee 
These redundancies can be solved in particular coordinates systems. Two common parametrizations of the null vectors are in the center of mass frame,
\be
z = (z^0,\mathbf{z}) = (1, \nhatbf), \qquad \text{with }\quad  \nhatbf^2=1,
\ee 
which we will discuss in detail below, or in terms of a null plane,
\be
z= (x^+,x^-,\vec{x}^\perp) = (1,y^2,\vec{y}^\perp)\, .
\ee
in light-cone variables with $y^2= \vec{y}^\perp \cdot \vec{y}^\perp$.
We define and frequently use the following cross ratios
\be
\label{eq:app_xratios}
\zeta_{ab}&= \frac{(z_a\cdot z_b) (P\cdot P)}{2 (P\cdot z_a) (P\cdot z_b)}\,,
\\ 
\chi_a &= -\frac{(Q\cdot z_a)}{(P\cdot z_a)} \,,
\ee
where $P^\mu = (p_1+p_2)^\mu$, and $Q^\mu =(p_1-p_2)^\mu$.

\paragraph{Center of mass frame.}
%
It is convenient to define the center of mass frame for the two particle incoming state. In standard Minkowski coordinates $(x^0,x^1,\dots,x^{d-1})$ we have
\be
p_1^\mu=\frac{\sqrt{s}}{2} (1,x\,\mathbf{\hat n}_0)\, ,
\qquad 
p_2^\mu=\frac{\sqrt{s}}{2} (1,-x\, \mathbf{\hat n}_0)\, ,
\ee 
where $\nhatbf_0$ is the unit vector characterizing the beam direction, and $x=\sqrt{s-4m^2}/\sqrt{s}$ is a dimensionless variable. This implies
\be
P^\mu  = \sqrt{s}(1,\mathbf{ 0})\,,
\qquad 
Q^\mu  = \sqrt{s} (0,x\, \mathbf{\hat n}_0)\,.
\ee 
In the center of mass frame, it is convenient to choose the detector positions as 
\be
z_a = (1,\mathbf{\hat n}_a) \, ,
\ee 
where $\mathbf{\hat n}_a$ are unit $d-1$ vectors parametrizing points on the $(d-2)$-dimensional celestial sphere. This has the physical interpretation that $\mathbf{\hat n}_a$ are the angles of the detectors on the celestial sphere. For the following, we pick $d=4$ for concreteness. 

There are two sets of coordinates on the celestial sphere that are particularly useful. We can parametrize points on the celestial sphere using Poincar\'e coordinates,
\be
z_a = (1,\nhatbf_a), 
\quad \text{with} \quad 
\nhatbf_a = \left(\frac{\bar{w}_a+w_a}{1+w_a \bar{w}_a},-\frac{i \left(w_a-\bar{w}_a\right)}{1+w_a \bar{w}_a},\frac{1-w_a \bar{w}_a}{1+w_a \bar{w}_a}\right)\, ,
\ee 
where $w_a,\bar w_a\in \C$ are complex conjugates. Alternatively, we can simply use angular coordinates:
\be
\nhatbf_a = (\sin \theta_a  \cos \phi_a ,\sin \theta_a  \sin \phi_a ,\cos \theta_a  )\,.
\ee
For our purposes, it will be convenient to use a combination of both sets of coordinates. 

Let us first determine the relative coordinates of the detector system and take into account the relative rotation with respect to the beam axis $\nhatbf_0$ afterwards. For the system of detectors we will use Poincar\'e coordinates. In terms of the complex conjugate parameters $(w_a,\bar{w}_a)$, we can express the cross-ratios in \Eq{eq:app_xratios} as
\be
\zeta_{ab} = \frac{\left(w_a-w_b\right) \left(\bar{w}_a-\bar{w}_b\right)}{\left(1+ w_a \bar{w}_a\right) \left(1+w_b \bar{w}_b\right)}\,.
\ee
Using translations on the celestial sphere, we choose $w_1=\bar w_1 =0$ for the relative position of the first detector, and $w_2=\bar w_2= \sqrt{\zeta}/\sqrt{1-\zeta}$ for the second detector. This choice ensures that $\zeta_{12}=\zeta$. Adding a third detector,
\be
w_3 = \frac{\sqrt{\zeta}}{\sqrt{1-\zeta}} z \, , \qquad
\bar w_3 = \frac{\sqrt{\zeta}}{\sqrt{1-\zeta}} \bar z \, .
\ee 
we rescaled $(w_3,\bar{w}_3)$ in order to have a single overall detector separation variable $\zeta$ that can be used to access e.g.~the multi-collinear limit.\footnote{We hope it is clear from the context that the celestial sphere coordinates $z,\bar{z}$ are \emph{not} to be confused with the four-vectors defined around \Eq{eq:covariant_basis_def}.} In summary, we have
\be
\nhatbf_1 &= (0,0,1), \nn \\ 
\nhatbf_2 &= (2\sqrt{\zeta(1-\zeta)},0,1-2\zeta),\nn \\ 
\nhatbf_3 &= \left( 
\tfrac{\sqrt{1-\zeta } \sqrt{\zeta } (z+\bar{z})}{1-\zeta  \left(1-z \bar{z}\right)},
i \tfrac{\sqrt{1-\zeta} \sqrt{\zeta }  \left(z-\bar{z}\right)}{1-\zeta  \left(1-z \bar{z}\right)},
\tfrac{1-\zeta  \left(1+z \bar{z}\right)}{1-\zeta  \left(1-z \bar{z}\right)}
\right).
\ee
In these variables, the three detector system is parametrized by $\zeta,z,\bar z$, with
\be
\zeta_{12}= \zeta,\qquad \zeta_{23}=\frac{(1-z)(1-\bar z)}{1-(1-z \bar z)\zeta} \zeta (1-\zeta), \qquad \zeta_{31}=\frac{z \bar z}{1-(1-z \bar z)\zeta} \zeta \, .
\ee 
In particular, the triple-collinear limit is taken as $\zeta\to0$, with
\be
\zeta_{12}= \zeta,\qquad \zeta_{23}=(1-z)(1-\bar z) \zeta + \cO(\zeta^2), \qquad \zeta_{31}=z \bar z \zeta + \cO(\zeta^2) \, .
\ee 
It is also useful to define the angle between detectors 1 and 2 as $\cos \theta = \nhatbf_1 \cdot \nhatbf_2$, so that 
\be
\label{eq:app_zeta_for_theta}
\zeta = \frac{1-\cos\theta}{2} = \sin^2(\theta/2) \, .
\ee
Having specified the relative coordinates of our detectors, we are left with orienting the entire detector system relative to the beam axis $\nhatbf_0$. We can do this by rotating the beam axis by angles $(\psi,\phi)$ relative to the first detector:
\be
\nhatbf_0 =  R_{xy}(-\phi)  \cdot R_{zx}(-\psi) \cdot \nhatbf_1 = \left( - \sin \psi \cos \phi ,\sin \psi \sin \phi,\cos \psi \right)\,.
\ee
Alternatively, we can fix the beam axis to $\nhatbf'_0 = \mathbf{\hat z}$, and rotate the system of detectors
\be
\nhatbf'_0 &= (0,0,1) \, ,
\nn \\ 
\nhatbf'_1 &= R_{zx}(\psi) \cdot R_{xy}(\phi) \cdot \nhatbf_1 
             = R_{zx}(\psi) \cdot \nhatbf'_0 \, ,
\nn \\ 
\nhatbf'_2 &= R_{zx}(\psi)  \cdot  R_{xy}(\phi) \cdot \nhatbf_2 
        = R_{zx}(\psi)\cdot R_{xy}(-\phi) \cdot R_{zx}(\theta)\cdot \nhatbf'_0 \, ,
\nn \\ 
\nhatbf'_3 &= R_{zx}(\psi)  \cdot  R_{xy}(\phi) \cdot  \nhatbf_3 \, 
\ee
Clearly, the $\nhatbf$ and $\nhatbf'$ systems are related by an overall rotation of the celestial sphere and are equivalent. The three-detector system together with the two-particle initial state is therefore characterized by five variables:~$\phi, \psi$, and $\zeta , z, \bar{z}$. 

Having discussed the beam axis, we can now evaluate the remaining cross ratios $\chi_a$ of \Eq{eq:app_xratios}
\be
\chi_1 &= x \cos \psi \, ,
\nn \\ 
\chi_2 &=
x \left((1-2 \zeta ) \cos \psi -2 \sqrt{1-\zeta } \sqrt{\zeta } \sin \psi \cos \phi \right) \,, 
\nn \\ 
\chi_3 &= x\left( \cos \psi - \frac{\sqrt{1-\zeta } \sqrt{\zeta } \sin \psi  \left(\left(z+\bar{z}\right) \cos \phi +i \left(z-\bar{z}\right) \sin \phi \right)}{1-\zeta  \left(1-z \bar{z}\right)}\right)\, .
\ee
Alternatively, we can trade $\zeta$ for $\theta$, using \Eq{eq:app_zeta_for_theta} and the
relation $2 \sqrt{1 - \zeta} \sqrt{\zeta} = \sin \theta$.

\section{Higher-point correlators}
 \label{app:multi-pt-kinematics}
%
%
Similar to the discussion of the 2-point correlator in \Sec{subsubsec:2pt_correlator}, we can determine the leading finite angle contribution to a $n$-point correlator, involving amplitudes with $m=n+1$ outgoing legs. For three-particle correlators we can set $n=3$ at the end as necessary. Starting from the general detector formula in \Eq{eq:observable_scattering_initial_state_def} and focusing on the first correction due to the additional exchange of a massless state, we discuss
\begin{align}
\mathbb{E}^{(1)}_n = \vcenter{\hbox{\genDetectorOne}} \,.
\end{align}
Repeating the steps above, we immediately have the following integral over the square of the underlying $k\to n+1$ amplitude;
\be
\mathbb{E}^{(1)}_n(p_i,z_a) &= \frac{1}{N_k} 
\int 
\prod_{i=1}^{n+1} \hat{d}^d q_i 
\prod_{a=1}^n V_{J_{La}}(q_a,z_a)  
\hat\delta^+(q_{n+1}^2)  \hat{\delta}^{d}(P-\sum^{n+1}_{j=1}q_j) 
\big|\cM_{k\to n+1}(p_i, q_j)\big|^2
\nn \\
&= \frac{1}{N_k \, (2\pi)^{nd}}
\int_0^\infty \prod^n_{a=1} d\b_a  \, \b_a^{-J_{La}-1} \, 
\hat\delta^+((P-\sum^n_{a=1} \b_a z_a)^2) \nn 
\\ &\hspace{2cm}
\times \left|\cM_{k\to n+1}\left(p_i,\, q_a {=} \beta_a z_a,\, q_{n+1}{=} P{-}\sum_a \b_a z_a\right)\right|^2 \, .
\ee 
Changing to $\a_a$ variables, we have
\be
\label{eq:gen_npt_detector_final}
\mathbb{E}^{(1)}_n(p_i,z_a) &= \frac{1}{N_k\, (2\pi)^{nd-1}} \frac{1}{P\cdot P}
\prod^n_{a=1} \left[\frac{2P\cdot z_a }{P\cdot P}\right]^{J_{La}}  
 \\ & 
\hspace{-1cm}
\times \int_0^\infty \prod^n_{a=1} d\a_a \, \a_a^{-J_{La}-1} \,
\theta(2{-}\sum^n_{a=1}\a_a) \, 
\delta\left(1{-}\sum^n_{a=1}\a_a{+}\sum^n_{a<b} \a_a\a_b\,  \zeta_{ab} \right) 
\big|\cM_{k\to n+1}(p_i, q_j^{**})\big|^2\, . \nn
\ee 
The amplitude squared is evaluated at the momenta
\be
q_a^{**} &= \frac{P\cdot P}{2P\cdot z_a} \a_a z_a \,,
\qquad \qquad 
q_{n+1}^{**} = P-\sum_{a=1}^n q_a^{**}\,.
\ee
Performing one of the $\a_a$ integrals, say $\a_n$, using the delta function fixes 
\be
\a_n = \a_n^*(\a_a) 
\equiv 
\frac{1-\!\!\sum\limits_{1\leq a<n} \!\! \a_a + \!\!\!\!\sum\limits_{1\leq a<b<n} \!\!\!\! \a_a \a_b\, \zeta_{ab}}{1-\!\!\sum\limits_{1\leq a<n} \!\! \a_a \, \zeta_{an}}.
\ee 
These integrals are in general much more complicated than in the collinear limit, as discussed in \Sec{sec:collinear}. The three-point energy correlator for generic kinematics has been computed in both QCD and $\mathcal{N}=4$ SYM \cite{Yan:2022cye,Yang:2022tgm}.

%
\section{Expressions for the DDC in the collinear limit}
\label{app:collinear ddc functions}
%
In section~\ref{sec:ddc collinear limit}, we computed the collinear limit of detector-detector correlators (DDC) of $\cD_{J_L}$ detector. Here we present the functions $h_2(y)$ and $h_0(y)$ that appear in the result~\eqref{eq:ddc collinear limit}, 
\be
h_2(y) = 
-\frac{
(1+y)}{2 y^3} 
&\Bigg[ \frac{ \Gamma \left(-J_{{L1}}-2\right) \Gamma \left(-J_{{L2}}-2\right)}{\Gamma \left(-J_{{L1}}-J_{{L2}}-4\right)}\left(y^4+4 y^3+6 y^2+4 y+2\right)
\nn \\ & \quad -\frac{ \Gamma \left(-J_{{L1}}-1\right) \Gamma \left(-J_{{L2}}-2\right)}{\Gamma \left(-J_{{L1}}-J_{{L2}}-3\right)} 2 \left(2 y^4+7 y^3+9 y^2+5 y+2\right)
\nn \\ & \quad + \frac{\Gamma \left(-J_{{L1}}\right) \Gamma \left(-J_{{L2}}-2\right)}{\Gamma \left(-J_{{L1}}-J_{{L2}}-2\right)} \left(14 y^4+42 y^3+45 y^2+20 y+6\right)
\nn \\ & \quad -\frac{ \Gamma \left(1-J_{{L1}}\right) \Gamma \left(-J_{{L2}}-2\right)}{\Gamma \left(-J_{{L1}}-J_{{L2}}-1\right)} 2 \left(14 y^4+35 y^3+29 y^2+8 y+2\right)
\nn \\ & \quad +\frac{ \Gamma \left(2-J_{{L1}}\right) \Gamma \left(-J_{{L2}}-2\right)}{\Gamma \left(-J_{{L1}}-J_{{L2}}\right)}\left(35 y^4+70 y^3+37 y^2-6 y+2\right)
\nn \\ & \quad -\frac{ \Gamma \left(3-J_{{L1}}\right) \Gamma \left(-J_{{L2}}-2\right)}{\Gamma \left(-J_{{L1}}-J_{{L2}}+1\right)}2 y \left(14 y^3+21 y^2+2 y-14\right)
\nn \\ & \quad +\frac{ \Gamma \left(4-J_{{L1}}\right) \Gamma \left(-J_{{L2}}-2\right)}{\Gamma \left(-J_{{L1}}-J_{{L2}}+2\right)}2 y \left(7 y^3+7 y^2-4 y-14\right)
\nn \\ & \quad -\frac{ \Gamma \left(5-J_{{L1}}\right) \Gamma \left(-J_{{L2}}-2\right)}{\Gamma \left(-J_{{L1}}-J_{{L2}}+3\right)}4 (y-2) y \left(y^2+2 y+2\right)
\nn \\ & \quad + \frac{\Gamma \left(6-J_{{L1}}\right) \Gamma \left(-J_{{L2}}-2\right)}{\Gamma \left(-J_{{L1}}-J_{{L2}}+4\right)}(y-2) y \left(y^2+2 y+2\right)
\Bigg]\,,
\ee
and 
\be
h_0(y) = \frac{
1}{4 y^4}
&\Bigg[
\frac{ \Gamma \left(-J_{{L1}}-2\right) \Gamma \left(-J_{{L2}}-2\right)}{\Gamma \left(-J_{{L1}}-J_{{L2}}-4\right)} \left(2 y^2+2 y+1\right) \left(y^4+4 y^3+6 y^2+4 y+2\right)
\nn \\ & \quad -\frac{ \Gamma \left(-J_{{L1}}-1\right) \Gamma \left(-J_{{L2}}-2\right)}{\Gamma \left(-J_{{L1}}-J_{{L2}}-3\right)} 4 \left(2 y^5+9 y^4+16 y^3+14 y^2+7 y+1\right) y
\nn \\ & \quad + \frac{ \Gamma \left(-J_{{L1}}\right) \Gamma \left(-J_{{L2}}-2\right)}{\Gamma \left(-J_{{L1}}-J_{{L2}}-2\right)} 14 \left(2 y^4+8 y^3+12 y^2+8 y+3\right) y^2
\nn \\ & \quad -\frac{ \Gamma \left(1-J_{{L1}}\right) \Gamma \left(-J_{{L2}}-2\right)}{\Gamma \left(-J_{{L1}}-J_{{L2}}-1\right)} 4 \left(14 y^5+49 y^4+62 y^3+31 y^2+9 y-2\right) y
\nn \\ & \quad +\frac{ \Gamma \left(2-J_{{L1}}\right) \Gamma \left(-J_{{L2}}-2\right)}{\Gamma \left(-J_{{L1}}-J_{{L2}}\right)} \left(70 y^5+210 y^4+227 y^3+100 y^2+44 y-4\right) y
\nn \\ & \quad -\frac{ \Gamma \left(3-J_{{L1}}\right) \Gamma \left(-J_{{L2}}-2\right)}{\Gamma \left(-J_{{L1}}-J_{{L2}}+1\right)} 4 \left(14 y^4+35 y^3+33 y^2+17 y+12\right) y^2
\nn \\ & \quad +\frac{\Gamma \left(4-J_{{L1}}\right) \Gamma \left(-J_{{L2}}-2\right)}{\Gamma \left(-J_{{L1}}-J_{{L2}}+2\right)} 2 \left(14 y^4+28 y^3+29 y^2+30 y+22\right) y^2 
\nn \\ & \quad -\frac{ \Gamma \left(5-J_{{L1}}\right) \Gamma \left(-J_{{L2}}-2\right)}{\Gamma \left(-J_{{L1}}-J_{{L2}}+3\right)} 4 \left(2 y^4+2 y^3+3 y^2+8 y+6\right) y^2
\nn \\ & \quad +\frac{ \Gamma \left(6-J_{{L1}}\right) \Gamma \left(-J_{{L2}}-2\right)}{\Gamma \left(-J_{{L1}}-J_{{L2}}+4\right)} \left(2 y^4+2 y^3+3 y^2+8 y+6\right) y^2
\Bigg]\,.
\ee
\bibliographystyle{JHEP}
\bibliography{eec_refs.bib}


\end{document}